\documentclass[9pt,bestpractices]{livecoms}

\usepackage{lipsum} 
\usepackage[version=4]{mhchem}
\usepackage{siunitx}
\usepackage{url}
\DeclareSIUnit\Molar{M}
\usepackage[italic]{mathastext}

\usepackage[colorinlistoftodos]{todonotes}
\usepackage[most]{tcolorbox}
\usepackage{enumitem,amssymb}
\usepackage{textgreek}
\usepackage{changepage}

\newcommand{\versionnumber}{1.0} 
\newcommand{\githubrepository}{\url{https://github.com/michellab/alchemical-best-practices}} 
\newcommand{\expect}[1]{\left\langle{#1}\right\rangle}
\title{Best Practices for Alchemical Free Energy Calculations [Article v \versionnumber]}
\author[1*]{Antonia S. J. S. Mey}
\author[7]{Bryce K. Allen}
\author[2]{Hannah E. Bruce Macdonald}
\author[2*]{John D. Chodera}
\author[1,10]{Maximilian Kuhn}
\author[1]{Julien Michel}
\author[3*]{David L. Mobley}
\author[11]{Levi N. Naden}
\author[4]{Samarjeet Prasad}
\author[2,8]{Andrea Rizzi}
\author[1]{Jenke Scheen}
\author[6*]{Michael R. Shirts}
\author[9]{Gary Tresadern}
\author[7]{Huafeng Xu}
\affil[1]{EaStCHEM School of Chemistry, David Brewster Road, Joseph Black Building, The King's Buildings, Edinburgh, EH9 3FJ, UK}
\affil[2]{Computational and Systems Biology Program, Sloan Kettering Institute, Memorial Sloan Kettering Cancer Center, New York NY, USA}
\affil[3]{Departments of Pharmaceutical Sciences and Chemistry, University of California, Irvine, USA}
\affil[4]{National Institutes of Health, Bethesda, MD, USA}
\affil[6]{University of Colorado Boulder, Boulder, CO, USA}
\affil[7]{Silicon Therapeutics, Boston, MA, USA}
\affil[8]{Tri-Institutional Training Program in Computational Biology and Medicine, New York, NY, USA}
\affil[9]{Computational Chemistry, Janssen Research \& Development, Turnhoutseweg 30, Beerse B-2340,Belgium}
\affil[10]{Cresset, Cambridgeshire, UK}
\affil[11]{Molecular Sciences Software Institute, Blacksburg VA, USA}
\corr{antonia.mey@ed.ac.uk}{ASJSM}
\corr{john.chodera@choderalab.org}{JDC}
\corr{dmobley@mobleylab.org}{DLM}
\corr{michael.shirts@colorado.edu}{MRS}

\orcid{Antonia S. J. S. Mey}{0000-0001-7512-5252}
\orcid{Bryce Allen}{0000-0002-0804-8127}
\orcid{Hannah E. Bruce Macdonald}{0000-0002-5562-6866}
\orcid{John D. Chodera}{0000-0003-0542-119X}
\orcid{Maximilian Kuhn}{0000-0002-2811-3934}
\orcid{Julien Michel}{0000-0003-0360-1760}
\orcid{David L. Mobley}{0000-0002-1083-5533}
\orcid{Levi N. Naden}{0000-0002-3692-5027}
\orcid{Samarjeet Prasad}{0000-0001-8320-6482}
\orcid{Andrea Rizzi}{0000-0001-7693-2013}
\orcid{Jenke Scheen}{0000-0001-9781-0445}
\orcid{Michael R. Shirts}{0000-0003-3249-1097}
\orcid{Gary Tresadern}{0000-0002-4801-1644}
\orcid{Huafeng Xu}{0000-0001-5447-0452}

\blurb{This LiveCoMS document is maintained online on GitHub at \githubrepository; to provide feedback, suggestions, or help improve it, please visit the GitHub repository and participate via the issue tracker.}
\pubDOI{10.XXXX/YYYYYYY}
\pubvolume{<volume>}
\pubyear{<year>}
\articlenum{<number>}
\datereceived{Month, Day, Year}
\dateaccepted{Month, Day, Year}

\begin{document}

\begin{frontmatter}
\maketitle
\begin{abstract}
Alchemical free energy calculations are a useful tool for predicting free energy differences associated with the transfer of molecules from one environment to another.
The hallmark of these methods is the use of "bridging" potential energy functions representing \emph{alchemical} intermediate states that cannot exist as real chemical species. The data collected from these bridging alchemical thermodynamic states allows the efficient computation of transfer free energies (or differences in transfer free energies) with orders of magnitude less simulation time than simulating the transfer process directly. 

While these methods are highly flexible, care must be taken in avoiding common pitfalls to ensure that computed free energy differences can be robust and reproducible for the chosen force field, and that appropriate corrections are included to permit direct comparison with experimental data.

In this paper, we review current best practices for several popular application domains of alchemical free energy calculations, including relative and absolute small molecule binding free energy calculations to biomolecular targets.
\end{abstract}
\end{frontmatter}



\section{What are alchemical free energy methods?}
\label{sec:intro}
Alchemical free energy calculations compute free energy differences associated with transfer processes, such as the binding of a small molecule to a receptor, the transfer of a small molecule from an aqueous to apolar phase~\cite{zwanzig1954hightemperature}, or the effects of protein side chain mutations on binding affinities or thermostabilities. 
 These calculations use non-physical\footnote{Here, the non-physical nature of the transformation is referred to as "alchemical", a term coined by Tembre and McCammon in Ref.~\cite{tembre1984ligandreceptor}.} intermediate states in which the chemical identity of some portion of the system (such as a small molecule ligand or protein sidechain) is changed by modifying the potential governing the interactions with the environment for the atoms being modified, inserted, or deleted. 

\begin{figure}
    \includegraphics[width=0.95\linewidth]{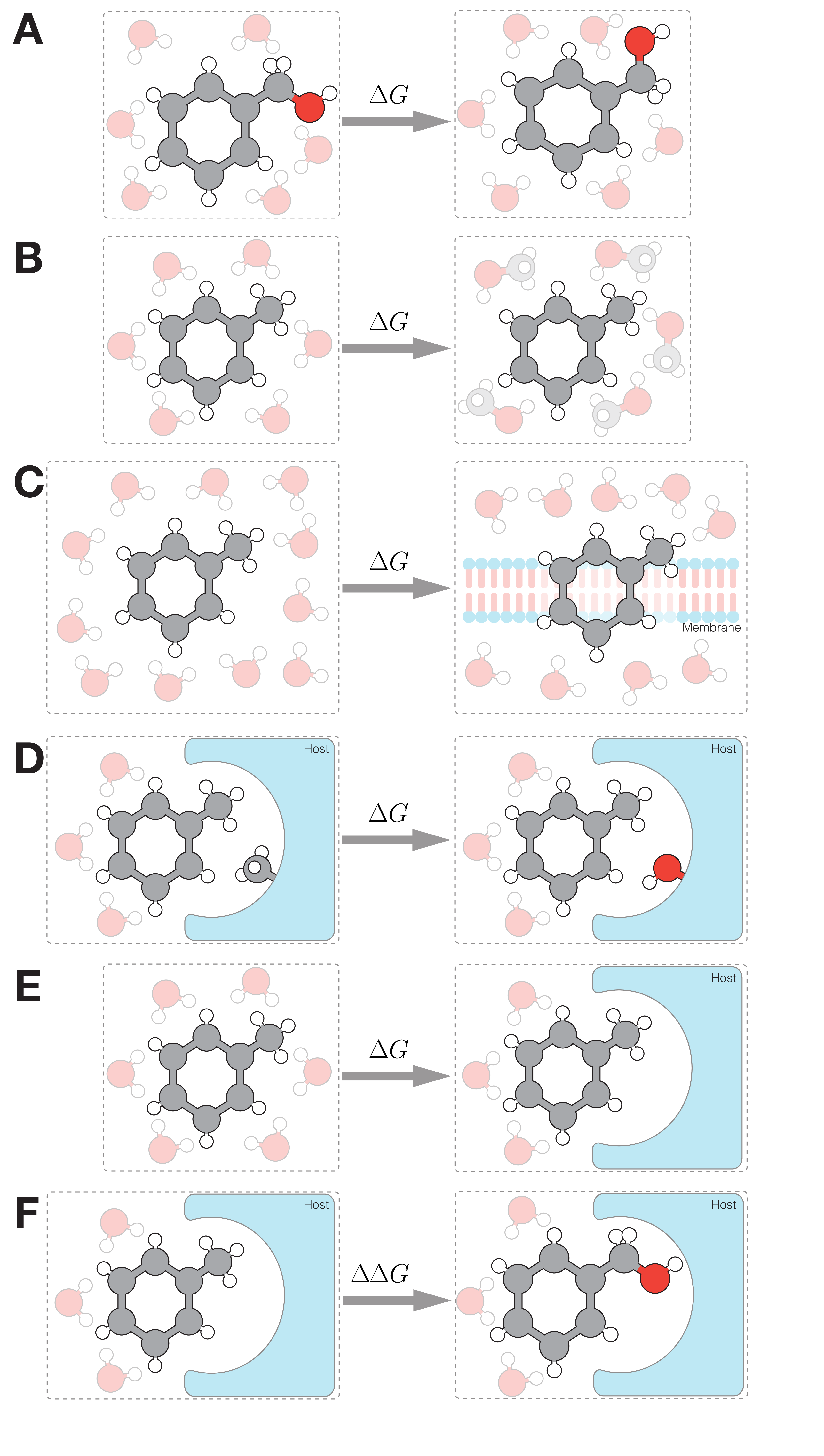}   
    \caption{\textbf{Illustration of common types of free energies differences that can be calculated using alchemical free energy methods.} \textbf{A}: Change in free energy due to a conformational change of the molecule across a high barrier. \textbf{B}: Partition coefficient such as $\log P$ or $\log D$ depend on a change in free energy between different phases; here, as an example the partition coefficient between methanol and water is shown. \textbf{C}: Free energy difference associated with the insertion of a molecule into a membrane. \textbf{D}: Effect of mutations of protein or host residues on free energies of binding. \textbf{E}: Absolute free energies of binding of a small molecule to a host (e.g. protein), \textbf{F}: Relative free energy of binding of one molecule with respect to another, here toluene and benzyl alcohol.
    \label{fig:fig_what_is_alchemy}
    }
\end{figure}

Fig.~\ref{fig:fig_what_is_alchemy} illustrates common free energy changes that may be difficult to compute with unbiased molecular dynamics methods, but are more tractable with alchemical methods.
In alchemical simulations, the introduction of intermediate \textit{alchemical states} that bridge the high-probability regions of configuration space between two physical endstates of interest, permits the robust computation of free energy for large transformations.
Alchemical calculations can be used in a variety of scenarios, such as: 
\begin{itemize}
\item computing the free energy of a conformational change for a molecule with a high barrier to interconversion (Fig.~\ref{fig:fig_what_is_alchemy} \textbf{A});
\item computing partition ($\log P$) or distribution ($\log D$) coefficients between environments (Fig.~\ref{fig:fig_what_is_alchemy} \textbf{B})~\cite{rustenburg2016measuring, bosisio2016blinded} 
\item determining partitioning between compartments into membranes (Fig.~\ref{fig:fig_what_is_alchemy} \textbf{C})~\cite{corey2019insights}. 
\end{itemize}

Furthermore, alchemical calculations are frequently used to estimate changes in free energies upon modifying a ligand or protein: 
\begin{itemize}
\item a protein residue can be alchemically mutated to probe the impact on binding affinity (Fig.~\ref{fig:fig_what_is_alchemy} \textbf{D})\cite{hauser2018predicting,aldeghi2018accurate} or changes in protein thermostability~\cite{seeliger2010protein,gapsys2016insights,gapsys2016accurate,aldeghi2019accurate}; 
\item the entire ligand can be alchemically transferred from protein to solvent in an absolute binding free energy calculation (Fig.~\ref{fig:fig_what_is_alchemy} \textbf{E})~\cite{mobley2007predicting,aldeghi2015accurate,aldeghi2017predictions}; 
\item small alchemical modifications can be made between chemically related ligands to estimate relative differences in binding free energies (Fig.~\ref{fig:fig_what_is_alchemy} \textbf{F})~\cite{wang2015accurate,mey2016blinded,song2019using,gapsys2020large,kuhn2020assessment}.
\end{itemize}

After an alchemical calculation is performed (which generally involves multiple simulations at a variety of alchemical states), the data must be analyzed to compute an estimate of the free energy for the transformation of interest.
 Early work used simple but statistically suboptimal estimators for this: free energy perturbation (FEP) used a simple (but highly biased) estimator based on the Zwanzig relation~\cite{zwanzig1954hightemperature} or numerical quadrature via thermodynamic integration (TI), for which the theory dates back the better part of a century but with the first computational applications emerging in the 1980's and 90's~\cite{kirkwood1935statistical, jorgensen1985monte, kollman1993free, wong1986dynamicsa, merz1989free}. %
 More recent developments have seen new, highly efficient statistical estimators that make better use of all the data, often building on the more efficient and less biased Bennett acceptance ratio (BAR)~\cite{bennett1976efficient}, producing multistate generalizations~\cite{shirts2008statisticallya} or removing the need for global equilibrium~\cite{wu2016multiensemble, mey2014xtram, wu2014statistically}.

Subsequent work in the 2000s led to improved implementations of alchemical methods in popular biomolecular simulation packages~\cite{shirts2003extremely,shirts2005solvation,vanderspoel2005gromacs, mermelstein2018fast, wang2015accurate, hedges2019biosimspace}. 
 This foundational work, combined with the methodological, technological, and hardware improvements of the last 5--10 years, has led to an explosion of interest and direct commercial application of these technologies~\cite{wang2015accurate, fratev2019improved, schindler2020largescale, cournia2017relative, sherborne2016collaborating, kuhn2020assessment}.

As the field of molecular simulation can now routinely access microsecond timescales with the aid of GPUs~\cite{salomon-ferrer2013routine}, and millisecond timescales appear to soon be within reach, accurate alchemical calculations on even more challenging problems will become reasonable to perform. 
In the meantime, today's users may find it difficult to get started with these complex calculations whilst also keeping up with the fast pace of change. 
This Best Practices guide provides current recommendations and tips for users of all experience. Updates and suggestions are welcomed via our GitHub repository.    

\section{Prerequisites and Scope}
\label{sec:pre}
This Best Practices guide focuses on providing a good starting point for new practitioners and a reference for experienced practitioners. 
 For this propose we provide a convenient checklist (Sec.~\ref{sec:checklist}) to help ensure all calculations comply with currently-understood best practices for alchemical simulation and analysis. Where the best practices are currently not certain, we highlight areas where further research is needed to identify an unambiguous recommendation.
 It can also serve as a set of best practices to ensure simulation robustness and reproducibility which reviewers may wish to consider as they evaluate papers.

 We assume that novice practitioners have at least moderate experience with molecular simulation concepts and use of simulation packages. 
 Furthermore, basic familiarity with the principles of molecular mechanics, molecular dynamics simulations, statistical mechanics, and the biophysics of protein-ligand association are essential. If you feel unfamiliar with some of these concepts, good starting points can be found in these references~\cite{braun2019best, grossfield2018best, klimovich2015guidelines, shirts2012best}. 

 While reading this Best Practices guide, it is important to bear in mind \emph{this is not a review} of all free energy calculation methods at the cutting edge of current research.
Instead this guide aims to answer the following questions:
\begin{itemize}
    \item Is my problem suitable for an alchemical calculation? 
     \item How do I select an appropriate alchemical protocol? 
     \item What software tools are available to perform alchemical calculations? 
     \item How should I analyze my data and report uncertainties? 
\end{itemize}

Some other background information may be needed depending on the nature of the alchemical project. For example, often, if binding poses are not known, docking calculations can be used to generate an initial small molecule binding pose to start alchemical simulations. This will require some basic familiarity on how to perform docking to generate reasonable simulation starting points~\cite{grinter2014challenges}. 

As some of the theoretical background can seem daunting, we do, however, provide a guide to the essential theory behind alchemical free energy calculations in Sec.~\ref{sec:theory}.
In the remainder of this paper, we will cover topics that are key to the preparation( Sec.~\ref{sec:prerequisites}), choice and use of correct protocols (Sec.~\ref{sec:simulation_protocol_choice}), and finally the best practices that should be used in the analysis of alchemical calculations (Sec.~\ref{sec:data_analysis}). 
Particular focus will be given to aspects of the molecular simulations which are unique to alchemical calculations---these include the calculation of transfer free energies (hydration free energies, partition coefficients, etc.), and binding free energies (absolute and relative).

While we try to address as many methods and practices as possible, the field of free energy calculations is broad, and there are many advanced topics that are left to future Best Practices documents focusing on specific issues. 
Below, we provide a non-exhaustive list of topics we have \emph{not} addressed, along with some references to provide starting points on these more advanced topics:
\begin{itemize}
\item covalent inhibition~\cite{lameira2019predicting}
\item free energies of mutation of protein side chains~\cite{gapsys2016accurate, aldeghi2018accurate}
\item nonspecific binding or multiple binding sites~\cite{gill2018binding}
\item approximate and often less accurate endpoint free energy methods such as MM-PBSA~\cite{genheden2015mm} and LIE~\cite{gutierrez-de-teran2012linear}
\item Free energy methods that extract the ligand using geometric order parameters and potential of mean force methods~\cite{heinzelmann2017attachpullrelease}
\item forcefield dependence for protein, ligand, ions, co-solvents, and co-factors. A number of different studies have looked at the influence of force fields and it is assumed the user has made an adequate choice for the system under study~\cite{loeffler2018reproducibility, vassetti2019assessment, lopes2015current}. 
\end{itemize}

For convenience we have also compiled a list of common acronyms and common symbols used throughout this paper.
\begin{tcolorbox}[title=Acronyms, colback=blue!10!white]
    {\bf CPU} --- Central Processing Unit\\
     {\bf BAR} --- Bennett Acceptance Ratio\\
     {\bf FEP} --- Free Energy Perturbation\\
     {\bf GPCR} --- G-Protein Coupled Receptor\\
     {\bf GPU} --- Graphics Processing Unit\\
     {\bf MBAR} --- Multistate Bennett Acceptance Ratio\\
     {\bf MCSS} --- Maximum Common Substructure\\
     {\bf MD} --- Molecular Dynamics\\
     {\bf RMSE} --- Root Mean Square Error\\
     {\bf MUE} --- Mean unsigned error\\
     {\bf SAR} --- Structure-Activity Relationships\\
     {\bf TI} --- Thermodynamic Integration
\end{tcolorbox}

\begin{tcolorbox}[title=List of Symbols, colback=green!10!white]
$L$ and $R$ --- generic names for ligand and receptor\\
$K_b^{\circ}$ --- binding  constant \\
$c^{\circ}$ --- standard state concentration \\
$U$ --- potential energy\\
$u$ --- reduced potential describing a thermodynamic state \\
$\Delta G$ --- Gibbs free energy (isothermal isobaric ensemble)\\
$\Delta A$ --- Helmholtz free energy (canonical ensemble)\\
$\Delta f$ --- reduced (dimensionless) free energy \\
$\Delta \hat{f}$ --- estimate from an estimator for the reduced free energy\\
$\Gamma$ --- conformation space accessible by simulations \\
$\vec{q}$ --- vector of a single configuration, i.e. $x$, $y$, $z$ coordinates of the simulation system\\
$k_B$ --- Boltzmann constant \\
$Z$ --- partition function \\
$p$ --- pressure \\
$\mu$ --- chemical potential (grand canonical ensemble)\\
$T$ --- temperature \\
$\beta \equiv (k_B T)^{-1}$ --- inverse thermal energy \\
$\vec{\lambda}$ --- alchemical progress parameter, which may be multidimensional \\
$g$ --- statistical inefficiency\\
$\mathcal{O}$ --- overlap matrix\\
$C_t$ --- discrete-time-normalized fluctuation auto-correlation function\\
$\tau _{eq}$ --- integrated auto-correlation time\\
$t_0$ --- equilibration time
\end{tcolorbox}

\section{Statistical mechanics demonstrates why alchemical free energy calculations work}
\label{sec:theory}
Why would you want to run an alchemical free energy calculation and why do they work?
In this section, we use the example of relative free energy calculations to sketch the theory of alchemical simulations and illustrate their utility.
The emphasis here is placed on bridging theoretical foundations and intuition.
A rigorous derivation of the standard (absolute) free energy of binding using the principles of statistical mechanics can be found in Gilson's classic work~\cite{gilson1997statisticalthermodynamic}.

\subsection{Simulating binding events of receptor-drug systems can be computationally expensive}
Suppose you want to compute the binding affinity, or free energy of binding, of a ligand $L$ to a receptor $R$, given by:
\begin{equation}
R+L \leftrightharpoons RL.
\end{equation}
The binding constant ($K_b^{\circ}$) is given by the law of mass action as the ratio of concentrations of product $[RL]$ and reactants $[R]$, $[L]$:
\begin{equation}
 K_b^{\circ} = c^{\circ}\frac{[RL]}{[L][R]}.
\end{equation}
The standard state concentration $c^\circ$ depends on the reference state, but it is usually set to $1\,\mathrm{mol}/\mathrm{L}$ assuming a constant pressure of $1\,\mathrm{atm}$ (see also Sec.~\ref{sec:standardstate-restraints}).
Thus, the Gibbs free energy of binding $\Delta G_{\mathrm{bind}}$ is given by:
\begin{equation}
    \Delta G_{\mathrm{bind},L} = -k_BT\ln K_b^{\circ},
\end{equation}
where $k_B$ is the Boltzmann constant and $T$ the temperature of the system.

\paragraph{The free energy of binding can be expressed as a ratio of partition functions}
A natural, though generally very computationally expensive, way to estimate the equilibrium constant is by directly simulating several binding and unbinding events and computing the probability of finding the receptor-ligand system in the bound state, $P(RL)$, or the unbound state, $P(R+L)$.
Assuming the volume change upon binding to be negligible, which is often the case at $1~\mathrm{atm}$ due to the incompressibility of water, then the Gibbs free energy $\Delta G_{\mathrm{bind}, L}$ is approximately equal to the Helmholtz free energy $\Delta A_{\mathrm{bind}, L}$, and we can simulate the system in a box of volume $V$ to obtain~\cite{jong2011determining}
\begin{equation}\label{eq:binding-free-energy-from-bound-unbound-probability-ratio}
    \Delta G_{\mathrm{bind},L} \approx \Delta A_{\mathrm{bind}, L} = - k_B T  \left( \mathrm{ln} \frac{P(RL)}{P(R+L)} + \ln \left( c^\circ N_{\mathrm{Av}} V \right) \right) \, ,
\end{equation}
where $N_{\mathrm{Av}}$ is the Avogadro number, and the last term corrects for the simulated concentration being different than the standard concentration.
Let $\Gamma_{\mathrm{bound}}$ and $\Gamma_{\mathrm{unbound}}$ be the set of receptor-ligand conformations $\vec{q}$ that we consider bound and unbound respectively.
The probability of a conformation $\vec{q}$ is given by the Boltzmann probability density function
\begin{equation}
    P(\vec{q}) = \frac{\exp\left( \beta U(\vec{q}) \right)}{\int_{\Gamma} \exp\left( \beta U(\vec{q}) \right) \, d\vec{q}} \, ,
    \label{eq:conf_probability}
\end{equation}
where $\beta = (k_B T)^{-1}$ is the inverse temperature, $U(\vec{q})$ is the potential energy of conformation $\vec{q}$, and the integration is over the set of all possible conformations accessible in the simulation box volume $\Gamma$, with $\Gamma_{\mathrm{bound}},\Gamma_{\mathrm{unbound}} \subset \Gamma$.
If the simulation is long enough, we expect the fraction of conformations $\vec{q}$ found in the bound state to converge to
\begin{equation}
    P(RL) = \int_{\Gamma_{\mathrm{bound}}} P(\vec{q}) \, d\vec{q} = \frac{\int_{\Gamma_{\mathrm{bound}}} \exp\left( \beta U(\vec{q}) \right) \, d\vec{q}}{\int_{\Gamma} \exp\left( \beta U(\vec{q}) \right) \, d\vec{q}} \, .
\end{equation}
After similar considerations for $P(R+L)$, we find that the ratio of visited bound and unbound conformations, in the limit of long simulations, should converge to
\begin{equation}\label{eq:bound-unbound-probability-ratio}
    \frac{P(RL)}{P(R+L)} = \frac{\int_{\Gamma_{\mathrm{bound}}} \exp\left( -\beta U(\vec{q}) \right) \, d\vec{q}}{\int_{\Gamma_{\mathrm{unbound}}} \exp\left( -\beta U(\vec{q}) \right) \, d\vec{q}} = \frac{Z(RL)}{Z(R+L)} \, ,
\end{equation}
where we have defined the \textit{configurational integral} or \textit{configurational partition function} as $Z(\mathrm{state}) \equiv \int_{\Gamma_{\mathrm{state}}} \exp\left(-\beta U(\vec{q})\right) \, d\vec{q}$.

\paragraph{Simulating binding events is computationally expensive}
While simulating binding events has been used to estimate binding affinities~\cite{jong2011determining,pan2017quantitative} or to get insights into the binding pathways and kinetics of receptor-ligand systems~\cite{teo2016adaptive,votapka2017seekr,doerr2014onthefly,plattner2015protein,dixon2018predicting}, the computational cost of these calculations is usually dominated by the rate of dissociation, which can be on the microsecond timescale even for millimolar binders~\cite{pan2017quantitative} and reaches the micosecond to second timescale for a typical drug~\cite{basavapathruni2012conformational,hyre2006cooperative}.
Depending on system size and simulation settings, common molecular dynamics software packages can reach a few hundreds of ns/day using currently available high-end GPUs~\cite{eastman2017openmm,kutzner2019more}, making these type of calculations unappealing and irrelevant on a pharmaceutical drug discovery timescale.
Other methods compute the free energy of binding by building potential of mean force profiles along a reaction coordinate~\cite{woo2005calculationa,velez-vega2013overcoming,limongelli2013funnel,heinzelmann2017attachpullrelease}, but these methods require prior knowledge of a high-probability binding pathway, which is not easily available, especially in the prospective scenarios typical of the drug development process.

\subsection{Alchemical free energy calculations yield predictions that do not require direct simulation of binding/unbinding events}

\begin{figure}
    \includegraphics[width=0.95\linewidth]{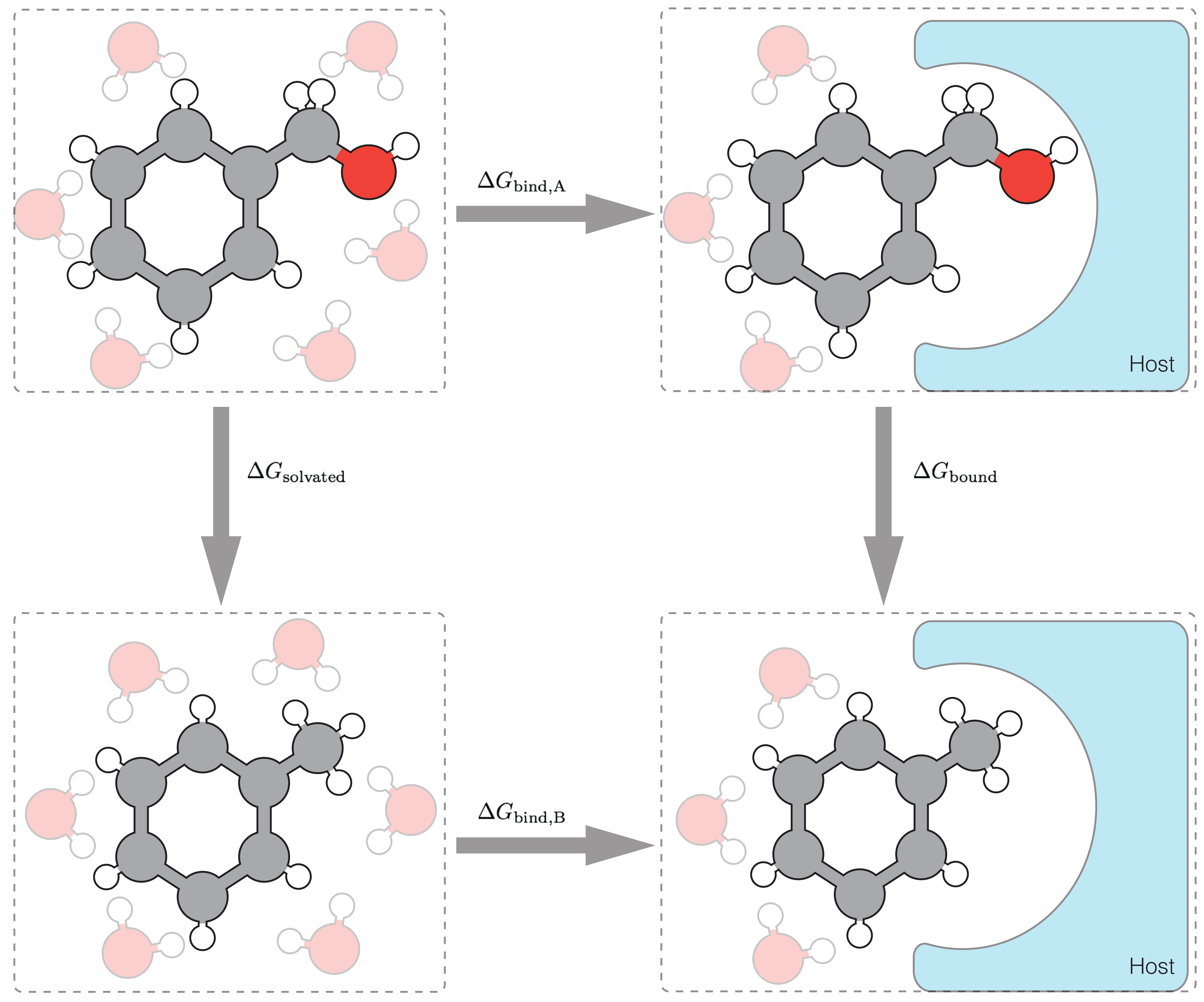}
    \caption{{\bf Thermodynamic cycle for computing the relative free energy of binding ($\Delta \Delta G$) between two related small molecules to a supramolecular host or a rigid receptor.}
    The relative binding free energy difference between two small molecules, $\Delta \Delta G_{\mathrm{bind}, A \rightarrow B} \equiv \Delta G_{\mathrm{bind}, B} - \Delta G_{\mathrm{bind}, A}$---here benzyl alcohol (top) to toluene (bottom)---can be computed as a difference between two alchemical transformations, $\Delta G_\mathrm{bound} - \Delta G_\mathrm{solvated}$, where $\Delta G_\mathrm{bound}$ represents the free energy change of transforming $A \rightarrow B$ in complex, i.e. bound to a host molecule, and $\Delta G_\mathrm{solvated}$ the free energy change of transforming $A \rightarrow B$ in solvent, typically water.}
    \label{fig:fig_binding_thermodynamic_cycle}
\end{figure}

In many cases, the quantity of interest is the change in binding affinity between a compound $A$ and a related compound $B$ (e.g., by modifying one the drug scaffold's substituents, see (Fig.~\ref{fig:fig_what_is_alchemy} \textbf{F}), which, by using Eq.~\ref{eq:binding-free-energy-from-bound-unbound-probability-ratio} and \ref{eq:bound-unbound-probability-ratio} is given by
\begin{equation}\label{eq:delta-delta-G-physical}
\begin{split}
    \Delta \Delta G_{\mathrm{bind}, AB} &= \Delta G_{\mathrm{bind}, B} - \Delta G_{\mathrm{bind}, A} \\
    &\approx -k_BT \left( \ln \frac{Z(RB)}{Z(R+B)} - \ln \frac{Z(RA)}{Z(R+A)} \right) \, .
\end{split}
\end{equation}
Note that the terms involving the standard concentration cancel out when we assume that the volume is identical for $A$ and $B$.
Predictions of $\Delta \Delta G_{\mathrm{bind}, AB}$ with non-alchemical methods generally require long simulations of both ligands, possibly through different binding pathways.
Alchemical relative free energy calculations avoid the need to simulate binding and unbinding events by making use of the fact that the free energy is a state function and exploiting the thermodynamic cycle illustrated in Fig.~\ref{fig:fig_binding_thermodynamic_cycle}.
This is apparent after rewriting Eq.~\ref{eq:delta-delta-G-physical} as
\begin{equation}\label{eq:delta-delta-G-alchemical}
\begin{split}
    \Delta \Delta G_{\mathrm{bind}, AB} &\approx -k_BT \left( \ln \frac{Z(RB)}{Z(RA)} - \ln \frac{Z(R+B)}{Z(R+A)} \right) \\
    &= -k_BT \left( \ln \frac{Z(RB)}{Z(RA)} - \ln \frac{Z(B)}{Z(A)} \right) \\
    &= \Delta G_{\mathrm{bound}} - \Delta G_{\mathrm{unbound}} \, ,
\end{split}
\end{equation}
where $\Delta G_{\mathrm{bound/unbound}}$ is the free energy of mutating $A$ to $B$ in the bound/unbound state.
Eq.~\ref{eq:delta-delta-G-alchemical} and Fig.~\ref{fig:fig_binding_thermodynamic_cycle} tell us that the difference in free energy of binding between toluene ($A$) and benzyl alcohol ($B$) can be computed by running two independent calculations estimating the free energy cost of mutating $A$ into $B$ in the binding pocket ($\Delta G_{\mathrm{bound}}$) and in solvent ($\Delta G_{\mathrm{unbound}}$), saving us the need to simulate the physical binding process of the two compounds.
In particular, the second line of Eq.~\ref{eq:delta-delta-G-alchemical} is a consequence of $\Delta G_{\mathrm{unbound}}$ being independent of the presence of the receptor in the simulation box as the definition of the unbound state assumes receptor and ligand to be at a sufficient distance for them to have no energetic interactions.
Note that, when $A$ and $B$ have different number of atoms, Eq.~\ref{eq:delta-delta-G-alchemical} implies the presence of a factor having units of volume entering both logarithms, which require unitless arguments.
However, the value of this factor is inconsequential as it cancels out in practice.

\paragraph{How are alchemical transformations performed in practice?}

In practice, the mutation of $A$ to $B$ is carried out by introducing one or more parameters $\vec{\lambda}$ controlling the potential energy function $U(\vec{q};\vec{\lambda})$ such that the potential of compounds $A$ and $B$ is recovered at two particular values $\vec{\lambda}_A$ and $\vec{\lambda}_B$.
Briefly, this is achieved by simulating a ``chimeric'' molecule composed of enough atoms to represent both $A$ and $B$.
A subset of the energetic terms in $U(\vec{q};\vec{\lambda})$ is then modulated by $\vec{\lambda}$ so that at $\vec{\lambda}_A$, the atoms that form molecule $A$ are activated and those belonging exclusively to $B$ are non-interacting ``dummy atoms'', while the opposite occur at $\vec{\lambda}_B$ (see Sec.~\ref{sec:relative-fe-protocol} for details).

We can rigorously account for fluctuations in other thermodynamic parameters such as changes in volume $V$ when simulating at constant pressure $p$ or changes in number of molecules $N_i$ of species $i$ at constant chemical potential $\mu_i$ (e.g., number of waters or ions) by introducing the \textit{reduced potential}~\cite{shirts2008statisticallya}
\begin{equation}\label{eq:reduced-potential}
u(\vec{q};\vec{\lambda}) \equiv \beta \left[ U(\vec{q};\vec{\lambda}) + p \, V(\vec{q}) + \sum_i \mu_i \, N_i(\vec{q}) + \cdots \right] \, .
\end{equation}
Here, the collection of thermodynamic and alchemical parameters $\{\beta, \vec{\lambda}, p, \mu, \ldots\}$ defines a \emph{thermodynamic state}.
In the context of alchemical calculations, in which the thermodynamic states vary only in their value of $\vec{\lambda}$, these are also referred to as \emph{alchemical states}.
The free energy of mutating $A$ to $B$ in any environment (e.g., binding site, solvent) can then be computed as
\begin{equation}\label{eq:delta-G-env}
    \Delta G_{\mathrm{env}} = - k_BT \ln \frac{Z(\vec{\lambda}_B)}{Z(\vec{\lambda}_A)} = - k_BT \ln \frac{\int_{\Gamma_{\mathrm{env}}} \exp\left( u(\vec{q}; \vec{\lambda}_B) \right) \, d\vec{q}}{\int_{\Gamma_{\mathrm{env}}} \exp\left( u(\vec{q}; \vec{\lambda}_A) \right) \, d\vec{q}} \, .
\end{equation}
While it is generally not feasible to compute the two partition functions $Z(\vec{\lambda})$, several estimators have been devised to robustly estimate the ratio of partition functions in Eq.~\ref{eq:delta-G-env} (see Sec.~\ref{subsec:estimators}) from a set of conformations usually collected with MD simulations from the thermodynamic states defined at $\vec{\lambda}_A$ and $\vec{\lambda}_B$ and intermediates thereof.

\paragraph{Why do alchemical calculations need unphysical intermediate states?}
While it is theoretically possible to estimate the ratio of partition functions from samples collected only at states $\vec{\lambda}_A$ and $\vec{\lambda}_B$, the efficiency of the free energy estimators rapidly decreases as the phase-space overlap between the two states also decreases~\cite{wu2005phasespaceb,wu2005phasespacec}.
Roughly, the phase-space overlap between two thermodynamic states measures the degree to which high-probability conformations (i.e., those with very negative potential energy) in one state are also high-probability conformations in the other state (see Sec.~\ref{sec:are-they-good} and Fig.~\ref{fig:fig_what_is_lambda}).

To solve the problem of having poor overlap between the states of interest, multiple intermediate alchemical states are introduced at values $\vec{\lambda}_A = \vec{\lambda}_0, \vec{\lambda}_1, \cdots, \vec{\lambda}_K = \vec{\lambda}_B$ so that each pair of consecutive states $\vec{\lambda}_k, \vec{\lambda}_{k+1}$ share good overlap.
Each intermediate state models a ligand that is neither $A$ nor $B$ but a mix of the two.
Many estimators (e.g., exponential reweighting (EXP)~\cite{zwanzig1954hightemperature} and Bennet's acceptance ratio (BAR)~\cite{bennett1976efficient,shirts2003equilibriuma}) can then be used to compute the free energy as
\begin{equation}
    \Delta G_{\mathrm{env}} = k_BT \sum_{k=0}^{K-1} \Delta f(\vec{\lambda}_k, \vec{\lambda}_{k+1})
\end{equation}
from samples collected at all the alchemical states $\{\vec{\lambda}_k \}$, where $\Delta f$ is the \emph{unitless free energy difference}
\begin{equation}
    \Delta f(\vec{\lambda}_k, \vec{\lambda}_{k+1}) = f(\vec{\lambda}_{k+1}) - f(\vec{\lambda}_k) = - \ln \frac{Z(\vec{\lambda}_{k+1})}{Z(\vec{\lambda}_k)} \, .
\end{equation}
While this strategy usually results in sampling thermodynamic states whose Boltzmann distributions are very similar, thus collecting information that is to some degree redundant, some estimators such as the Multistate Bennett acceptance ratio (MBAR)~\cite{shirts2008statisticallya} can exploit similarities between states to improve the precision of the estimates. This is achieved by using the conformations sampled at all alchemical states $\{\vec{\lambda}_k \}$ to compute the free energy difference $\Delta f(\vec{\lambda}_i, \vec{\lambda}_{j})$ between any pair of states $i,j$ (see Sec.~\ref{subsec:estimators}).

\paragraph{How do absolute free energy calculations differ from relative?}

While absolute and relative free energy calculations have subtle differences in their practical applications (e.g., use of restraints, handling of the standard state), the fundamental ideas and concepts of relative free energy approaches remain unaltered in other types of alchemical calculations.
Absolute binding, hydration, and partition free energies still use thermodynamic cycles that enable computing transfer free energies without actually simulating the physical transfer from one environment to another.

The main difference in these approaches lies instead in the thermodynamic cycle to which this strategy is applied.
For example, a typical thermodynamic cycle for an alchemical absolute binding free energy calculation is represented in Fig.~\ref{fig:fig_absolute_thermodynamic_cycle}.
In this case, two independent calculations compute the free energy of removing the interactions between the ligand and its environment in solvent or in the binding site respectively through a series of intermediate states in which the energy terms are only partially deactivated.

\section{What can be expected from alchemical simulations?}
\label{sec:step0}
When starting an alchemical free energy project, a key first step is to decide whether free energy calculations are really the right tool. Particularly, count the cost of your
project: Can you even hope to tackle the problem with available resources and, if successful, will it
be worth it in terms of human and computational cost?

\subsection{How accurate are alchemical free energy calculations?}
\label{subsec:expectation}
Current alchemical free energy calculations involving small molecules seem to achieve, in favorable cases, root mean square (RMS) errors around 1-2 kcal/mol depending on force field, system, and a variety of other factors such as simulation time, sampling method, and whether the calculations employed are absolute or relative. A small selection of example datasets and case studies can be found in Sec.~\ref{sec:benchmark} at the end of this document.
However, the domain of applicability is a significant concern~\cite{sherborne2016collaborating, cournia2017relative}, especially for relative calculations, which typically require a high quality and usually experimental bound structure of a closely
related ligand as a starting point. Additional factors such as slow protein or ligand rearrangements, uncertainties in ligand binding mode, or charged ligands can make these calculations far less reliable and more of a research effort.

It is worth noting that the accuracy of free energy calculations is highly variable across different protein targets, and likely across different ligand chemotypes as well.
For instance, FEP+ with OPLS3 achieves an RMSE of 0.62 kcal/mol for a set of 21 compounds binding to JNK1 kinase, but an RMSE of 1.05 kcal/mol for a set of 34 compounds binding to P38$\alpha$ kinase~\cite{harder2016opls3}.
Furthermore, perturbations for the same chemotype in different pockets of the BACE enzyme gave varied errors~\cite{keranen2017acylguanidine}. Here the errors refer to the difference in $\Delta G$ derived by fitting the $\Delta \Delta G$'s to the known experimental binding free energies~\cite{wang2015accurate}. This prompts us to consider another important aspect. It is important to be clear on what error to report: $\Delta G$ after shifting by a constant to minimize the RMSE, unshifted $\Delta G$, $\Delta \Delta G$ of computed edges, or $\Delta \Delta G$ of all edges. (See recommendations for reporting best practices, Sec.~\ref{sec:plot_data}.) Additionally, as it is possible to perform calculations on a set of ligands using different pairwise comparisons of molecules, the performance of the method may be biased based on which pairs of comparisons are performed. Additionally, it is possible that the error associated to the relative free energy between a two ligands that was not directly computed (however can be deduced using thermodynamic paths involving other ligands) will likely be more uncertain \url{https://github.com/jchodera/jacs-dataset-analysis}.
Given the need to understand the performance of the system with alchemical free energy calculations, we recommend that retrospective studies for a particular target and a particular chemical series be performed for each application case.

\subsection{How reproducible are alchemical free energy calculations?}
\label{subsec:reproducible}
Finite computing resources necessarily limit the generated number of uncorrelated samples of potential energy surfaces, and therefore alchemical free energy calculations only give free energy estimates to within finite precision. An important consideration is how reproducible alchemical free energy calculations are in practice. In simple cases such as absolute hydration free energies of small organic molecules, or relative hydration free energy calculations between structurally similar small organic molecules, it should be possible to obtain highly precise estimates with a given software package (with a sample standard deviation under 0.01 kcal/mol)~\cite{rizzi2019sampl6}.
For more complex use cases such as protein-ligand binding free energies the repeatability is often substantially worse~\cite{rizzi2019sampl6}. A good practice is to perform two or three runs of the same perturbation to assess precision with a given protocol. The sample standard deviation will give a crude estimate of the reliability of the estimates, and whether the precision is sufficient for the problem at hand. When practical, a more stringent test is to use different input coordinates for each repeat run. 

Note that these issues concern calculations carried out with a single software package, but simulation package variations can introduce additional issues. Such issues of reproducibility of free energy calculations across different simulation packages have attracted attention recently. Greater variability is expected due to methodological differences such as integrators, thermostats, barostats, treatment of long-range electrostatics, and potentially other factors. For absolute and relative hydration free energies of small organic molecules a variability of ca. 0.2 kcal/mol between popular simulation packages has been reported~\cite{loeffler2018reproducibility}. In the recent SAMPL6 SAMPLing challenge a larger variability of 0.3 to 1.0 kcal/mol was noted in the computed absolute binding free energies of host/guest systems even though the study sought to use identical input and simulation parameters~\cite{rizzi2019sampl6} and, in many cases, single-point energies were identical or nearly so. Further work is needed to ensure reproducibility of alchemical free energy calculations across different software implementations to guarantee that force-field development efforts lead to transferable potential energy functions. 

\subsection{Is my problem suitable for alchemical free energy calculations?}
\label{subsec:suitability}
Before even planning free energy calculations to study binding to a
particular target, it is important to assess what is known about the
system and its timescales and its suitability for free energy
calculations, as well as the \emph{purpose} of the calculations and
the amount of available computer resources. In some cases, predicting accurate binding free energies for a particular target might be
\emph{more} challenging than simply measuring them! This is
often the case when dealing with database screening problems, where
compounds might be easily and quickly available commercially for
testing and free energy calculations could consume far more resources. Free energy calculations thus typically only
appeal when (slow or costly) synthesis would be required or experiments are otherwise cost-prohibitive.

Sometimes free energy calculations can provide answers that are not
readily available from experiments. For example, type II kinase
inhibitors selectively bind to different kinases in the so-called
DFG-out conformations~\cite{schindler2000structural}. The selectivity of such
inhibitors may be attributed either to their differential binding to
different kinases in the DFG-out conformations, or to different
stability of the DFG-out conformations of different kinases. 

Let
$K_C$ be the equilibrium constant between DFG-in and DFG-out
conformations of one kinase, and $K_D^\ast$ be the dissociation
constant of a type II inhibitor against this kinase, the apparent
binding constant of this inhibitor against this kinase is then
\begin{equation}
  K_D = K_D^\ast \frac{1 + K_C}{K_C}
  \label{eqn:conformational-binding}
\end{equation}

Since binding experiments cannot resolve $K_D^\ast$ and $K_C$ individually, such experiments cannot address the basis of selectivity of the type II inhibitors. Absolute binding free energy calculations, in contrast, can take advantage of the slow kinetics of DFG-in/out conversion, and estimate the conformation-specific binding constant $K_D^\ast$, thus yielding clues as to the source of selectivity.

\subsection{Is the expected accuracy of the computation sufficient?}
\label{subsec:accuracy}
The requisite level of accuracy is another important consideration. If the
goal is to guide lead optimization when many compounds will be
synthesized, free energy calculations can be appealing even with
accuracies in the 1--2 kcal/mol range~\cite{mobley2012perspective}, but if the number of compounds to be synthesized is very small, this accuracy may not be enough to provide much value.

Here we provide a simple estimate of the value provided by alchemical
free energy calculations in lead optimization. Let $P(\Delta\Delta
G)$ be the probability distribution of the changes in the binding free
energies of a new set of molecules during one round of lead
optimization, and let $P(\Delta\Delta G^\dagger|\Delta\Delta G)$ be the
conditional probability of the binding free energy change computed by
the free energy calculations, $\Delta\Delta G^\dagger$, given the actual
change $\Delta\Delta G$. The latter conditional probability can be modeled
by a normal distribution
\begin{equation}
  P(\Delta\Delta G^\dagger|\Delta\Delta G) = \frac{1}{\sqrt{2\pi\sigma^2}}
  \exp\left(-\frac{(\Delta\Delta G^\dagger - \Delta\Delta G)^2}{2\sigma^2}\right),
  \label{eqn:free-energy-distribution}
\end{equation}
where $\sigma$ signifies the accuracy of free energy calculations.
Here we assume that there is no systematic bias in the free energy
calculations, i.e., on average, the free energy change computed by
free energy calculations agrees with the actual free energy change.

In lead optimization guided by free energy calculations, we will likely only
synthesize and experimentally test molecules that are predicted to
have favorable free energy changes. We are thus interested in how
often that a molecule predicted to bind stronger actually turns out to
bind stronger. In other words, we are interested in the conditional
probability:
\begin{equation}
  P(\Delta\Delta G<0|\Delta\Delta G^\dagger<0).
  \label{eqn:true-positive}
\end{equation}

For illustrative purposes, consider a proposed set of new molecules, and assume that the changes proposed in these molecules yield a set of relative binding free energies that follow a normal
distribution. That is, assume that the standard deviation in the relative binding free enrgies for the changes represented is $RT\ln 5$
(corresponding to a 5-fold change in the binding affinities), and that
1 in 10 new molecules have increased binding affinity ($\Delta\Delta G
\leq 0$). Under such assumptions, the conditional probability in
Eq.~\ref{eqn:true-positive} can be easily computed. 

If the accuracy of free energy calculations is $\sigma = 1$ kcal/mol, $P(\Delta\Delta
G<0|\Delta\Delta G^\dagger<0) = 0.35$, which means that out of every
10 molecules selected for predicted favorable free energy change, on
average 3.5 molecules will have actual favorable free energy change.
In other words, selection by free energy calculations yields 3.5 times
more molecules of improved affinities than selection without free
energy calculations under these assumptions.
  
Available computational resources and timescales of motion also factor
into this initial analysis. An individual free energy calculation
involves simulations at many different intermediate states (perhaps
20-40 or more) and each of these must typically be long enough to
capture the relevant motions in the system. If such motions are
microsecond events or longer, the computational cost of running 20-40
microsecond or longer simulations for each of $N$ ligands will likely be
prohibitive for most users with today's hardware. Alternatively, if key motions are fast and minimal (as is often assumed in practice), only much shorter simulations may be necessary. 

\subsection{Can I afford the calculation?}
\label{subsec:affordability}
Furthermore, are available computational resources sufficient that throughput will be reasonable compared to needs of experimental collaborators working on
this system? How many ligands ($N$) can you afford to handle given
your computational resources? As cloud computing becomes more available, in-house GPU clusters may not be necessary if calculations are not run on a regular basis.
This analysis should be done up front as part of ``counting the cost''
of involvement in a particular project. In some cases, the analysis may conclude that free energy calculations will not be feasible for the proposed problem.
Here, by ``cost'', we refer not just to financial cost of the calculations relative to experiments, but also time -- can the calculations be run faster than experiments are done? How will the relevant resource and opportunity costs factor in? Both computation and experiment require human time, supplies (of different sorts), and equipment. In the extreme limit, for example, it would not make sense to spend a month running a binding free energy calculation if the equivalent experiment could be done in a day with resources already on hand. Such issues should be considered before deciding to conduct binding free energy calculations.

\subsection{Is an exploratory study what I want?}
\label{subsec:exploration}
An additional consideration is how much is known about your particular
target, ligand binding modes in the target, and any relevant motions
-- essentially, has it been studied enough to know whether it might be
suitable for free energy calculations? It is important to know if the system has hardly been studied, because should the initial calculations perform poorly, the effort may turn into an attempt to understand the relevant sampling, force field, or system preparation problems.

If you are unsure whether your project is feasible, as mentioned above, one recommended option is to conduct a short exploratory study to assess tractability for a small
number of ligands. This can be sufficient to get an initial
idea of feasibility and accuracy of the calculations for the
proposed target~\cite{schindler2020largescale}.

\section{How should alchemical simulations be applied to drug discovery?}
\label{sec:drugdiscovery}
Many practitioners expect alchemical methods to provide valuable guidance for drug discovery, and to exhibit accuracy superior to most alternative approaches for suitable targets~\cite{kuhn2017prospective}. Successful application in industry may require considerable knowledge of the ``domain of applicability'' of free energy calculations -- where they work well and where they will not~\cite{sherborne2016collaborating}. Successful application also requires robust protocols for preparing, submitting and analysing alchemical calculations. In this regard, the issues mentioned in the previous section such as understanding the suitability and timescales to capture the structure activity relationships (SAR), and performing up-front tests of performance are all relevant to drug discovery applications. Without venturing too far into details of system setup, which is beyond the scope of this article, we highlight some critical factors affecting accuracy and successful application. 

\subsection{Capturing experimental conditions}
\label{subsec:exp_condition}
The calculations aim to capture the alchemical change from one ligand to another as accurately as possible. Therefore, it is necessary to consider details of the experimental setup, such as pH. Biological assays are usually run at neutral pH but this is not always the case. For example, some enzymes exhibit pH-dependent activity and assays may thus be done in conditions other from neutral pH. Therefore, computational protein and ligand preparation protocols should reflect experimental pH. 

The formal charge and/or tautomeric state of the small molecules can change within a series of analogs, necessitating care in treatment. Additionally, medicinal chemistry efforts might deliberately modify the pKa of a series to modify drug properties, requiring explicit efforts to incorporate these changes into alchemical calculations.

To ensure modeling matches experiment, we also need to accurately prepare and simulate the same system -- which requires understanding what protein construct is used in the bioassay. For instance, does the X-ray structure that is to be used for the calculations match the construct used for screening (i.e. only the catalytic domain vs. full length, monomer vs. dimer, etc.)~\cite{perez-benito2018predicting}? Also, were certain co-factors or partner proteins required in the bioassay? 

\subsection{Is my binding mode accurate?}
As also mentioned, good performance of alchemical calculations requires an accurate representation of the ligand binding mode, usually from a high quality X-ray crystal structure. If more than one structure is available, the modeler should pay attention to choose the most suitable. The quality of the structure can be a concern, and the reader is referred to work of Warren et al. for a detailed discussion of choosing optimal structures for structure-based modeling~\cite{warren2012essential}.

It is also useful to study the structure activity relationship and understand the expected impact of any mutations on the binding site, such as whether side chain movement in the protein will be required, and whether there is evidence of this in any alternative X-ray structures of the same protein. Often, only one protein and water configuration is used for a series of alchemical calculations, so this needs to be capable of accommodating the smallest through to largest ligands in a way that allows stable and well behaved simulations. This can provide a practical limit on the alchemical changes that are feasible, though a simple work-around can be to separate compounds into sub-series for different calculations. 

If multiple structures are available there is some evidence the higher affinity complex can give better performance~\cite{perez-benito2019predicting}, at least in some cases. However, ligands and proteins can also undergo unexpected changes in binding mode for related ligands, which can make these issues more complex to deal with~\cite{mey2016blinded}.

\subsection{Input setup and scale of calculations}
In a drug discovery setting it is normal to consider dozens (or more) of ligands and it is necessary to align them in the binding site. There is no detailed study of how different alignment approaches may affect results, but the user should be aware of some practical considerations. Tools are available to compare the ligands and build the hybrid topologies that define the changes between one ligand and another~\cite{loeffler2015fesetup,hedges2019biosimspace,gapsys2015pmx}. In simple terms, providing poor alignment to these tools will make this job harder. Docking with restraints is often beneficial in this regard. Particularly, fixing the 3D spatial position of the scaffold using maximal common substructure (MCSS) restrained docking can help provide well aligned input for the topology generation. Nevertheless, in this case careful attention is still needed to ensure consistency of alignment for identical substituents. Another alternative is to manually edit the same core and add/modify the changing substituents. This provides assurances that coordinates for the non-perturbed portion of the structure remain identical and aromatic substituents, for instance, have consistent dihedral angles. However, it is not feasible for many compounds and therefore automation is desirable. 

Finally, the role of water in ligand binding is not always well understood and it can be crucial to capture the changes in binding site solvation during ligand binding. Can crystallographic waters be retained? Do they clash with some of the larger ligands used in the alchemical perturbation? See Sec.~\ref{subsec:binding} for different strategies that can be applied to dealing with waters. Generally, before launching large numbers of alchemical free energy calculations it is always recommended to test the system using classical MD simulations and limited numbers of alchemical perturbations. Metrics such as ligand and protein RMSD and RMSF can be inspected, along with visual inspection of simulations, to ensure the system is stable and likely to be suitable for alchemical calculations. 

Running binding free energy calculations in a drug discovery application will typically require the use of software or tools to facilitate the large number of calculations. Commercial implementations such as FEP+, OpenEye Tools, or Flare allow for a fast setup and deployment to GPU hardware in minutes, but may have limited ability to customize calculations~\cite{wang2015accurate,kuhn2020assessment}. Commercial tools can be expensive in some cases, but non-commercial tools are becoming more straight forward to use to run alchemical free energy calculations~\cite{gapsys2015pmx, loeffler2015fesetup, song2019using, gapsys2020large, jespers2019qligfep, hedges2019biosimspace, kuhn2020assessment}.

For relative free energy calculations, various graph topologies or maps of calculations are possible, and choices may depend on the target application. For instance, if the goal is to accurately assess the relative binding energy of a small number of compounds, possibly with challenging synthesis, the map of perturbations should contain as many connections between compounds as affordable. However, when running calculations on hundreds of compounds a so called \emph{star-map} (see Fig.~\ref{fig:fig_types_of_networks} \textbf{A}) can be used that just contains one connection per compound: perturbing every compound to a central ligand, typically the crystal structure ligand~\cite{konze2019reactionbased}. In this way the top-ranking examples can be readily identified and submitted to additional calculations in a second round. Alternatively, if the goal is to achieve the smallest possible error with minimal computational expense, certain graph topologies provide benefits~\cite{yang2020optimal, xu2019optimal}

\subsection{Making predictions, understanding errors}
\label{subsec:predictions}
For prospective drug discovery applications there are several other considerations including understanding likely errors and taking selection bias into account. 

It is crucial when proposing compounds for synthesis to have some idea of the underlying error or uncertainty in the predictions. A retrospective assessment can give an indication of prospective performance for similar molecules~\cite{ciordia2016application}. Beyond this, several parameters provide useful indicators of performance. For example error estimates provided by free energy estimators that are too large can highlight poorly converged simulations~\cite{perez-benito2019predicting}. Hysteresis, within cycles in the perturbation network or between forward and backward perturbations can be checked~\cite{wang2013modeling} to indicate problematic perturbations involved in cycles connecting many compounds (See also Secs.~\ref{sec:relative-fe-protocol} and~\ref{sec:are-they-good}). Once synthesis and testing of compounds is complete a standard strategy is to look back at how the calculations performed. In this regard it is important to consider the issue of selection bias upfront. It is tempting to only synthesize the compounds predicted to be most active, thus a narrow range of calculated activity is tested that imposes limits on the statistical assessment of performance, ideally example molecules from across the range of predicted activity can be assessed or corrections can be applied based on previous recommendations~\cite{abel2017critical}. For a more detailed discussion on checking the robustness of your alchemcial free energy calculation see also Sec.~\ref{sec:are-they-good}.

In summary, the successful use of alchemical calculations not only in, but particularly for drug discovery requires working in the domain of applicability, using a high quality X-ray structure of the target bound to compounds in the series, and testing the approach retrospectively to ensure the system setup is well-behaved. Always assess your confidence in the resulting predictions and communicate this when discussing with experimentalists. Consider performing repeat calculations for at least some of the perturbations in the study. 
 There are many accounts of success of alchemical calculations, the methods show good performance towards the goal of binding energy prediction. However, it is important to have realistic expectations. 

Structure based drug design projects are often capable of improving potency relatively quickly, even with only limited application of computational approaches and the range of activity narrows to just two-to-three log units. It may seem hard to have impact with substantially different, more potent, stand-out compounds in this scenario, but binding energy prediction can still be extremely useful for ensuring activity is maintained as other properties are optimized. An interesting cost benefit analysis has shown the value of activity prediction, see discussion above and articles such as~\cite{mobley2012perspective}. 
From a drug discovery point of view, alchemical calculations are expanding their domain of applicability, and there are reports of success using homology models~\cite{cappel2016relative} and GPCRs~\cite{deflorian2020accurate,lenselink2016predicting} for instance, as well as enabling charge change and scaffold hopping~\cite{chen2018accurate, wang2017accurate}, but these systems are undoubtedly more difficult. In the meantime, the use cases are expanding to resistance prediction, selectivity prediction , solubility prediction – an exciting future for alchemical calculations~\cite{hauser2018predicting, albanese2020structure, mondal2019free}. 

%
%
\section{Simulation prerequisites}
\label{sec:prerequisites}
Alchemical free energy protocols as discussed below (Sec.~\ref{sec:simulation_protocol_choice}) are defined for a specific type of free energy calculation, i.e. a free energy of binding or a free energy of hydration. Different types of simulations require different choices for ligands, solvent, and host molecules (in the case of the estimation of free energies of binding).

\subsection{Free energies of binding}
\label{subsec:binding}
In principle, in the limit of sufficient conformational sampling, the free energy changes estimated from an alchemical free energy calculation should be independent of the system's initial coordinates. However, in practice, because simulations are of finite duration (typically 1-100 ns per state at present), this is only true for certain classes of alchemical free energy calculations such as relative or absolute free energies of hydration of small and relatively rigid organic molecules. Protein-ligand complexes typically exhibit slowly relaxing degrees of freedom that significantly exceed the duration of an alchemical free energy calculation, and host-guest calculations can be susceptible to these issues as well, depending on timescale and system. It is therefore generally important to carefully select input coordinates to obtain satisfactory results. 
The following questions may be relevant before diving into the simulation setup.

\begin{itemize}
    \item Do I have one or multiple good receptor structures? (e.g. a good resolution X-ray crystal of the protein target)
    \item Do I have information on one or all of the ligand binding sites (e.g. a X-ray structure)
    \item Should I include buried waters, or other small molecules that can be found in an X-ray structure?
    \item Are my ligands part of a congeneric series? (i.e. simple R group substitutions around the same scaffold)
    \end{itemize}

\paragraph{Are there good X-ray structures available?}
As with any simulation, care should be taken in selecting available X-ray structures in the Protein DataBank~\cite{berman2003announcing}. In some cases it may be wise to choose multiple starting structures to account for variability in receptor conformations as well as the accuracy of available X-ray structures. Typically, clustering of receptor structures can be used to identify different receptor conformations near the binding site, as well as assessing relevant side chain placements from the X-ray structure, see for example~\cite{mey2016blinded}. In terms of set up and other choices, following general best practice guidelines is advisable~\cite{braun2019best}.

Many free energy calculations focus on a congeneric series of ligands, which can make these calculations suitable for relative free energy protocols (see Sec.~\ref{sec:simulation_protocol_choice}). For relative calculations, some care has to be taken selecting binding poses for these ligands. Generally, a common assumption for a congeneric series is that the binding mode is conserved. Therefore, if an X-ray structure of one of the ligands is available, this should be used to position the ligands in the putative binding site in an energetically reasonable conformation without steric or electrostatic mismatch with the receptor. Checking the X-ray structure versus the experimental electron densities is important, as the position of part of the ligand or important sidechains may be based on the interpretation of the crystallographer rather than the available electron density, especially in cases of missing density. For example, looking at a cyclohexane ring density, a chair conformation is vastly more likely than that of a boat and, if a boat conformation is present in the structure, it may be worth inspecting the density to ensure it adequately supports this choice. 

\paragraph{Are you prepared to deal with any binding mode challenges?}
Generally, binding modes within congeneric series are conserved~\cite{wacker2010conserved}, however, exceptions exist~\cite{brandt2011congeneric,nazare2005probing}, as discussed in more detail in Sec.~\ref{sec:multiple_binding_modes}. Certain functional groups may be particularly prone to this due to symmetries or near symmetries. One such issue involves a 180 degree flip in the dihedral angle of an aromatic ring, or five-membered ring leading to a different spatial position of ortho- or meta- substituents that otherwise should overlap within a series. The 180 degree flip of the ring may not occur enough during simulations (due to steric obstructions) to overcome bias due to the starting configuration. Another scenario may be equatorial and axially substituted saturated rings (e.g. cyclohexane derivatives). This situation may be addressed by explicitly modelling different binding modes of the same ligand and combining later computed free energy differences for different binding modes into a relative free energies of binding~\cite{kaus2015how}.

\paragraph{Have you considered sterioisomers and enantiomers?}
Congeneric series can contain stereoisomers or enantiomers which can bind very differently, resulting in large errors if treated incorrectly. For racemates, the relative abundance of each stereoisomer is normally not known. Therefore, the experimental activity associated with just one stereoisomer/enantiomer is more uncertain. However, the modeling typically uses just the bioactive conformation that best fits the active site. Clearly this introduces potential for larger errors compared to experiment. Nevertheless, if all compounds in the congeneric series are racemic, originating from similar synthetic procedures with an expected similar abundance of stereoisomers, then the differences may cancel and the trend in calculated and observed binding energies may be robust. Despite this, we can see that care and further testing is needed in this scenario, and the quality of the predictions may suffer. Additionally, unexpected changes in what stereoisomer binds experimentally, if they occur, could pose significant challenges for modelling efforts.

\paragraph{Conserved binding site waters can play an important role in binding free energies}
Binding site water molecules may form water mediated protein-ligand interactions which can pose challenges whenever exchange with bulk water is slow compared to simulation timescales. This happens typically in buried binding sites. Overlaying multiple protein X-ray structures can identify conserved or additional water molecules that can be useful to include in calculations. In cases where water molecules are known to play an important role in the binding, software implementations that use water sampling facilitated by Grand Canonical Monte Carlo methods may be useful, i.e. consider pre-solvation methods such as GCMC steps~\cite{michel2010prediction}. Other tools such as WaterMap or open source equivalents (SSTMap, GIST, and others) can be used to define water structure for systems with no experimental evidence of water sites~\cite{wang2011ligand}. Well known protein systems with water mediated ligand interactions are for example: HSP90 which formed part of the D3R grand challenge 2015~\cite{mey2016blinded}, A2A~\cite{brucemacdonald2018ligand}, MUP~\cite{ross2015water}, ~\cite{deflorian2020accurate}, and others~\cite{michel2009energetics}.

\paragraph{Protonation states depend on the pH of the experimental assay}
Care should be taken when preparing ligands and proteins to match the pH of the experimental assay, if known. As mentioned above in Sec.~\ref{subsec:exp_condition}, the pH of the assay can differ from neutral pH and will determine the protonation states of the proteins and ligands. Since the pKa of reference amino acid sidechain residues is known, but can vary in the protein environment, many different tools have emerged for predicting sidechain pKa in proteins, such as the H++ server, ProPKa, APBS, and Maestro ~\cite{anandakrishnan2012automating, sondergaard2011improved, jurrus2018improvements, 2020schrodinger}. Strongly acidic (Glu, Asp) or basic (Arg, Lys) sidechains can reliably predicted to be ionised, but care is still needed as the local environment can modify expected ionization states (for instance the catalytic Asp dyad in proteases). Histidine is notoriously more difficult to predict as its pKa suggests it ionizes closer to the experimental pH range. For ligands often the pKa needs to be determined, if not known experimentally. There are many different available tools for this purpose, but common choices may be propKa~\cite{olsson2011propka3,sondergaard2011improved}, Chemicalize (\url{https://chemicalize.com/welcome}), or Maestro~\cite{2020schrodinger}. Still, accurate pKa prediction for small molecules remains a challenging problem, even with dedicated tools~\cite{isik2018pka}. While often it can be assumed that the protonation state of a ligand and protein will remain the same as a ligand binds, some care needs to be taken with systems where the protonation state may change upon binding~\cite{onufriev2013protonation}. BACE, for example, famously undergoes a protonation state change on ligand binding.

\paragraph{Congeneric series often need alignment}
Input coordinates for a congeneric series may be generated by docking calculations, or by ligand alignment using MCSS algorithms. The latter tends to produce alignments that are more conserved and more consistent free energy changes across a dataset, but will struggle to yield reasonable results for relative binding free energy calculations that involve a significant binding mode rearrangement. This may also lead to steric clashes with the receptor coordinates of the reference ligand if structural rearrangements are needed to accommodate different members of the congeneric series. Small steric clashes may be resolved during subsequent simulation equilibration prior to data collection, but there is a risk that the complex relaxes to an alternative metastable state. 

An additional consideration arises for single topology relative free energy calculations. In this class of alchemical free energy calculations it is necessary to generate a molecular topology that may describe the initial and final states of the perturbation (see Fig.~\ref{fig:fig_topology}). In cases where the end-states have high topological similarity and high structural overlap this is relatively straightforward and typically handled by use of MCSS calculations. In situations where the end-state topologies differ significantly, or there is relatively little spatial overlap between the two end-states some user intervention may be necessary to produce a satisfactory input topology.

If the binding site location is uncertain but the structure of the receptor is well defined and plausible binding sites are identified, it may be more useful to choose an absolute free energy protocol to compute the standard free energy of binding of the ligand to a set of binding sites. This requires the user to prepare input files describing the bound conformation in different putative binding sites~\cite{evoli2016multiple}. The apparent binding free energy of the ligand may be obtained by combining the individual binding site free energies, which also indicate where the ligand is more likely to bind. In this case a docking program can generate initial structures. Different commercial and none commercial tools are available, such as rDock, Autodock Vina, Glide, or Flare, to name a few~\cite{ruiz-carmona2014rdock, trott2010autodock, friesner2004glide, kuhn2020assessment}. 

If the putative binding sites are not apparent, for instance due to significant induced-fit effects, it may be challenging to obtain meaningful free energies of binding. One may have to account for the free energy cost of forming a binding site in the target receptor which may not be feasible on alchemical simulation timescales.

\subsection{Free energies of hydration or partition coefficients}
\label{subsec:hydration}
Preliminary considerations necessary for using free energy methods to compute partition coefficients are generally more straight forward. For example, a 3D minimised structure of the solute can be generated with a simple tool such as openBabel and solvated to prepare the input to compute a free energy of hydration~\cite{oboyle2011open}. However, in these cases a careful choice of forcefield, as well as water models or organic solvents is essential. See for example~\cite{bosisio2016blinded,rustenburg2016measuring} for a good discussion of these choices. And, while sampling problems might seem to be a non-issue for small molecules, this is not always the case; e.g. even the hydroxyl orientation on neutral carboxylic acids can occasionally pose a challenge~\cite{klimovich2010predicting, lim2019assessing}.

\section{What simulation protocol should I choose?}
\label{sec:simulation_protocol_choice}
Alchemical free energy calculations can be grouped into two main categories, ``absolute'' (see Fig.~\ref{fig:fig_absolute_thermodynamic_cycle}) and ``relative''\footnote{The distinction is a bit of a misnomer, since both compute ratios of partition functions relative to another state and in that sense are relative, while neither computes an absolute free energy.} (see Fig.~\ref{fig:fig_binding_thermodynamic_cycle}), which differ in whether they compute properties for a single molecule (absolute) or compare properties of different, usually closely related, molecules (relative).
To use binding as a concrete example, in absolute binding free energy calculations, we compute the binding free energy of a ligand to an individual receptor relative to a standard reference concentration.

In contrast, in relative binding free energy calculations, we compare the binding free energy of two related inhibitors to determine the potency difference.

\subsection{Absolute and relative free energy calculations have some differences}
Many of the issues around simulation setup and protocol choice for alchemical calculations are common, but there are some differences between absolute and relative calculations. We will consider protocol differences before treating the common elements.

\subsubsection{Choices unique to relative free energy calculations}
\label{sec:relative-fe-protocol}

\paragraph{Topologies} 
A critical first step in relative calculations is to select an approach to these calculations, determining whether to use a \emph{dual topology}, \emph{single topology}, or \emph{hybrid topology} approach to relative calculations.

\begin{figure}
    \includegraphics[width=0.95\columnwidth]{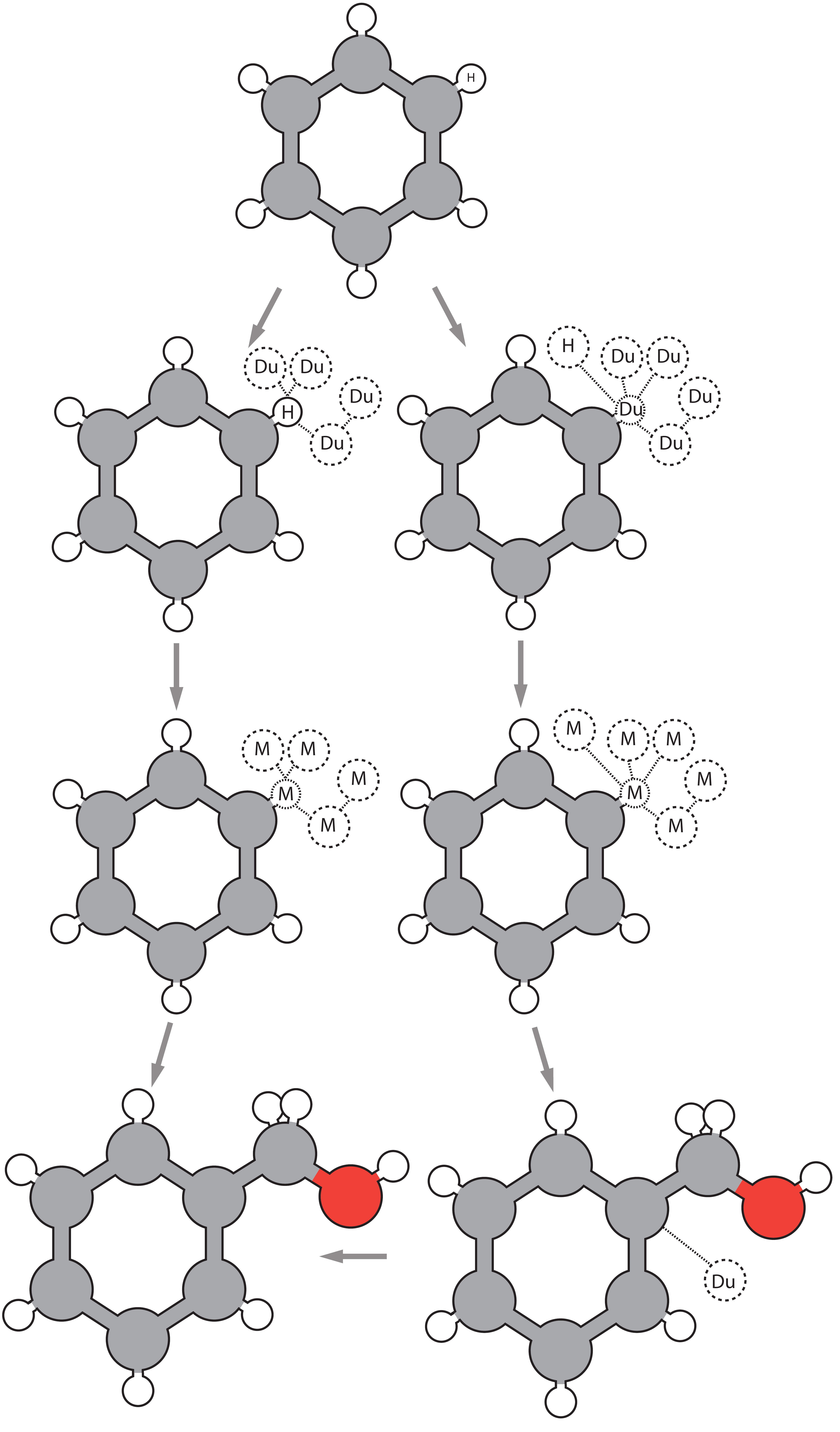}
    \caption{\textbf{Two common topologies for alchemical calculations: single and dual topology.} \textbf{Left}: A single topology uses softcore potentials to convert from one type of atom to an other. Dummy atoms (Du) are used when there is no corresponding maximum common substructure match between the two molecules for certain atoms. \textbf{Right}: The dual topology does not convert one species to another, but only converts between Du atoms and an interacting species, but usually uses softcore potentials for this. The 'mixed' intermediate atoms are used in both dual and single topology approaches. Only the way the transformation occurs and the end states differ. Following the arrow along the left and right illustrate the differences. Figure adapted from \url{http://www.alchemistry.org/wiki/Constructing_a_Pathway_of_Intermediate_States}}
    \label{fig:fig_topology}
\end{figure}

The distinction between these can be illustrated by considering a hypothetical transformation from molecule A to molecule B, where both atoms share a common substructure but differ in their substituents; e.g. consider a transformation of benzene to benzyl alcohol Fig.~\ref{fig:fig_topology}.
In this case the common substructure is the benzene ring, though the substructure may be larger depending on how it is defined, as we discuss below.

In single topology calculations, the overall transformation is set up to involve as few additional atoms as possible, so benzene would be typically changed into benzyl alcohol by changing one of the hydrogens into a carbon. This site will also be the future home of two additional hydrogen atoms bound to the new carbon, so these must initially be present as non-interacting atoms called ``dummy atoms'', which retain their bonded interactions but do not interact with the rest of the system.  Bond parameters as well as partial charges between the changing atoms are adjusted accordingly between the initial and final state. 
Thus, in a single topology calculation, atoms may change their type, ensuring minimal  dummy atoms are created. This is illustrated in the left arm of Fig.~\ref{fig:fig_topology}. 

In contrast, in a dual topology alchemical free energy calculation, no atoms are allowed to change type~\cite{shirts2012best}. This means that the benzene to benzyl alcohol transformation involves starting with benzene plus the non-interacting dummy atoms making up the hydroxy methyl group, then passing through an intermediate state where some atoms are partially interacting--- particularly, those atoms which are becoming dummy atoms or ceasing to be dummy atoms~\cite{mobley2014blind}. The transformation finally culminates in a state where benzyl alcohol is present along with the additional dummy atom which was previously a corresponding hydrogen of the benzene. Fig.~\ref{fig:fig_topology}'s right branch depicts how such a dual topology works. 

Hybrid topology calculations have seen much less use but essentially consist of two absolute free energy calculations in opposite directions at the same time, turning one molecule's interactions with the environment off, while turning the other molecule's interaction on~\cite{jiang2019computing, rocklin2013separated}.

Most existing software implementations currently use single or dual topology approaches; for example, AMBER TI uses a dual topology approach, while BioSimSpace uses a single topology approach. Please make sure to check what approach is used with your software package of choice, or whether it supports your choice of approach (GROMACS, for example, supports both). 
To our knowledge efficiency differences have not been thoroughly explored, though conventional wisdom suggests that fewer dummy atoms are better, as introducing or removing atomic sites is usually more difficult, requiring more intermediate steps ~\cite{liu2013lead, mobley2012perspective}.

\paragraph{Atom mapping}
Once a particular approach to the topology is selected, a crucial next step is to identify the common atoms which will not be perturbed.
Rigorously, this process typically comprises a MCSS search of the molecules involved to identify the common substructure---though the parameters of the MCSS search will differ depending on whether single or dual topology calculations are planned.
Specifically, with a single topology approach in mind, atom types are allowed to change, so a permissive MCSS search can be done, whereas with dual topology a more strict search is required.
\begin{figure}
    \includegraphics[width=0.95\linewidth]{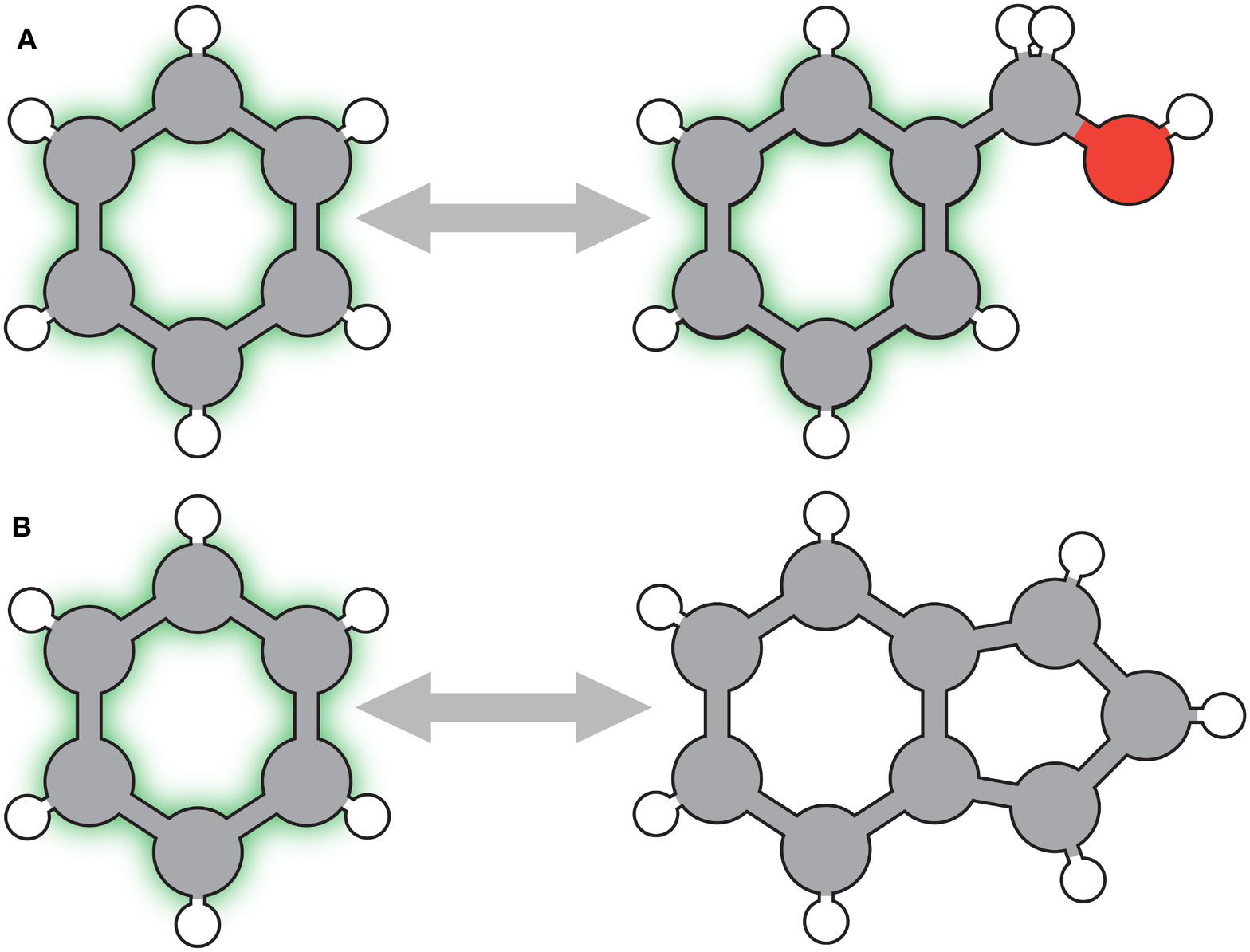}
    \caption{\textbf{Illustration of maximum common substructure matches} MCSS is shown in green for when (\textbf{A}) a restrictive MCSS match is used and in (\textbf{B}) ring breaking is allowed.}
    \label{fig:fig_mcss}
\end{figure} 

There are different tools that allow the generation of MCSS matches as well as single topology input. A large number of software tools can compute MCSS matches using different cheminformatics packages. Some rely on RDKit~\cite{rdkit2019Dec}, such as LOMAP~\cite{liu2013lead}, FESetup~\cite{loeffler2015fesetup} and partially BioSimSpace~\cite{hedges2019biosimspace}, while others such as fkckombu~\cite{kawabata20143d} are standalone tools. Schr\"{o}dinger's FEP+ planning tool was originally based on a version of LOMAP, and it also uses MCSS matching as well as 3D considerations to plan the network of single topology calculations between molecules~\cite{wang2015accurate}. 

MCSS searches can be relatively time consuming, so if the goal is to assess a library of ligands to identify promising pairs for relative calculations, it can be helpful to use faster approaches such as shape similarity to perform an initial similarity assessment and then use MCSS only to identify final mappings for relative calculations~\cite{raymond2002maximum,klabunde2012mars,jones2009elucidating}. The MCSS approach, though relatively standard, takes into account only topological similarity. It is possible that changes in binding mode could actually require a different choice of mapping, so in some cases mappings may need to be planned differently depending on 3D positioning of atoms in space.

Single topology relative calculations, and calculations based on substructure searches, only work if in fact the ligands share a common substructure, e.g. are part of a congeneric series, see Fig.~\ref{fig:fig_mcss}.
If no common substructure is shared, then alternative dual or hybrid topology free energy calculations are needed, where one would co-localize a pair of compounds in a binding site, exclude their interactions with one another, and compute the relative binding free energy by turning one molecule on from being dummy atoms while turning the other off.
To our knowledge no general pipeline for such calculations yet exists and this would likely remain a research problem. Using an absolute free energy approach instead seems more promising in such a case. 

\paragraph{Ring breaking and forming.} Relative free energy calculations for ring breaking and forming are particularly challenging/problematic (see Fig.~\ref{fig:fig_mcss} \textbf{B}), in part because relative calculations rely on the free energy contributions of dummy atoms canceling between different legs of the thermodynamic cycle, which may not be true whenever dummy atoms are involved in rings~\cite{liu2015ring}.
Some approaches have attempted to address this~\cite{clark2019relative} but a general solution is not yet in mainstream use, though FEP+ implements one solution.

\paragraph{Perturbation maps}
Based on the input ligand series, a perturbation map or network can be planned. Recent heuristics have shown the more connected the perturbation network the better, however, there is a way to optimize network structure while minimizing the number of perturbations that need to be computed reducing the resulting computational cost~\cite{yang2020optimal,xu2019optimal}. Sometimes the introduction of intermediates that are not part of the original congeneric series are essential to avoid ring breaking, or deal with perturbations that would otherwise result in large numbers of atoms being inserted or deleted. Some commercial tools have good underlying heuristics but may fail with complicated input, needing user validation in particular when dealing with chiral compounds. 

\begin{figure}
    \includegraphics[width=0.95\columnwidth]{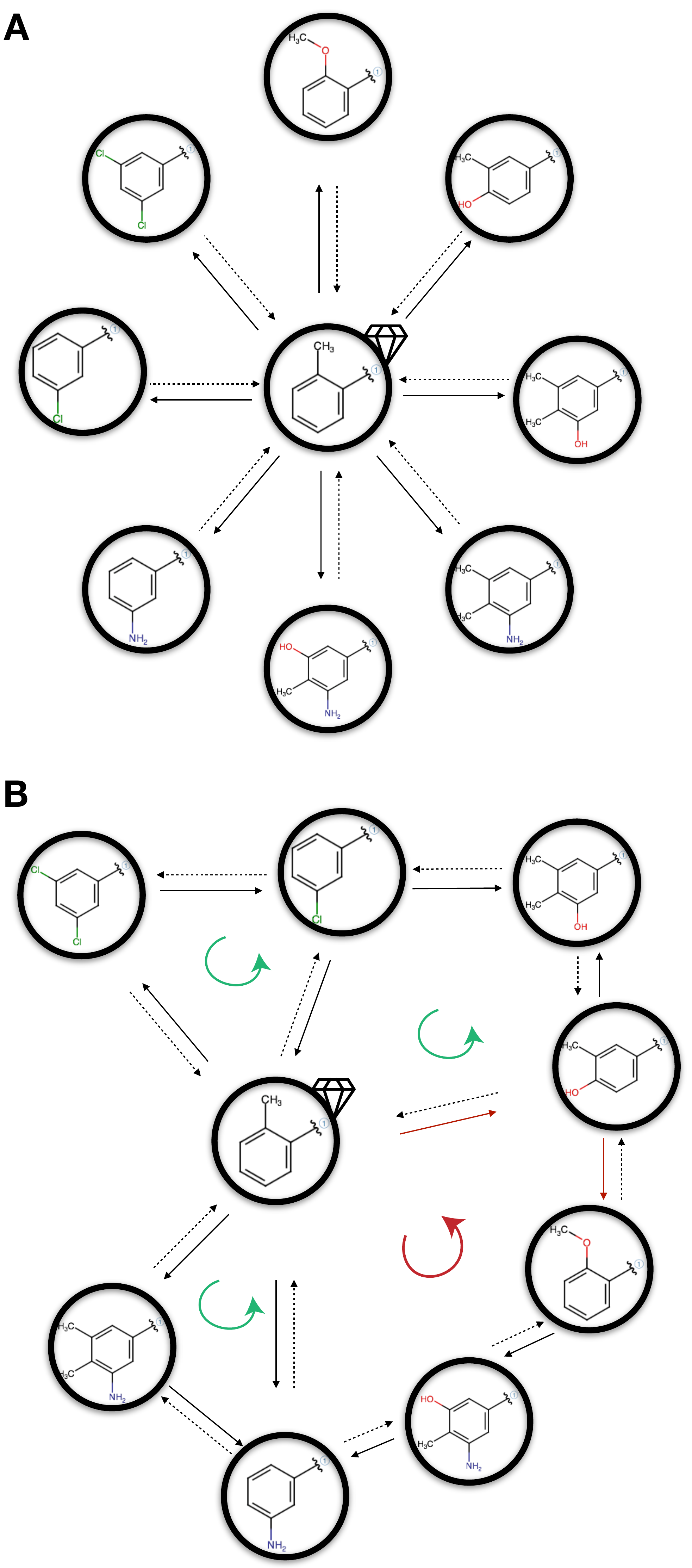}
    \caption{\textbf{Examples of perturbation networks} (\textbf{A}) Star shaped network with the crystal structure in the center. (\textbf{B}) Network with cycle closures (see more on this in Sec.~\ref{sec:are-they-good}). Arrows indicate the direction of the perturbation. Fully converged binding free energy calculations yield binding free energy changes which sum to zero around any closed cycle. However, in practice errors may not sum to zero around closed cycles, providing a way to look for potential sampling problems. Here in (B), green cycles indicate cycles with hypothetically good cycle closure, red those with poor cycle closure. The red arrow indicates a poorly converged simulation that would give rise to bad cycle closures. The diamond indicates the use of a crystallographic binding mode.}
    \label{fig:fig_types_of_networks}
\end{figure} 

In some cases, during the lead optimization stage, or for very large datasets that would benefit from rougher initial free energy ranking, or in cases where perturbations would be rather large a star shaped network as seen in Fig.~\ref{fig:fig_types_of_networks} \textbf{A} is used. However, adding redundancy into the network means that a better error analysis can be carried out, by looking at cycle closure errors as discussed in sec.~\ref{sec:are-they-good}, with an example given in Fig.~\ref{fig:fig_types_of_networks} \textbf{B}.

Methods in experimental design have been applied to the construction of the perturbation maps. Yang et al.~\cite{yang2020optimal} optimized the perturbation map by selecting a fixed number of calculations from the pairwise perturbations so that the resulting set of calculations minimize the total variance. Xu~\cite{xu2019optimal} optimized the perturbation map by allocating different amounts of simulation time to different pairwise perturbations so as to minimize the total variance, given the total simulation time of all the perturbation calculations. Both approaches lead to substantial reduction in the statistical error of the estimated free energies.

\paragraph{Constraints and relative free energy calculations}
One issue which requires particular care is the use of constraints.
Commonly, bonds involving hydrogen are constrained to a fixed length using algorithms such as SHAKE or LINCS, allowing the use of longer timesteps~\cite{krautler2001fast}.
However, in single topology relative free energy calculations, the atoms involved might be mutated to other atom types -- for example, in a mutation of methane to methanol, one hydrogen might become an oxygen atom.
Typical molecular dynamics engines are not set up to recognize this change, or at least not to correctly include contributions to the free energy from changing constraints/constraint length, so results for a transformation would usually be erroneous.
At present the most general solution to this problem is simply to avoid the use of constraints (and thus use a smaller timestep if necessary, usually of around 1 fs) in any relative free energy calculation involving a transformation of a constrained bond, as done by GROMACS.

\subsubsection{Absolute free energy calculations must handle the standard state and use restraints}
\label{sec:standardstate-restraints}

\begin{figure}
    \includegraphics[width=0.95\linewidth]{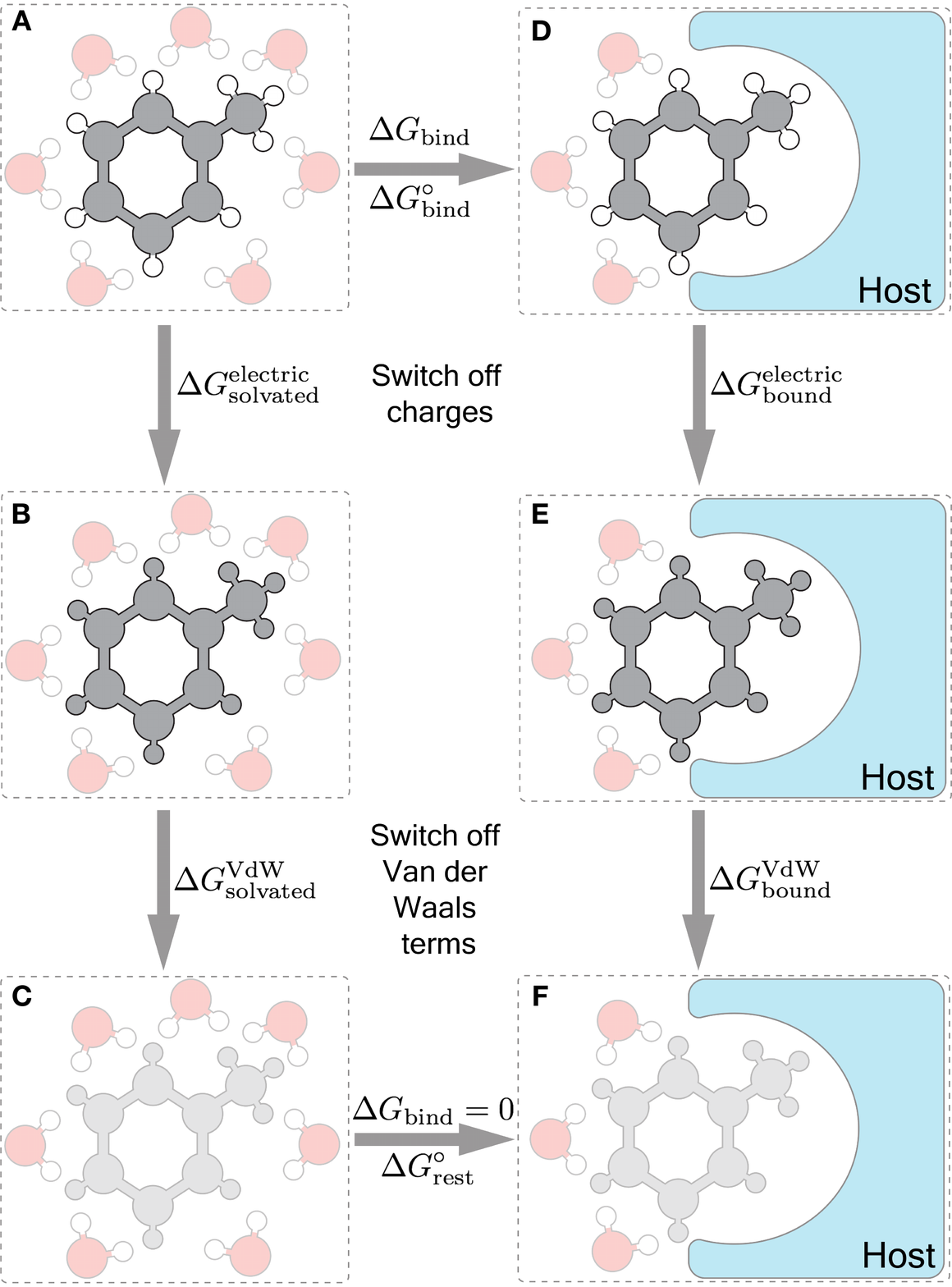}
    \caption{\textbf{Thermodynamic cycle required for an absolute free energy calculation -- absolute free energy of binding example} The fully interacting ligand in water (\textbf{A}), has its charges turned off to pass to (\textbf{B}) followed by turning of van der Waals terms, resulting in a non-interacting ligand in water in (\textbf{C}). Restraints are used on the fully interacting ligand in the binding site of a protein or host molecule (\textbf{D}). The next step is to turn off the charges again (\textbf{E}) followed by the van der Waals interactions resulting in a non-interacting complex state (\textbf{F}). Free energyes can be computed as $\Delta G_{bind} = (\Delta G^{\mathrm{elec}}_{\mathrm{solv}}+ \Delta G^{\mathrm{VdW}}_{\mathrm{solv}})-(\Delta G^{\mathrm{elec}}_{\mathrm{bound}}+ \Delta G^{\mathrm{VdW}}_{\mathrm{bound}}$).
    }
    \label{fig:fig_absolute_thermodynamic_cycle}
\end{figure}

Absolute free energy calculations involve completely removing the interactions between the ligand or solute and its environment, taking it to a non-interacting state that may or may not retain intramolecular non-bonded interactions.
This non-interacting state can then be shifted between environments (from the protein to water, or from one solution to another) without changing its free energy other than that due to the changing volume of the simulations, and then interactions can be restored in the new environment.

Absolute free energies are by definition reported with respect to a specific reference or standard state, which effectively determines the arbitrary point at which the free energy is 0.
The role of the standard state is particularly evident from the expression of the binding free energy between a receptor $R$ and ligand $L$
\begin{equation} \label{eq:DGfromKAB}
    \Delta G = -k_BT ~ \ln \left( c^{\circ} K_b \right)  = -k_BT ~ \ln\left( c^{\circ} \frac{[RL]}{[L][R]} \right) \, .
\end{equation}
Here, the reference state concentration $c^{\circ}$ converts the binding constant $K_b$ into a dimensionless quantity expressed in reference concentration units.
It should be noted that ignoring the term $c^{\circ}$ is equivalent to assuming a reference concentration of 1~D$^{-1}$, where D are the units used to express $K_b$, and would thus cause the value of $\Delta G$ to vary with the choice of the units.
It is convenient to define a standard state at a constant pressure of 1~atm and where each chemical species (i.e., A, B, and AB) in the reaction solvent has a concentration of $c^{\circ}$~=~1~M~=~1~molecule/1660~\r{A}$^3$ but do not interact with other molecules of $R$, $B$, or $RL$.

\paragraph{Handling the standard state in absolute free energy calculations.}
For solvation free energy calculations, handling the standard state is typically straightforward, and treating it correctly simply means ensuring that the non-interacting solute is taken to the same (or equivalent) final reference state in both environments, e.g. that the transformation involves a 1 M to 1 M equivalent transfer free energy (where the non-interacting solute still occupies essentially the same volume as the solute in the interacting system).
So typically in such cases no special care is required to ensure the correct standard state, as long as the \emph{experimental} data being analyzed uses the same standard state.
If this is not the case, a simple entropic correction is needed.

For binding, however, the situation is more complex and requires special care.
Because the simulations are typically performed using restraints and at concentrations that are different from 1 M, the expression of the free energy requires the following correction~\cite{gilson1997statisticalthermodynamic} (see an example of such a thermodynamic cycle in Fig.~\ref{fig:fig_absolute_thermodynamic_cycle})
\begin{equation}\label{eq:restraint-correction}
    \Delta G^{\circ}_{\mathrm{restr}} = -k_BT ~ \ln \left( \frac{c^{\circ}}{V_L} \right) -k_BT ~ \ln \left( \frac{\xi_L}{8 \pi^2} \right),
\end{equation}
where $V_L$ and $\xi_L$ are respectively the volume of the translational and rotational degrees of freedom of the non-interacting ligand in the simulation box.
When no restraints are used, the non-interacting ligand is free to translate and rotate in the simulation box (i.e., $V_L = V_{\mathrm{box}}$ and $\xi_L = 8\pi^2$), and the rotational term is zero.
A sufficiently thorough exploration of the simulation box by the non-interacting ligand is, however, required for the formula to be valid.
This is typically hard to achieve as the exploration process is governed by diffusion.
The addition of a restraint limits the volume available to the non-interacting ligand, thus speeding the convergence of the sampling.
In addition, when enhanced sampling methods such as Hamiltonian $\lambda$ exchange are used (see Sec.~\ref{sec:sampling_schemes}), the use of a restraint is typically necessary as it keeps the ligand in the binding site in the interacting state (see also Sec.~\ref{sec:weak-binders}) and generally reduce the round-trip time of replicas.
When restraint are employed, the values of $V_L$ and $\xi_L$ are restraint-dependent, but for commonly employed restraints, these can be usually easily computed analytically or numerically by solving the relevant integral.

\paragraph{Several choices of restraints are possible.}
In practice, a variety of types of restraints are common, from simple harmonic distance restraints between the ligand and the protein~\cite{mobley2006use}, to flat-bottom restraints which work similarly but only exert a force if the ligand leaves a specific region~\cite{chen2007can}.
Because these restraints do not limit the rotational degrees of freedom of the ligand, the rotational term entering the correction in Eq.~\ref{eq:restraint-correction} is zero.

Alternatively, a set of restraints proposed by Boresch have also commonly been employed, where all six rigid-body degrees of freedom governing the orientation of the ligand relative to the receptor are restrained~\cite{boresch2003absolutea, leitgeb2005alchemicala}.
Further restraints, such as on the overall ligand RMSD have also been used~\cite{woo2005calculationa}.

In principle, all of these forms will yield correct binding free energies in the limit of adequate sampling if their effects and connection to the standard state are correctly handled, but they have different strengths and weaknesses.
For example, with more involved restraints, sampling at intermediate $\vec{\lambda}$ values will usually not need to be as extensive but more computational effort must go to computing the restraining free energy.
Additionally, such restraints would typically keep the ligand from exploring alternative binding modes. This  restriction may be undesirable when using Hamiltonian $\lambda$ exchange or expanded ensemble techniques where allowing the ligand to exchange binding modes when it is non-interacting could provide sampling benefits~\cite{wang2013identifying}.
More specifically, flat-bottom restraints might allow a ligand to explore multiple binding sites, harmonic restraints multiple binding modes within a site, while Boresch restraints a single binding mode within a single site.
See additional discussion of the possibility of multiple binding modes in Sec.~\ref{sec:multiple_binding_modes} below.

Many choices of restraints involve selecting reference atoms.
Again, in principle this choice is unimportant given adequate simulation time but practical considerations may be important.
The choice is likely especially important with Boresch-style restraints, where some relative placements of reference atoms are likely to be numerically unstable; additionally, ligand reference atoms should likely be in a part of the molecule which defines the binding orientation well, rather than in a floppy solvent-exposed tail, for example.

\subsection{Absolute and relative calculations deal with some of the same issues}
\subsubsection{Handling weak binders and high dissociation rates}\label{sec:weak-binders}
In binding free energy calculations, only the conformations in which the receptor and ligand form a bound complex should be sampled from the bound states (Sec.~\ref{sec:theory}). Determining what the bound states actually are can be challenging for weakly bound ligands. For tightly bound ligands, virtually all reasonable definitions of the bound state will lead to be equivalent free energies, since the partition function will be dominated by a relatively small number of low-energy poses. For weak binders, this simplification breaks down.  In fact, the correct bound state may depend on the type of experiment performed.  For example, isothermal titration calorimetry (ITC) or surface plasmon resonance (SPR) measurements effectively define a binding state that includes all ligand comformations that are complexed with the protein, regardless of where on the protein they bind. In contrast, fluorescence polarization competition assays measure binding to only a single location, where the ligand of interest displaces a competing binder. Therefore, care must be taken to ensure that a reasonable definition of the binding site is used~\cite{wang2013identifying}.
In absolute calculations, this applies to the fully interacting state in the complex leg of the thermodynamic cycle (top-right state in Fig.~\ref{fig:fig_absolute_thermodynamic_cycle}), while in relative calculations this must be true at both end states of the complex leg (top- and bottom-right states in Fig.~\ref{fig:fig_binding_thermodynamic_cycle}).
In principle, this requires defining which conformations are considered to be bound before running the calculation, but it is common practice to start the simulation with the ligand already placed in the binding site and rely on kinetic trapping to maintain the bound complex.
This strategy, however, can fail when the dissociation rate of the ligand has the same or smaller order of magnitude than the length of the simulation.
This is typical of weak binders such as fragments binding shallow pockets with $\mu$M-mM affinities~\cite{georgiou2017pushing,pan2017quantitative}.
In this case, using a flat-bottom or harmonic restraint between receptor and ligand in the bound state(s) can prevent dissociations~\cite{georgiou2017pushing,rizzi2019sampl6}.
We stress that this is normally avoided as it generally introduces bias in the free energy estimate, which is why the restraint is usually activated only in the intermediate states in absolute calculations.
The bias can be corrected through reweighting schemes~\cite{rizzi2019sampl6}, but this post-processing step can be avoided if a flat-bottom restraint is used and the ligand never hits the potential wall during the simulation in the bound state.
It is important to note that the spring constant and/or radius parameters of the restraint effectively determine which conformations are considered to be bound.
As a consequence, these parameters must be tuned to the system so that only the binding site is accessible to the ligand.
Again, this step is particularly important for weak binders as their free energy of binding is known to be more sensitive to the definition of the binding site~\cite{gilson1997statisticalthermodynamic}.

In absolute calculations, this restraint can substitute or be added to the restraint used to handle the standard state correction (Sec.~\ref{sec:standardstate-restraints}).
In the latter case, to compute the standard state correction analytically, the bound-state restraint must be turned off in the decoupled state.
Alternatively, a flat-bottom restraint can be activated also in the decoupled state as long as the second restraint (e.g., a harmonic or Boresch restraint) prevents the ligand to hit the wall of the flat-bottom potential~\cite{rizzi2019sampl6}.
Finally, even for tight binders, dissociation events can be enhanced by methods such as Hamiltonian replica exchange~\cite{sugita2000multidimensionala,chodera2011replica,wang2013identifying} and expanded ensemble~\cite{lyubartsev1992newa,li2007simulated}, especially in absolute free energy calculations using harmonic or flat-bottom restraints.
In the latter case, dissociations can be averted simply by increasing the spring constant and/or reducing the radius of the restraint potential to prevent the exploration of ligand conformations outside the binding site in the decoupled state (bottom-right state in Fig.~\ref{fig:fig_absolute_thermodynamic_cycle}) that could be propagated to the bound state.

\subsubsection{Changes in net charge can be challenging/problematic.}
If the net charge of the system changes as the alchemical variable changes during the  calculation, this can pose major challenges.
Specifically, finite-size effects can introduce significant charge-dependent artifacts into computed binding free energies, in part because typical schemes for long-range electrostatics (including PME and reaction field) do not handle free energy contributions from such changes effectively or as they would be handled in a hypothetical macroscopic bulk solution~\cite{lin2014overview, ohlknecht2020correcting, rocklin2013calculating}.

There are two main potential solutions to avoid artifacts due to changes in net charge: Correcting for the introduced artifacts, or avoiding changing the net charge.

Many relative free energy planning tools have been set up to avoid changing the net charge of the systems considered, including LOMAP~\cite{liu2013lead} and Schr\"{o}dinger's FEP+~\cite{wang2015accurate}. Absolute free energy calculations can also potentially avoid changing the charge of the system by making a charge perturbation of equal and opposite sign elsewhere in the system; for example, as a charged ligand is removed, a charged counterion of opposite sign could also be removed, or one of the same sign could be inserted. This is sometimes referred to as an "alchemical ion" approach for dealing with the needed charge change, and is also employed by the Yank free energy package~\cite{wang2013identifying}.
Charge corrections have also been explored, and are potentially a viable solution to this problem~\cite{mey2018impact} where artifacts introduced by finite-size effects are corrected numerically~\cite{chen2018accurate, ohlknecht2020correcting}. However, application of such corrections typically remains less common than the use of a co-alchemical ion.

When free energy calculations \emph{do} need to change the charge of a ligand or solute, the literature does not yet seem to indicate what approach should be preferable, so considerable care should be taken.
We are not yet aware of a careful comparison of charge corrections versus other approaches such as decoupling an ion at the same time, so in our view the issue of proper handling of charge mutations in the context of alchemical calculations remains a research problem.

\subsubsection{The importance of the alchemical pathway
\label{sec:important_path}}
\begin{figure}
    \includegraphics[width=0.95\linewidth]{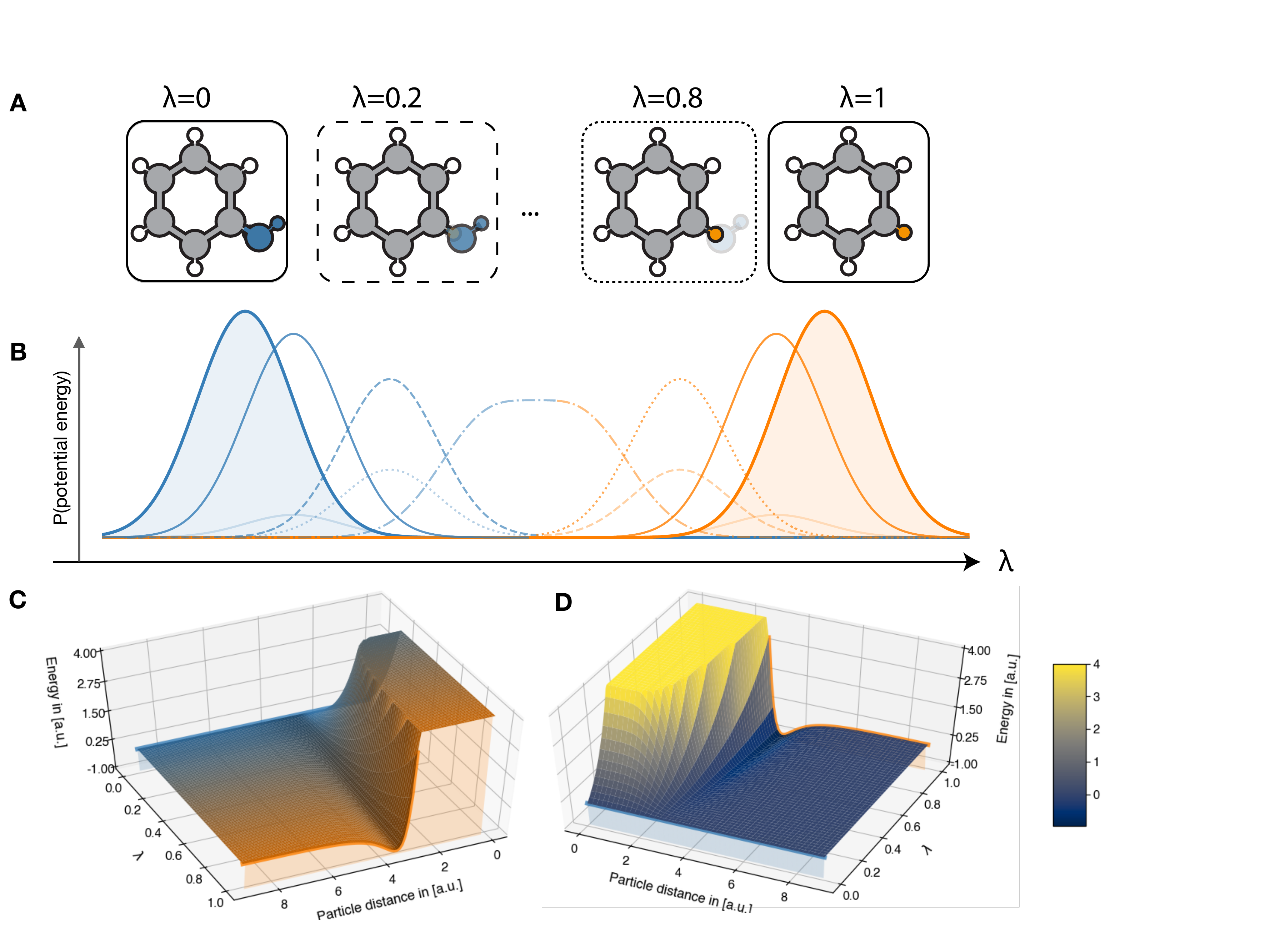}
    \caption{Alchemical intermediates are created by making the potential energy depend on an additional variable $\vec{\lambda}$ that interpolates between the chemical endpoints. In (\textbf{A}), at $\vec{\lambda}=0$ the molecule is a fully interacting phenol and at $\vec{\lambda}=1$,  a fully interacting benzene. (\textbf{B}) shows an illustration of the probability distribution of the potential energies as the switching function takes values of $\vec{\lambda}=0$ to $\vec{\lambda}=1$. Intermediates states are required for a sufficient overlap in potential energies to estimate a free energy difference between $\vec{\lambda}=0$ and $\vec{\lambda}=1$.
    Soft-core potentials provide one of the most efficient families of intermediate pathways, with a $\vec{\lambda}$ dependence. In (\textbf{C}) the potential energy surface is coloured according to $\vec{\lambda}$ with blue being $\vec{\lambda}=0$ and $\vec{\lambda}=1$ orange. In (\textbf{D}) the potential is coloured according to the potential energy. Note how as $\vec{\lambda}$ approaches 0, the energy smoothly approaches zero at all $r$, a necessary requirement for efficient and stable calculations.  }
    \label{fig:fig_what_is_lambda}
\end{figure}

Both absolute and relative calculations must choose an alchemical pathway connecting initial and final states. In principle, because of the path independence of the free energy, any arbitrary pathway will give the correct free energy change, but the choice of pathway will greatly affect the efficiency of the calculations. 

Some choices are particularly crucial---for example, transformations involving insertions or deletions of atoms should employ a soft-core potential path for Lennard-Jones or other interactions with repulsive interactions that go to infinite energy at small radius~\cite{beutler1994avoiding, beutler1994molecular,gapsys2012new}.

The key consideration for choosing alchemical pathways is that the intermediate states that a given pathway produces should sample configurational ensembles that change as slowly as possible as $\vec{\lambda}$ changes, while still managing to go from the initial state to the final state as $\vec{\lambda}$ goes from 0 to 1.

Another way of stating this is that intermediate states should sample molecular configurations that are as similar as possible to their neighboring states. The more similar the configurations are between intermediate states, the lower the statistical uncertainty is in the estimate of free energy between intervals. This can be proven directly from the BAR and MBAR formulas~\cite{bennett1976efficient,klimovich2015guidelines}, though the exact same principles apply for TI. For a 'good' path to work and give a sequence of states with maximally similar configurations, sufficient similarity in potential energies is required. Fig.~\ref{fig:fig_what_is_lambda} \textbf{A} and \textbf{B} illustrate this. Fig.~\ref{fig:fig_what_is_lambda} \textbf{A} shows in a pictorial way a soft-core potential can be applied across different $\vec{\lambda}$s. Fig.~\ref{fig:fig_what_is_lambda} \textbf{B} illustrates the potential energy distributions at the different $\vec{\lambda}$ intermediates, with sufficient overlap between neighboring $\vec{\lambda}$ states to ensure that reweighting estimators such as MBAR can be used for analysis (see Sec.~\ref{subsec:estimators}). The actual transformation is best handled with soft-core potentials of the form shown in Fig.~\ref{fig:fig_what_is_lambda} \textbf{C} and \textbf{B}, with more details given below. 

So what are the options to adjust the potentials between the two end states based on $\vec{\lambda}$? The simplest possible alchemical pathway is a \textit{linear} pathway:
\begin{equation}
U(\vec{q},\vec{\lambda}) = \vec{\lambda} U_0(\vec{q}) + (1-\vec{\lambda})U_1(\vec{q}) \end{equation}

so-called because the dependence on $\vec{\lambda}$ is linear. This clearly satisfies the basic requirement that it gives the initial endpoint potential energy $U_0(\vec{q})$ when $\vec{\lambda}=0$ and final endpoint energy $U_1(\vec{q})$ when $\vec{\lambda}=1$. 

For many energy terms, \textit{as long as a repulsive core remains on}, this is a very good approach. For example, it can be shown that if van der Waals repulsions are left on, then the linear approach is very nearly the optimal path possible for changing, removing, or inserting the electrostatic energy terms, with the path being within about 10--20\% of the minimum possible uncertainty~\cite{naden2015linear} for a fixed amount of simulation, as well as being nearly optimally efficient for van der Waals attractive terms with repulsion terms turned on~\cite{naden2014linear}. Although we are not aware of any quantitative tests for dipolar or higher multipole terms, theoretically it should behave equally well for those systems.

However, this approach ends up being terrible for removing or adding repulsive potentials that go to infinity quickly at or near the origin. One way to look at this is to examine how low $\vec{\lambda}$ values must go to reduce the energy at $0.5\sigma$ (the atomic size parameter) down to 1 $k_BT$, where thermal fluctuations make it possible for other atomic sites to penetrate routinely that deep. If we are trying to go from a particle being present, and desire to make it disappear alchemically, then if the repulsive terms are of the form $\epsilon\frac{\sigma}{r}^{12}$, then if $\epsilon$ was 1 $k_BT$ at the temperature of interest, and we start with the particle present, then solving for $(1-\vec{\lambda})(1 kB_T)\left(\frac{\sigma}{\frac{1}{2}\sigma}\right)^{12} = 1 k_B T$ we get $\vec{\lambda} = 1-2^{-12} \sim 0.999976$. At this point,  we have gone virtually all the way to the end of the transformation, but there is still an impenetrable post in the middle of our simulation! This is not very much like the desired final state of no interactions between the particle and its environment. We can play around with a few ways of modifying this, like simulating many more intermediate states near $\vec{\lambda}=1$. However, various analyses have shown that this is not a very good strategy~\cite{pham2011identifying, beutler1994avoiding, zacharias1994separationshifted, blondel2004ensemble, gapsys2012new}.

What we need instead is a function that smoothly gets rid of this infinity. A large number of schemes have been tried~\cite{beutler1994avoiding, zacharias1994separationshifted, blondel2004ensemble, pham2011identifying, pham2012optimal, naden2014linear, donnini2005molecular}, but the most common strategy that appears to be the best practice is to use a "soft-core" potential, of the form:

\begin{equation}
    U(\vec{r_{ij}},\vec{\lambda}) = 4\epsilon_{ij} \vec{\lambda} \left(\frac{1}{(\alpha(1-\vec{\lambda}) + (r_{ij}/\sigma_{ij})^6)^2} -  \frac{1}{\alpha(1-\vec{\lambda}) + (r_{ij}/\sigma_{ij})^6}\right)
    \label{eq:softcore},
\end{equation}

where $r_{ij}$ is the distance between two particles $i$ and $j$, $\epsilon_{ij}$ and $\sigma_{ij}$ are the Lennard-Jones parameters corresponding to the interaction between particles $i$ and $j$, and $\alpha$ is a constant (0.5 is optimal for the specific functional form shown above). This functional form has exactly the property we are looking for: it recovers the Lennard-Jones potential when $\vec{\lambda}=1$, and the other endpoint ($\vec{\lambda}=0$), it is exactly zero for all $r_{ij}$ everywhere, and as $\vec{\lambda}$ goes to zero, the $\alpha(1-\vec{\lambda})$ term lowers the infinite energy in the core. There are several different variants of the same functional form~\cite{zacharias1994separationshifted, beutler1994avoiding,pham2011identifying}, but the one given in eq.~\ref{eq:softcore} is easy to understand and implement and fairly numerically stable. This functional form is shown in \textbf{C} and \textbf{D} of Fig.~\ref{fig:fig_what_is_lambda}.

It has been shown that more complicated forms are not significantly more efficient than eq.~\ref{eq:softcore}~\cite{pham2012optimal}. We therefore recommend using the softcore potential given in eq.~\ref{eq:softcore}, unless there is a compelling reason otherwise. Using a similar equation to eq.~\ref{eq:softcore} may be acceptable in most circumstances if that is what is supported in your chosen software. However, if you are inserting or removing entire atomic sites, we heavily recommend against using the linear approach; it will be very difficult to get correct and converged results. 

So far in this section, we have discussed optimal ways of disappearing or appearing Lennard-Jones interaction sites and turning on and off electrostatics terms. What about performing both transformations at the same time? We can not turn off the electrostatics linearly at the same time we turn off the Lennard-Jones terms, as it would leave infinitely large attractive and repulsive electrostatic terms "bare" at small $\vec{\lambda}$, resulting in the simulation crashing. It \textit{is} possible to apply the same soft core approach to the Coulomb interaction and this is indeed done in a number of implementations, in which case it is important that the Coulomb interaction is softened more rapidly than the Lennard-Jones interaction to avoid charge penetration issues, which  can be tricky to ensure for all types of perturbations ~\cite{steinbrecher2011softcore}. 

A safe but potentially more computationally expensive approach is to perform the transformations in sequence; first, turning off all electrostatics for atoms that must be removed, inserting and removing Lennard-Jones sites (both the insertion and removal can be done simultaneously), and then turning electrostatics for the introduced particles on. Again, If there are no removals or introduction to atomic sites, then it is reasonable to change the interactions in the first and third steps  linearly. 

Other issues, such as whether absolute calculations should retain or remove intramolecular non-bonded interactions
through either annihilation~\cite{hermans1997inclusiona, mann2000modelinga, boresch2003absolutea, wang2006absolutea, mobley2006use} or decoupling~\cite{fujitani2005directa, mobley2006use} must be considered. Reasonable efficiency can be often obtained with either choice even if some are somewhat better or worse than others, and there is no consensus on which is better in most given situations. Our recommendation is to leave the intramolecular interactions on during the transformation for simplicity if there are no other known issues with this approach. The key thing to watch out for is whether the total potential energy, and therefore the intermediate ensembles sampled, change smoothly from beginning to end. These problems can be diagnosed by noticing lack of configuration space overlap between different simulations (see Sec.~\ref{sec:are-they-good}).

Relative calculations introduce additional choices, such as the order in which to modify nonbonded interactions.
A common process in single topology relative calculations is to first remove electrostatic interactions of any atoms which will be deleted, then modify other non-bonded interactions, then restore electrostatic interactions of any atoms which are being inserted. Although this is a simpler path to understand cognitively and can take advantage of the soft-core potential from eq~\ref{eq:softcore}, this can lead to more intermediate steps and thus be more computationally expensive.
Other schemes, such as simultaneously changing electrostatic and Lennard-Jones interactions with electrostatic ``soft-core'' potentials~\cite{steinbrecher2007nonlinear} may be implemented with fewer intermediate but could require fine-tuning of electrostatic and Lennard-Jones softcore parameters to avoid numerical instabilities. 
At the time of writing, there has not been conclusive evidence to suggest one approach is better in general than the other, so discretion should be left up to the user as to what is viable from both hardware resources, and what the simulation software supports.

A key additional consideration in choosing the alchemical pathway is the choice of spacing of intermediate states.
The spacing depends to some extent on the choice of analysis method, though states should essentially be spaced equidistant in the relevant thermodynamic length~\cite{crooks2007measuringa, sivak2012thermodynamic}.
For BAR/MBAR techniques this means that states should be spaced so that the statistical uncertainties between neighboring states be equal;~\cite{pham2012optimal, shenfeld2009minimizing}
Some schemes to adaptively optimize the spacing of intermediate states based on initial exploratory simulations have been proposed~\cite{hayes2017adaptive}. For molecules changing in dense solvent, then the best path is roughly independent of molecule size and shape, so what works for one molecular transformation is likely to be relatively efficient for another~\cite{monroe2014converging}.

\subsubsection{Which sampling scheme will work best for my problem?}
\label{sec:sampling_schemes}
\begin{figure}
    \includegraphics[width=0.88\columnwidth]{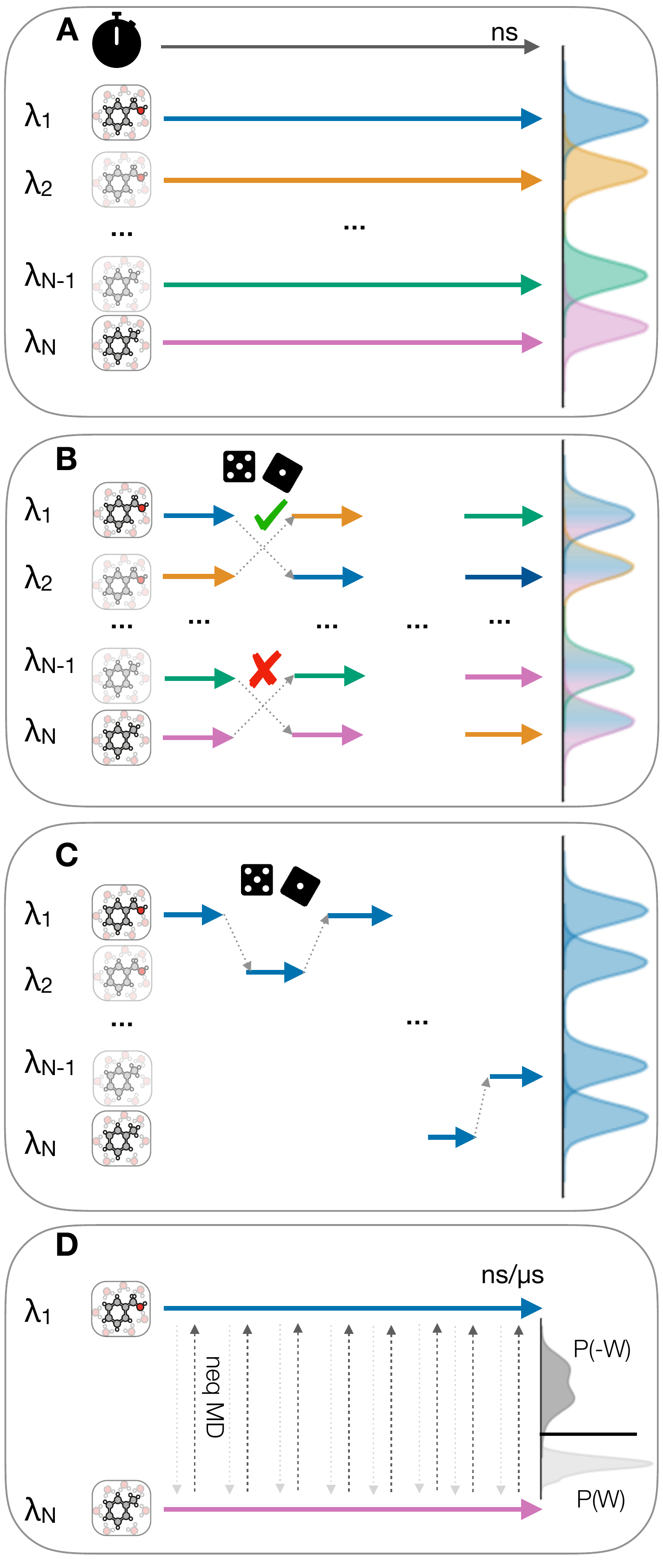}
    \caption{\textbf{Four most common sampling strategies.} (\textbf{A}): Multiple replicas in parallel at different lambda states. Each arrow symbolises an independent $\vec{\lambda}$ simulation. (\textbf{B}): Hamiltonian replica exchange scheme. Each arrow represents a short simulation interval before an exchange through Metropolis Hastings acceptance (dice) is attempted. A tick means an accepted exchange a cross a rejected exchange. (\textbf{C}): Single replica scheme sampling from all $\vec{\lambda}$ states. After a short simulation time symbolised by the arrow, the lambda-state is attempted to change until all N lambda states will be sampled. (\textbf{D}): Non-equilibrium sampling scheme, where two equilibrium simulations at the end-states are run as indicated by the blue and pink arrow. Non-equilibrium simulations are attempted at intervals to switch between the two end-states.}
    \label{fig:fig_sampling_scheme}
\end{figure}

Though all alchemical simulations must sample from multiple $\vec{\lambda}$ states, different approaches can be used to achieve this. Fig.~\ref{fig:fig_sampling_scheme} illustrates the four most common schemes. The simplest approach involves running an independent simulation at each of the predefined $\vec{\lambda}$ values (see Fig.~\ref{fig:fig_sampling_scheme} \textbf{A}). This type of scheme is currently used for AMBER TI calculations~\cite{song2019using} and for Sire as implemented in BioSimSpace~\cite{hedges2019biosimspace}. However, if these simulations can be run simultaneously with communication between them, a simple extension allows mixing between these replicas. In this approach, the simulation at each $\vec{\lambda}$ can undergo periodic exchanges with neighboring $\vec{\lambda}$ values. This form of replica exchange (Hamiltonian replica exchange) is based on ideas developed from Monte Carlo simulations of spin glasses by Swendsen and Wang~\cite{swendsen1986replica}. With the Metropolis-Hastings acceptance criterion for exchanges, the generated ensemble of all replicas still samples from the Boltzmann distribution, thus this approach has been used in many different contexts for molecular simulations~\cite{sugita2000multidimensionala,sugita1999replicaexchangea, woods2003developmenta, jiang2010free}. The basic idea of the replica exchange scheme is shown in Fig.~\ref{fig:fig_sampling_scheme} \textbf{B}. It is supported in various software packages that provide alchemical implementations, such as GROMACS~\cite{aldeghi2015accurate}, FEP+~\cite{wang2015accurate}, and NAMD~\cite{jiang2019computing}. A third approach borrows ideas from simulated tempering~\cite{marinari1992simulateda}. In this scheme a single replica rapidly explores all of $\vec{\lambda}$ space by working out optimal weights that allow switching between different intermediate $\vec{\lambda}$ values, as seen in Fig.~\ref{fig:fig_sampling_scheme} \textbf{C} . This approach is also referred to as self adjusted mixture sampling~\cite{lyubartsev1992newa, li2007simulated, tan2017optimally} and while promising, has so far only been supported in OpenMM Tools~\cite{andrearizzi2019choderalab}. The last approach makes use of non-equilibrium simulations~\cite{aldeghi2018accurate}. In this approach, only end state $\vec{\lambda}$ replicas ($\vec{\lambda}$=0, $\vec{\lambda}=1$) are simulated at equilibrium; intermediate information is generated from non-equilibrium simulations that rapidly transition between end-states. This approach is available in GROMACS and appears to be coming online in several other packages. A schematic of this approach is shown in Fig.~\ref{fig:fig_sampling_scheme} \textbf{D}. 

Currently, we recommend using Hamiltonian replica exchange type sampling schemes (Fig.~\ref{fig:fig_sampling_scheme} \textbf{B}). If these are not available in the code of choice, running independent simulations at different $\vec{\lambda}$ values can be acceptable, especially when conformational sampling is fast (Fig.~\ref{fig:fig_sampling_scheme} \textbf{A}). Single replica schemes and non-equilibrium schemes are not as established yet, but are very promising.

\subsubsection{How long should I run my simulation for and what information should be saved?}
\label{sec:sim_length_information_kept}
Before launching alchemical free energy calculations it is wise to consider how convergence and completion will be assessed. Different conditions on when to stop alchemical free energy calculations should be determined, and this may require several iterative checks and therefore modifications to the calculation protocol.
One useful metric to use for termination is the expected or desired uncertainty of a desired free energy estimate, though care must be exercised should the uncertainty estimate prove unreliable.
In particular, if the rate of change in the free energy estimate is significant when this condition is met, the simulation may not be locally converged, and more sampling may be necessary to determine a stable free energy estimate which is no longer changing significantly over time. 
However, this is not the only metric which should be used, as the uncertainty only captures the information about the sampled phase space, not necessarily the entirety of the phase space.  
For example, convergence of relative free energy calculations in predictive simulations where the entire phase space is not known in advance, requires sampling the different kinetically stable states~\cite{mobley2012perspective}. 
This highlights the importance of choosing the correct thermodynamic path to ensure you sample the required thermodynamic states as discussed in Sec.~\ref{sec:important_path}.

The condition of minimizing the statistical uncertainty of different free energy estimators below a sufficient threshold should be one metric monitored over the simulation. This can be done through the uncertainty estimator built into certain analysis tools such as MBAR, or can be done though more general statistical tools like bootstrap sampling. 
A target statistical uncertainty should be chosen at the onset of the simulation to avoid excessively long simulations, or falling into the trap of running until the free energy estimate is "good enough," which is subjective and has no defined criteria. This could be a fixed value such as $0.20 \mathrm{kcal/mol}$, or a functional quantity such as "below $0.5 \mathrm{kcal/mol}$ and $10\%$ of the free energy estimate." The user does not need to monitor this information in real-time and can choose to run simulations for fixed duration (either time or number of samples) and run analysis on the data collected thus far. If more samples are needed, the simulations can be resumed, or, started again in different initial conditions. 

\begin{figure}
    \centering
    \includegraphics[width=0.95\linewidth]{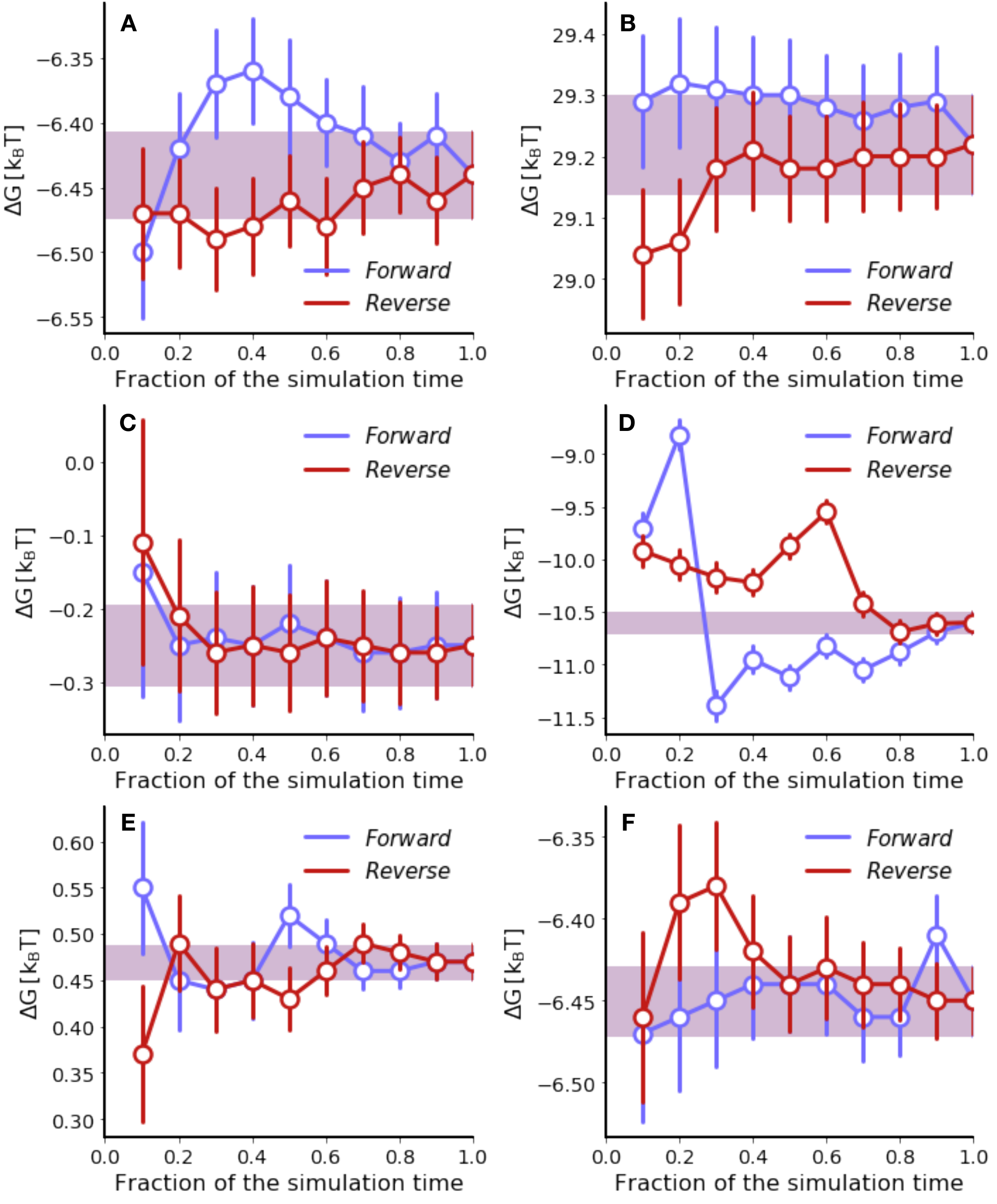}
    \caption{\textbf{Free energy (in k$_{B}$T) for two different relative binding free energy perturbations.} 
    Each plot shows the estimated free energy change using a varying fraction of total simulation time (up to 5 ns total). 
    Subplots (\textbf{A}), (\textbf{C}), and (\textbf{E}) show a three step protocol for a perturbation involving 3 perturbed atoms, while (\textbf{B}), (\textbf{D}), and (\textbf{F}) shows the same protocol for a perturbation involving 10 perturbed atoms. The first step of the protocol is the decharging then removing van der Waals interactions and then recharging. The difference in energy between the forward (blue) and reverse (red) free energy calculations at the midpoint of the simulation time gives an indication of the overall convergence of the simulation, with differences over 1 k$_{B}$T indicating poor convergence.}
    \label{fig:convergence_forward_reverse}
\end{figure}

Convergence in other alchemical observables should also be monitored to determine if the defined phase space has been sufficiently sampled and enough decorrelated samples have been drawn. These additional observables include, but are not limited to, the variance in $\frac{dU}{d\vec{\lambda}}$ across all $\vec{\lambda}$ values, calculating the variance in free energy using bootstrap analysis, and comparing differences in free energies calculated using different percentages of the simulation in both the forward and reverse directions ( see Fig. ~\ref{fig:convergence_forward_reverse}).

Each of these metrics have demonstrated promising results for diagnosing when a simulation has a convergence issue beyond simple convergence of uncertainty estimate. 
Results obtained from calculations with convergence issues should be checked for errors or run for longer before any confidence should be placed in conclusions drawn from their analysis.
In relative calculations that share similar binding modes, for example, and do not induce large conformational changes when in complex with protein, the need to sample exhaustively to converge estimates in free energy differences is often not necessary due to the locality of sampling changes in the molecular topology and shared phase space of the core atoms.
However, even subtly induced changes in protein binding configuration will require more sampling or cause local convergence to a free energy estimate that has high error.
The confidence a user should have in a free energy estimate is significantly improved when both the uncertainty of the free energy estimate is low, and when other observables have reached a convergence.

\begin{figure}
    \centering
    \includegraphics[width=0.8\linewidth]{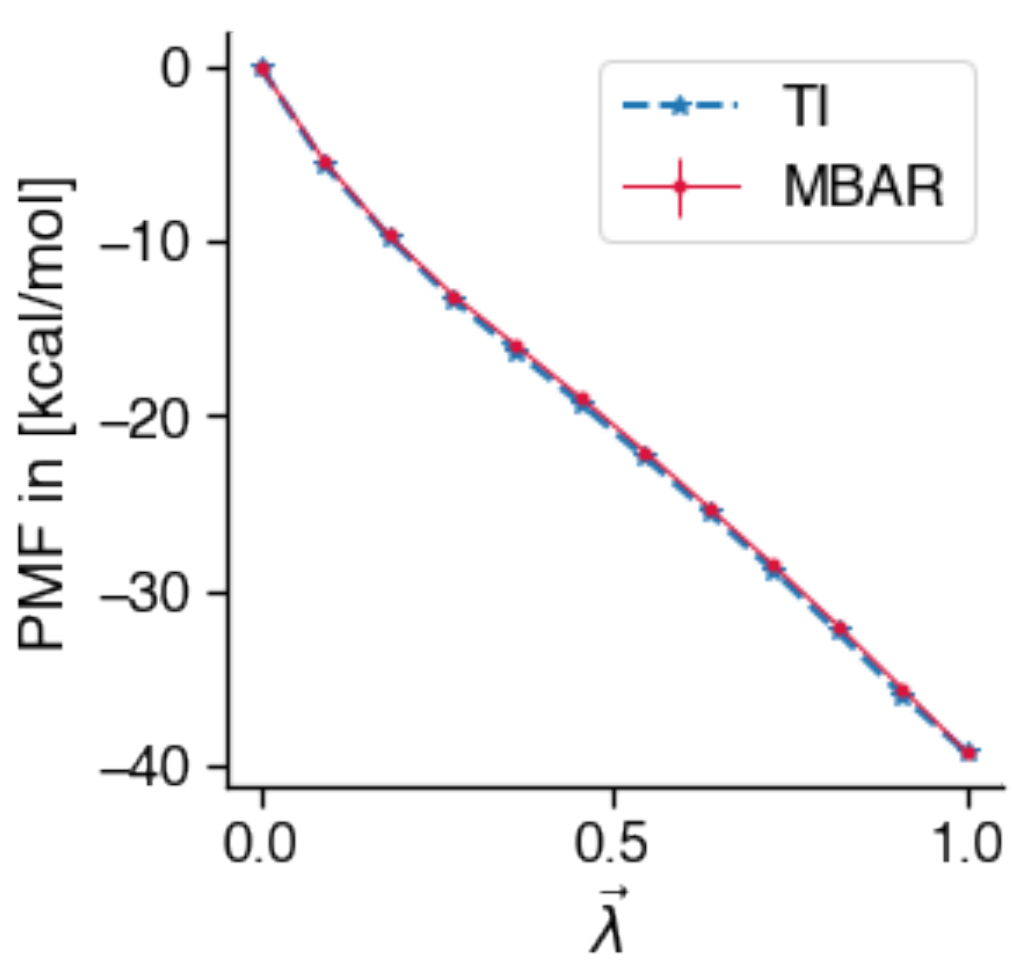}
    \caption{{\bf Potential of mean force with respect to $\vec{\lambda}$ for TI and MBAR}
    The estimated PMF for a bound calculation of a Tyk2 ligand pair of the Wang et al.~\cite{wang2015accurate} with respect to $\vec{\lambda}$ estimated from TI and MBAR and showing agreement within errorbars. 
    }
    \label{fig:pmf}
\end{figure}

The uncertainty in the free energy, for example, has multiple ways to be estimated, e.g. through standard error propagation methods (including MBAR's estimator, which is based on the same principles as standard error propagation), through bootstrap methods, through multiple independent runs, etc. 
Independent of how the property is estimated, its important to remember that they are \textit{estimations of the property}, not the true underlying property itself. 
These estimators are usually consistent estimators, meaning they will converge to the true answer in the limit of sufficient sampling, not necessarily unbiased ones though.
As such, it is a good idea to subject different estimators to the same data to see if they yield either the same estimate (within error and bias), or if they fluctuate wildly. See for example the potential of mean force with respect to $\vec{\lambda}$ estimated from a bound simulation of a Tyk2 ligand pair of Wang et al.~\cite{wang2015accurate} for both the MBAR and TI estimators, as seen in Fig.~\ref{fig:pmf}.
This is not a perfect method as some estimators, such as exponential averaging, will converge significantly more slowly, relative to more accurate estimators like MBAR. 
Therefore, it is a good idea to apply the estimators to different fractions of the data to see if the main estimator of free energy you have chosen is stable.

Each method requires different data from the simulation be collected. If, for instance, the free energy estimator selected is thermodynamic integration, then values of $\frac{dU}{d\vec{\lambda}}$ at uncorrelated data points must be collected. Once a combination of knowing what type of simulation you will run, which alchemical topology you will simulate, what alchemical path you will simulate along, and what your stopping conditions are, then you are ready to enumerate the information you should capture. Below is a sample of the minimal information you need for a set of common estimators (discussed in more detail in Sec.~\ref{subsec:estimators}):

\begin{itemize}
    \item Thermodynamic Integration (TI) requires $\frac{\partial u(\vec{q})}{\partial\vec{\lambda}}$.
    \item Exponential Averaging (EXP) needs \textit{either} $\Delta u_{k,k+1}(\vec{q})$ or $\Delta u_{k,k-1}(\vec{q})$, depending on the direction its being evaluated in.
    \item Bennett Acceptance Ratio (BAR) needs \textit{both} $\Delta u_{k,k+1}(\vec{q})$ and $\Delta u_{k,k-1}(\vec{q})$.
    \item Weighted Histogram Analysis Method (WHAM) and Multistate Bennett Acceptance Ratio (MBAR) both need the complete set of $\Delta u_{k,j} \, \forall \, j=\{1...K\}$. WHAM must have this information binned.
\end{itemize}

The potential derivative required for TI should generally be calculated during the simulation; only under very rare circumstances~\cite{naden2015linear} can it post-processed by a code that does not evaluate the derivatives. Many codes already have options for doing this.
If that option is unavailable, you can estimate it through finite difference (if sufficient information is collected), but this will introduce significant error, and is generally not a best practice. The BAR estimator may be a better, and simpler choice at that point as you will have at least the same level of information. 
The potential energy differences required for EXP, BAR, MBAR, and WHAM can be calculated either during the simulation or in post-processing. It is recommended to calculate the potential differences in code when possible to avoid extra overhead and possible errors produced by running the simulation twice, and to reduce the amount of stored information. 
Although TI must usually be calculated in code, as it requires the derivative, there is one condition under which it actually has the fastest computation time. 
If the alchemical path you have chosen is a linear alchemical path, then you get $\frac{du}{d\vec{\lambda}} = u_0(\vec{q}) - u_1(\vec{q})$, which is the difference between the initial and final states. 
However, because of the problems with linear paths already discussed in this paper, this simplification is rarely that useful.

Free energy information should generally be saved more frequently than coordinate data, approximately at the rate that uncorrelated samples are produced.  
The on-disk size of the data for free energy estimation is often significantly smaller than full atomic coordinates, so the information should be collected frequently. 
However, the information should not be collected \textit{every} time step, as most free energy techniques are operated at equilibrium, and need equilibrated \textit{and decorrelated} samples for an unbiased estimate.
A sample collected every time step will likely result in most samples being discarded due to decorrelation routines in the analysis. However, if it is computationally cheap and disk space is plentiful, do save often. One may safely assume that the correlation time is greater than 100-200 fs even for relatively simple systems such as small molecules in solvent, so saving no more frequently than every 50-100 steps is recommended. 
How decorrelation impacts calculations, and how to compute it is discussed in Sec.~\ref{sec:decorrelating-samples}. 

\subsubsection{Multiple or uncertain binding modes may require considerable care}
\label{sec:multiple_binding_modes}
In a discovery setting, new ligands can have unknown or at least uncertain binding modes~\cite{kaus2015how, plountprice2000analysis,mobley2009binding,calabro2016elucidation}, complicating binding free energy estimation.
This uncertainty is because it is usually not desirable to estimate a binding affinity for a ligand which already has an available bound structure, since such a compound has already been tested.
To deal with prospective ligands with unknown binding modes, discovery projects commonly assume that modifications of functional groups on a common scaffold result in a consistent binding mode across all members of a series.
This is not necessarily always the case~\cite{kaus2015how}, as reviewed elsewhere~\cite{mobley2009binding} and in some cases unexpected binding mode changes can be the origin of apparent non-additivity in structure-activity relationships~\cite{calabro2016elucidation}.
Binding modes also tend to be particularly variable in the case of fragments, which often may have multiple relevant binding modes~\cite{steinbrecher2015accurate}.

Absolute free energy calculations for dissimilar ligands can have particular challenges because the (potentially incorrect) assumption of consistent binding modes across a series of similar ligands is likely to be even less robust than the in the case of relative calculations.
This means that researchers performing absolute binding free energy calculations will have to pay particular attention to generating reasonable putative binding modes.

In some cases, it is tempting to simply use docking techniques to generate initial bound structures for starting molecular dynamics simulations.
However, timescales for binding mode interconversion are usually slow compared to MD/free energy timescales, meaning that simulations started from different potential binding modes are likely to yield disparate computed binding free energies~\cite{mobley2006use, palma2012computation, mobley2012perspective, gill2018binding} .
And docking techniques are good at identifying sterically reasonable potential binding modes, but still perform relatively poorly at identifying a single dominant binding mode \emph{a priori}.

It is worth highlighting a recent SAMPL blind challenge on HIV integrase as an illustration of this. 
Many submissions, using state-of-the-art methods, had difficulty even predicting which \emph{binding site} ligands would bind in (most submissions placed more than half of the ligands into the incorrect binding site), and even given correct binding sites, the binding mode within each site was also quite difficult to predict~\cite{mobley2014blind}.
The best performing submission for predicting binding modes actually ended up being a human expert (aided by computational tools) with more than 10 years of experience on the particular target~\cite{voet2014combining}, rather than a fully automated approach.
While free energy calculations on this set had some success, many of the failures actually ended up being cases where the binding mode selected as input for free energy calculations was later found to be incorrect~\cite{gallicchio2014virtual}, highlighting the importance of these issues.

One approach which has shown some success is to retain diverse potential binding modes from docking, perform short MD simulations of these to identify distinct stable binding modes, and then consider these in subsequent calculations~\cite{gallicchio2014virtual, mobley2006use,rocklin2013blind, boyce2009predicting, mobley2007predicting}.

Routes to handle multiple potential binding modes are different depending on whether absolute or relative calculations are selected, unless a method is available to estimate the relative populations of different stable binding modes in advance (e.g. such as the BLUES approach currently in development~\cite{gill2018binding}), in which case this approach could be applied to assist both types of calculations.

\paragraph{Handling multiple potential binding modes within absolute calculations.}
Within absolute binding free energy calculations, multiple potential binding modes can be handled by two main strategies: Consider each binding mode separately (a separation of states strategy) or sample all binding modes within a single simulation~\cite{mobley2012perspective}.
This couples to the choice of restraints selected, as some restraints will allow transitions between binding modes and even binding sites (Sec.~\ref{sec:standardstate-restraints}), and others do not.

Sampling all potential lignad binding modes within a single free energy calculation is usually impractical without some form of enhanced sampling or at least Hamiltonian replica exchange~\cite{wang2013identifying} because barriers for binding mode interconversion result in kinetics which are too slow compared to simulation timescales~\cite{mobley2006use, palma2012computation,mobley2012perspective, gill2018binding}.
Hamiltonian exchange, coupled with appropriate restraints, can allow the ligand to relatively rapidly exchange between potential binding modes when non-interacting, accelerating sampling of binding modes~\cite{wang2013identifying}. However, it is not always clear that this is desirable, since this also increases the size of the configuration space which must be sampled even if the binding mode is known.

Separation of states provides a simple though potentially expensive alternative, where each stable binding mode is considered separately with a binding free energy calculation restricted to that binding mode, and then (as long as the binding modes are non-overlapping) the resulting component binding free energies can be combined into a total~\cite{mobley2006use, mobley2012perspective}.
This approach necessitates a separate binding free energy calculation for each potential binding mode, however, so it can be computationally quite costly.
If relative populations of different stable binding modes were available from some other technique, it could make this separation of states approach considerably more efficient~\cite{mobley2012perspective, gill2018binding}.

\paragraph{Handling multiple potential binding modes within relative calculations.}
Multiple potential binding modes pose particular problems for relative free energy calculations, as having multiple starting structures for these calculations could yield substantially different calculated relative binding free energies for the same transformation due to kinetic trapping, and, without additional information (specifically, the free energy of binding mode interconversion or, equivalently, the relative populations of different binding modes) it becomes impossible to sort out which of the multiple answers is in fact the correct relative binding free energy.

To deal with this, some practitioners have actually computed relative binding free energies of different binding modes of the same ligand~\cite{palma2012computation}.
For example, a perturbation which adds a methyl to an aromatic ring of a larger ligand might yield one result if the methyl points in one direction, and a different value if it points in the other due to slow ring motions~\cite{lincoff2016comparing, sasmal2020sampling}.
One could compute the free energy of turning off the methyl group in one orientation and turning it back on in the other orientation to obtain the free energy difference between the two potential binding modes.
While this approach has precedent, it is relatively difficult to automate at present and requires considerable care.

Overall, this likely means that relative free energy calculations will be susceptible to problems resulting from uncertainty in ligand binding modes until more robust approaches are available to determine dominant binding modes, or the relative populations of different potential binding modes, in advance.

\section{Data analysis}
\label{sec:data_analysis}
Once data has been collected from alchemical intermediates, it must be analyzed to produce an estimate of the free energy change (and its associated statistical uncertainty) for each leg of the thermodynamic cycle.
While a number of different estimators are available that will give consistent results under optimal circumstances, some approaches are recommended over others due to their robustness and ability to provide information on poor convergence.

\subsection{Detecting the boundary between equilibrated and production regions}
\label{sec:automatic-equilibration-detection}
Much of the infrastructure for analyzing alchemical free energy calculations relies on the concept of asymptotically unbiased estimators, which produce unbiased estimates of the free energy when fed very long simulations~\cite{shirts2005comparison}.
In reality, free energy calculations are often initiated from highly atypical initial conditions (such as a protein-ligand geometry obtained from docking and subjected to a heuristic solvent placement scheme), and simulations are of a finite length dictated by available computational resources and computing demands.
As a result, these estimators can produce significantly biased estimates if fed the entirety of simulation data generated without further processing~\cite{chodera2016simple}.
\begin{figure}
    \centering
    \includegraphics[width=0.95\linewidth]{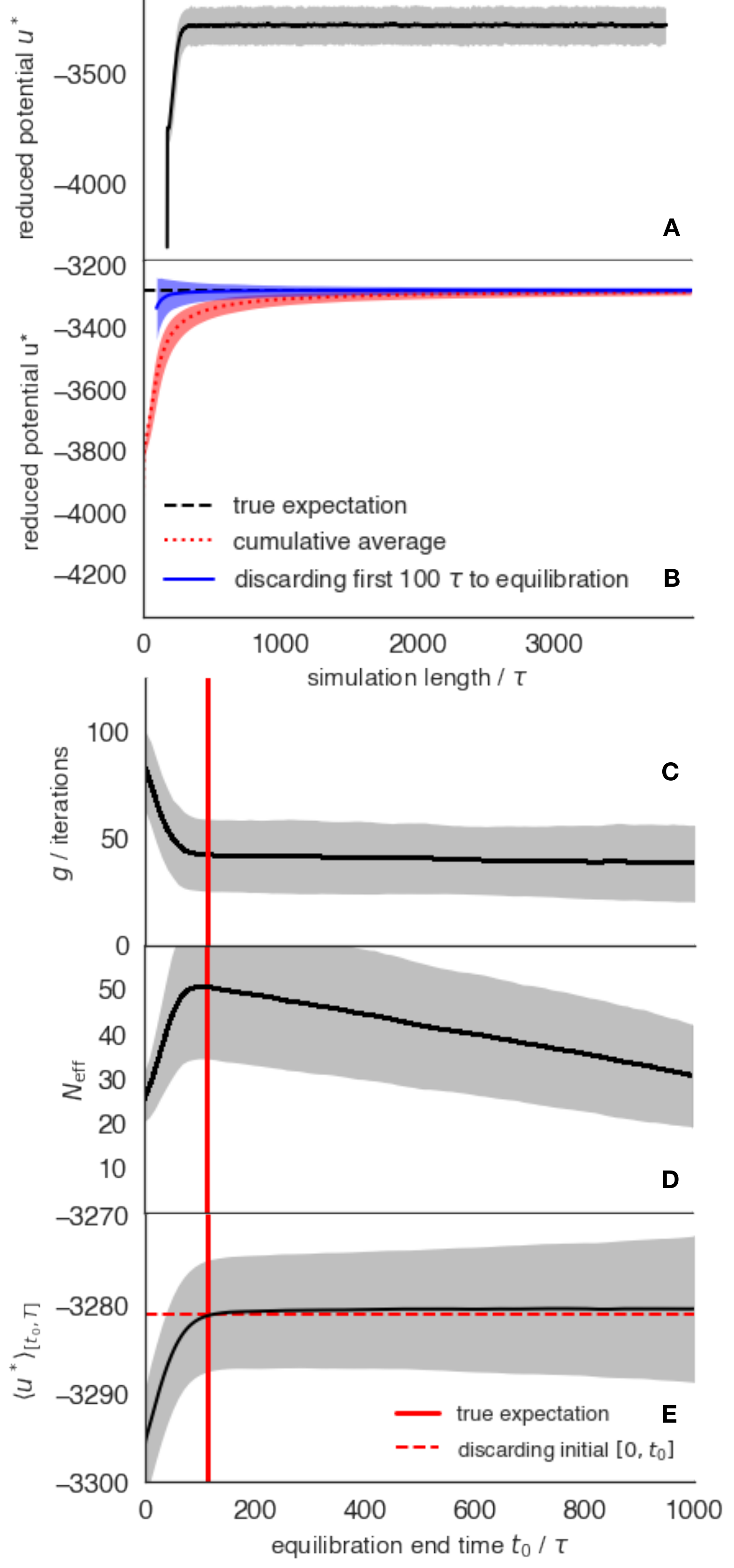}
    \caption{{\bf Automatic partitioning into equilibration and production regions.}
    (\textbf{A}) The average (black line) standard deviation (shaded region) of the reduced potential $u^*$ over many independent replicate simulations started from the same initial conditions show a significant initial transient change before relaxing to the true average potential energy (\textbf{B}). A cumulative average (red) of the entire simulation data demonstrates simulation bias not seen when initial simulation data is omitted (blue). Using an automated approach to detect equilibration of the boundary $t_0$ using statistical inefficiency $g$ (\textbf{C})  for an effective simulation interval (\textbf{D}). (\textbf{E}) The optimal equilibration boundary $t_0$ is selected to maximize the number of uncorrelated samples.
    \emph{Figure adapted from~\cite{chodera2016simple}.}
    }
    \label{fig:automatic-equilibration-detection}
\end{figure}

To minimize this effect, an initial portion of the simulation is often discarded to \emph{equilibration}~\cite{braun2019best}, with the idea of removing the most heavily biased initial portion of simulation data but retaining the unbiased \emph{production} region that represents a stationary Markov chain process sampling from the desired equilibrium target distribution.
Because the simulation time required for the atypical initial sampler state to relax toward equilibrium is a property of the specific system being simulated and the specific initial conditions selected, it is simplest to collect data for the whole process and use an automated algorithm to select how much data should be discarded to equilibration in a post-processing step.

A simple approach to automatically partitioning simulation data into equilibration and production regions is described in~\cite{chodera2016simple} (illustrated in Fig.~\ref{fig:automatic-equilibration-detection}).
Suppose we have a simulation of length $T$ consisting of correlated data.
Here, the goal of the post-processing step is to select the equilibration boundary $t_0 \in [0, T]$ so as to \emph{maximize} the number of effectively uncorrelated samples remaining in the production region $N_{[t_0,T]}$, which is defined as
\begin{eqnarray}
N_{[t_0,T]} &=& \frac{T - t_0}{g_{[t_0,T]}}
\end{eqnarray}
where $g_{[t_0,T]}$ is the \emph{statistical inefficiency} of a timeseries $a_t$, described in more detail below.
Conveniently, this procedure also produces the information necessary to decorrelate the simulation data for estimating the free energy differences, a requisite next step in analysis.
This approach is implemented within the MBAR~\cite{kylebeauchamp2019choderalab} and alchemlyb~\cite{daviddotson2019alchemistry} packages, and is highly recommended for standard practice.

For additional discussion of working with correlated data and autocorrelation analysis, please refer to the work on Best Practices for Quantification of Uncertainty and Sampling Quality in Molecular Simulations~\cite{grossfield2018best}.

\paragraph{Computing the timeseries for equilibration detection}
Typically, the timeseries of note $a_t$ analyzed in automated equilibration detection is the negative logarithm of the probability density ($\pi(x_t; \vec{\lambda}$)) sampled by the MCMC algorithm (up to an irrelevant additive constant).
For simple independent simulations that sample $x_t \sim \pi(x ; \vec{\lambda})$, this is given by the reduced potential
\begin{eqnarray}
a_t &\equiv& - \ln \pi(x_t; \vec{\lambda}) + c = u(x_t; \vec{\lambda}) .
\end{eqnarray}

Note that the use of the effective reduced potential is not guaranteed to pick up on all slow relaxation processes that may be coupled to the alchemical free energy, but the simplicity of its computation means it is generally appropriate for most cases.

\paragraph{Cautions in automating equilibration detection}
For simulations that are simply not long enough to contain a large number of samples from true equilibrium either because they are very short or contain slow processes, this procedure cannot completely remove the bias.
In such cases, this approach simply selects the final portion of the the simulation, which may be contained in a single substate of conformational space, and may itself lead to biased estimates. 
This situation can be detected if the equilibration boundary $t_0$ is a significant fraction of the total simulation length $T$, with a good rule of thumb being that $T \gtrsim 20 t_0$.
If this is not possible, advanced analysis techniques that assume only local equilibrium (rather than global equilibrium) such as the TRAM estimators~\cite{mey2014xtram,wu2016multiensemble,nuske2017markov} may be more appropriate, but are beyond the scope of this paper. 

\subsection{Decorrelating samples for analysis}
\label{sec:decorrelating-samples}
\paragraph{Computing the statistical inefficiency}
Most estimators require an uncorrelated set of samples from the equilibrium distribution to produce (relatively) unbiased estimates of the free energy difference and its statistical uncertainty.
To do this, the production region of the simulation is generally \emph{subsampled} with an interval approximately equal to or greater than the \emph{statistical inefficiency} $g \ge 1$ to produce a set of uncorrelated samples that can be fed to the estimator machinery~\cite{chodera2016simple},
\begin{eqnarray}
g &\equiv& 1 + 2 \tau_\mathrm{eq} \label{eq:statistical-inefficiency-definition}
\end{eqnarray}
where $\tau_\mathrm{eq}$ is the integrated autocorrelation time, formally defined as
\begin{eqnarray}
\tau_\mathrm{eq} &\equiv& \sum_{t=1}^{T-1} \left(1 - \frac{t}{T}\right) C_t \label{eq:integrated-autocorrelation-time-definition} , 
\end{eqnarray}
with the discrete-time normalized fluctuation autocorrelation function $C_t$ defined as
\begin{eqnarray}
C_t &\equiv& \frac{\expect{a_n a_{n+t}} - \expect{a_n}^2}{\expect{a_n^2} - \expect{a_n}^2} . \label{equation:autocorrelation-definition}
\end{eqnarray}
The basic concept is that $\tau_\mathrm{eq}$ corresponds to the single-exponential decay time for the autocorrelation process that generates samples, so the statistical inefficiency $g$ measures the approximate temporal separation between two effectively uncorrelated samples (where two exponential relaxation times are presumed to be sufficient).

Robust estimation of $C_t$ for $t \sim T$ is difficult due to growth in statistical error, so common estimators of $g$ make use of several additional properties of $C_t$ to provide useful estimates (see \emph{Practical Computation of Statistical Inefficiencies} in~\cite{chodera2016simple} for a detailed discussion).

We recommend using the robust statistical inefficiency computation routines available within the MBAR~\cite{kylebeauchamp2019choderalab} and alchemlyb~\cite{daviddotson2019alchemistry} packages.

\paragraph{Subsampling data to generate uncorrelated samples}
Once the statistical inefficiency $g$ has been estimated, it is straightforward to subsample the correlated timeseries simulation data to produce effectively uncorrelated data that can be fed to the free energy estimators.
Suppose the correlated timeseries is $\{a_t\}_{t=1}^T$; we can form a new timeseries of $N_{\mathrm{eff}} \approx T / g$ effectively uncorrelated samples by selecting a subset of indices $\{ \: t = \mathrm{round}((n-1) \, g) \: | \: n \in \mathrm{range}(1,\ldots ,N) \: \}$ where $\mathrm{round(x)}$ denotes rounding to the nearest integer.

If independent simulations are used, the alchemical state $\vec{\lambda}$ may have a significant impact on the correlation time, and these simulations should be subsampled independently using a separate estimate of the statistical inefficinecy $g$ for each alchemical state.
If coupled simulations are used (such as a Hamiltonian replica exchange simulation), the replicas should undergo equivalent random walks in alchemical space, and the replicas can be can be subsampled with the same $g$ to generate an equal number of uncorrelated samples at each alchemical state.
Conveniently, the approach described above for automated equilibration detection produces an appropriate estimate of $g$ over the production region for automating this process.

\paragraph{Cautions and considerations}
Reliable estimation of the statistical inefficiency is difficult, and estimates will not generally be as precise (in a relative error sense) as averages.
To ensure there is sufficient data available for reliable decorrelation and estimation of free energy differences, it is recommended that the effective number of uncorrelated samples $N_{\mathrm{eff}} \ge \approx 50$ if the BAR or MBAR estimators (discussed below in sec~\ref{subsec:estimators}) are used; the number may need to be much higher with alternate estimators.

\subsection{Estimators for free energy differences}
\label{subsec:estimators}
Free energy differences between two different states differing in the energy function are directly related to the
ratio of probabilities of those states.
As can be noted, the partition functions in Eq.~\ref{eq:conf_probability} are simply the total accumulated probabilities for all possible configurations of the system. Virtually all of the ways to estimate this free energy are based in converting this ratio of integrals to something that can be measured in one (or several) simulations.  

\paragraph{The Zwanzig relationship (EXP)}
The simplest method for calculating free energy differences from simulations is the so-called \textit{Zwanzig
relationship}~\cite{zwanzig1954hightemperature}, also called one-sided exponential re-weighting (EXP), or simply free energy perturbation, though this final term is sometimes used to encompass all ways of calculating free energy differences.

The (reduced) free energy difference $\Delta f_{01}$ between an initial state 0 and a final state 1 defined by two different potential energy functions 
$u_0(\vec{q})$ and $u_1(\vec{q})$ over coordinate space $\vec{q}$ can be calculated as:
\begin{eqnarray}
\Delta f_{01} & = & \ln \expect{e^{-(u_1(\vec{q}) - u_0(\vec{q}))}}_0 =  \ln \expect{e^{-\Delta u(\vec{q})} }_0
\end{eqnarray}\label{eqn.zwanzig}
and the average is over all samples from the simulation performed with $u_0$. In the case of NVT (canonical) sampling and assuming the masses do not change, then $u$ is simply $U/k_BT$, and $f$ is $F/k_BT$, but it can be generalized to other ensembles with the proper definition of $f$ and $u$.
Described in words, we take the samples generated during our run with the potential energy function $u_0(\vec{q})$ and calculate what the difference in energy would be if we switched to potential energy function $u_1(\vec{q})$, and average the exponential of the negative energy difference to get the negative of the exponential of the free energy difference. The original distributions, P($u_0$) as generated at $\vec{\lambda}=0$ and P($u_1$) would look like those seen in Fig.~\ref{fig:fig_sampling_scheme} \textbf{A}-\textbf{C} on the left hand side. Reevaluating requires almost no extra code functionality to perform; one need only to save a full precision trajectory, and run an unmodified molecular simulation code using the $u_1$ in order to calculate the new energies of stored snapshots. The analysis can be written in a line of code. We note that this method is even more general, in that the instantaneous work to change the potential energy function from $u_0$ to $u_1$ can be replaced by the non-reversible work $W$ to make the same change under the same equilibrium conditions at either end state~\cite{jarzynski1997nonequilibrium,jarzynski1998equilibrium,crooks2000pathensemble}, though we do not go into all of the details of non-equilibrium transformations here, and refer the reader to more advanced treatments~\cite{maragakis2008bayesian,oberhofer2005biased,procacci2015unbiased,shirts2003equilibriuma,ytreberg2004singleensemble}.

Although the Zwanzig equation is formally correct (as long as the two states considered sample the same phase space volume, which is true for standard molecular models), it has some very important numerical issues that mean that it often performs badly for standard free energy calculations, even for small molecules~\cite{shirts2005comparison,lu2003appropriate}. One can show that if the standard deviation of the difference $\Delta u(\vec{q}) = u_1(\vec{q})-u_2(\vec{q})$ over all sampled $\vec{q}$ is large (which in this case, means only several times $k_BT)$, then very few samples contribute to the average, and the answer will be both biased and extremely noisy~\cite{lelievre2010free}. Essentially, the method is dominated by contributions of rare snapshots~\cite{jarzynski2006rare, wu2005phasespaceb, wu2005phasespacec}.

\paragraph{The Bennett Acceptance Ratio (BAR)}
If we have the differences in the potential energy sampled from the distribution defined by $u_0$ to the state defined by $u_1$, and we also have the differences in potential energies from the distribution sampled by $u_1$ to the state defined by $u_0$, we can obtain a significantly improved estimate of the
free energy difference compared to that obtained by EXP. 
This estimate was first derived by Bennett and is hence generally called the Bennett Acceptance Ratio (BAR). It is solved by finding the reduced free energy $f_{ij}$ that satisfied the following implicit equation:
\begin{eqnarray}
 \sum_{i=1}^{n_i} \frac{1}{1 + \exp[\ln(\frac{n_i}{n_j}) + u_{ij}(\vec{q}) - f_{ij})
 ]} \nonumber \\
 =\sum_{i=1}^{n_j} \frac{1}{1 + \exp[\ln(\frac{n_i}{n_j}) - u_{ij}(\vec{q}) + f_ij)]},
\end{eqnarray}
where $n_i$ and $n_j$ are the number of samples from each state. More recent derivations show that this formula is the maximum likelihood estimate of the free energy difference given sets of samples from the two states~\cite{shirts2003equilibriuma}. 

Many studies have demonstrated both the theoretical and practical superiority of BAR over EXP in molecular
simulations~\cite{shirts2005comparison,lu2003appropriate}, and BAR converges to EXP in the limit that all samples are from a single state~\cite{bennett1976efficient,bennett1976efficient,shirts2003equilibriuma}. BAR also requires significantly less overlap between the configurational space of each state to converge than EXP, though some overlap must still exist.

The Bennett acceptance ratio is only defined between two states. Usually, the endpoints of interest in a free energy calculation are sufficiently different that we will need a chain of states that gradually change the potential energy function from $u_0$ to $u_1$, as discussed in Sec.~\ref{sec:important_path}. You can simply carry out BAR between each pair of states $\Delta f_{1 \rightarrow N} = \Delta {f_{1\rightarrow 2}} + \Delta {f_{2\rightarrow 3}} +  \ldots + \Delta f_{N-1\rightarrow N}$.

There is one important thing to note about the uncertainty estimates when summing multiple free energies together to calculate an overall free energy estimate. Although BAR itself gives a free energy estimate that is asymptotically correct in and is much less biased than the uncertainty estimate for EXP, the uncertainties in $\Delta {f_{i-1\rightarrow i}}$ and $\Delta {f_{i\rightarrow i+1}}$ are not uncorrelated, because they both involve the energies $u_i(\vec{q})$. The variances of each of the free energies will \textit{not} propagate as variances usually do (in quadrature) into the variance of the overall free energy. Instead, some other method for propagating the uncertainty, such as bootstrapping~\cite{grossfield2018best} must be used.

\paragraph{Thermodynamic integration (TI)}
By taking the derivative of the free energy with respect to the
variable $\vec{\lambda}$, we find that:
\begin{equation}
\frac{df}{d\vec{\lambda}} = \frac{d}{d\vec{\lambda}} \left[-\ln \int \frac{\exp{-u(\vec{\lambda},\vec{q})}}{Z(\vec{\lambda})} d\vec{q}\right] = \expect{\frac{du(\vec{\lambda},\vec{q})}{d\vec{\lambda}}\
}_{\vec{\lambda}} .
\end{equation}
And than we can then numerically integrate $df/d\vec{\lambda}$ over an alchemical transformation, using a range of different well-established techniques, to obtain:
\begin{equation}
\Delta f    = \int_{0}^{1} \expect{\frac{du(\vec{\lambda},\vec{q})}{d\vec{\lambda}}}_{\vec{\lambda}}  d\vec{\lambda}.    
\end{equation}
This approach to calculating the free energy is called thermodynamic integration (TI). Averaging over $\expect{\frac{du}{d\vec{\lambda}}}$ requires fewer uncorrelated samples to reach a given level of relative error
than averaging $e^{-u(\vec{q})}$, as the distribution of values is usually narrower, with a more Gaussian shape to the distribution. Rather than being limited by overlap, as in the case of BAR and MBAR (see below), we are instead limited by the bias in the numerical quadrature, which must be minimized sufficiently to be beneath the level of statistical noise.

Various numerical integration schemes are possible, but the trapezoid
rule provides a simple and robust scheme. All types of numerical integration can be written as:
\[ \Delta f \approx \sum_{k=1}^{K} w_k
\expect{\frac{du(\vec{\lambda},\vec{q})}{d\vec{\lambda}}}_{k}, \] where the weights $w_k$ correspond to a particular choice of numerical integration.
Researchers have tried a large number of different integration schemes~\cite{resat1993studies,jorge2010effect,shyu2009reducing}. However, many other integration choices require specific choices of $\vec{\lambda}$ to minimize bias, which makes them unsuitable when the intermediates
have widely-varying levels of uncertainty. For example, integrating a cubic spline interpolation provided negligible benefits over a simple trapezoid rule~\cite{paliwal2011benchmark}. For starting researchers, we therefore recommend the simple trapezoid rule scheme, as it allows for maximal flexibility in which values of $\vec{\lambda}$ are simulated. As fitting to higher order polynomials can have numerical instabilities for some energy functions, and because alternate functional forms might only be appropriate with some types of transformations, expertise and experience is required to perform such numerical integration modifications. In practice, adding 2-3 more intermediate states is typically sufficient to match the performance of these more complicated numerical quadrature schemes.

One drawback of TI is that it requires derivatives with respect to $\vec{\lambda}$ to be calculated directly in the code. Unfortunately, many problems of interest require using pathways (such as the soft-core pathways, for removing repulsive interactions) that are not linear, as we discuss, making this more complex. Still, if the code of interest does compute $\frac{du}{d\vec{\lambda}}$, then TI is perhaps the simplest method to use, as it involves a very little post-processing effort.

\paragraph{The multistate Bennett acceptance ratio (MBAR)}
One can generalize Bennett's logic from two states to multiple states to obtain a free energy estimator that uses energy differences between configurations at all intermediate states to compute free energy differences between all states. MBAR gives a system of implicit equations for the free energies $f_i$:
\begin{equation}
f_i = - \ln \sum_{n=1}^{N} \frac{\exp(-u_i(\vec{q}_n))}{\sum_{k=1}^K N_k \exp(f_k-u_k(\vec{q}_n))},
\end{equation}
where there are $N_k$ samples from each of $K$ states, with $\sum_k N_k=N$ the total number of samples. Thus, we need to evaluate the energy function $u_i$ for all samples obtained at all states in the transformation. The equations can be solved by a number of different standard routines. We note that there are only $K-1$ independent equations, so only $K-1$ of the free energies are independent variables, and one of the $f_i$ must be specified (usually, without loss of generality, setting it to zero).

MBAR is provably the lowest variance asymptotically unbiased estimator of the free energy given the energies of the samples~\cite{tan2004likelihood}, which means that BAR is also the lowest variance estimator for the free energy difference between only 2 states, as it is mathematically exactly the same as MBAR in this case. MBAR also provides an uncertainty estimate, derived from standard error propagation methods for implicit functions, which has been shown to be highly accurate as long as there are sufficient samples at each state~\cite{paliwal2011benchmark}.

MBAR can also be thought of as the Zwanzig estimator of the free energy to state $i$ where the sampled distribution is the \textit{mixture distribution} of all the other samples thrown together in one ``pot'', defined by $p_m(\vec{q}) = N^{-1} \sum_k N_k \exp(f_k-u_k\vec{q})$, which is the weighted average of all the individual normalized probability distributions from the simulations that are performed.~\cite{shirts2017reweighting}.

\paragraph{Recommendations}
\begin{itemize}
\item We recommend MBAR if all energy differences are available. It is the lowest variance unbiased free energy estimate given samples from multiple states.
\item BAR is essentially just as good as MBAR for highly optimized $\vec{\lambda}$ intermediates. Specifically, if the $\vec{\lambda}$s are chosen such that intermediate states have moderate overlap with their neighbors (i.e. between $i$ and $i+1$ and between $i$ and $i-1$, they will \textit{not} have significant overlap with their next nearest neighbors $i+2$ and $i-2$. Thus MBAR does not actually get significant information from these energy differences, so one might as well not even calculate them, and just perform BAR between nearest neighbors.~\cite{paliwal2011benchmark} 
\item TI usually gives similar values as MBAR implemented with sufficient numbers of intermediates, but quadrature errors are hard to estimate beforehand  can occur if one is not careful.~\cite{paliwal2011benchmark}
\item WHAM is an approximation to MBAR, and there are no compelling reasons it should be used. If careful, it is not necessarily much worse than the other methods, but it always introduces some degree if binning error.
\item Other variants, especially ones that adaptively determine the free energies can be useful in certain circumstances but beyond the scope of a Best Practices article.
\end{itemize}

\subsection{Uncertainty estimation}
\label{subsec:uncertainty}
It is important to consider the variation in your computed free energies from your equilibrium simulations, in order to obtain an estimate of uncertainty of the obtained value for the free energies of interest. A recent best practices paper by Grossfield et al.,~\cite{grossfield2018best} provides substantial detail on how to estimate uncertainties from molecular simulations and is a good starting point for this topic. 
In general, the quantification of different error metrics depends on both data generation and analysis methods used from the ones discussed above. 

The computation of free energies using TI (Sec.n~\ref{subsec:estimators}) is straightforward and the trapezoidal rule is often recommended since it allows unequal spacing of $\vec{\lambda}$ states, which is required to minimize the variance in the free energy estimate, but in principle any good numerical integration method can be used. 
The determination of regions of high curvature when estimating the integral is helpful to determine regions of phase space where more sampling and/or more $\vec{\lambda}$ states are necessary to obtain the best approximation of the integral. Plotting $\vec{\lambda}$ with respect to the gradients at each of the $\vec{\lambda}$ values can be be a helpful diagnostic. 
Additionally, computation of the overall variance of TI requires the calculation of the overall variance of integration, rather than each individual $\Delta$G$_{i,i+1}$ and assuming variances add independently. 
Therefore, $\mathrm{var}$($\Delta$f) = $\sum_{i=1}^{K}w_{k}^2 \mathrm{var}(\frac{du}{d\vec{\lambda}})_{k}$.

For alchemical changes that result in smooth, low curvature sets of $\expect{\frac{dU}{d\vec{\lambda}}}$, a relatively small number of $\vec{\lambda}$ states is necessary for sufficient accuracy and low variance in the free energy estimate. 
Depending on the difficulty of the perturbation, the bias introduced by discretization of the integral can become large due to increased curvature, and more $\vec{\lambda}$ intermediate states become necessary to reduce error.
It is recommended that researchers verify that a sufficient number of states are included such that the free energy is essentially invariant to the number of lambda intermediate states chosen. Good heuristics or measures to assess the 'difficulty' of a given perturbation is still an ongoing research topic. 

Compared with TI, the MBAR method (Sec.~\ref{subsec:estimators}) discussed above provides uncertainty estimation directly from solving a set of linear equations to compute the variances between all states. 
The number of states and amount of sampling should be optimized to minimize the uncertainty in the MBAR free energy estimate, while balancing other key considerations such as computational expense. 

If possible, it is advisable to analyze the same set of simulations with different estimators, providing an opportunity for synergy. If different estimators agree the free energy estimate is more reliable than if there are differences between methods that are larger than 1 kcal/mol and would indicate poor convergence. 

Uncertainty can also be assessed for a particular perturbation by repeating calculations with slight changes in initial configurations, forcefield parameters, and different random seeds in the MD engine. 
The assessment of variability in free energy calculations due to repeating simulations has been previously reported~\cite{aldeghi2019accurate,paliwal2011benchmark,mey2016blinded,mey2018impact}, and large variance in free energies estimated from simulations with different random seeds should be flagged as issues with convergence. 

For relative binding free energy calculations, additional sensitivity analysis can be performed by changing the initial configurations of non-core regions of the perturbation topology and determining if this change in configurations results in a large differences in the computed relative free energy, indicating poor sampling of ligand configuration.
The proposed changes in configuration are increasingly relevant if no experimental evidence is available to reduce uncertainty in where the changing atoms should be positioned.

In addition to statistical uncertainty and sampling, a variety of other factors can impact results from binding free energy calculations. In addition to the choice of initial configuration, results can depend on the choice of force field for the protein/receptor, water, and small molecule(s), so rerunning calculations with different choices of force field can also be used to assess how sensitive results and conclusions are to these particular choices. Other factors, like system preparation (choice of protonation state, tautomer, counterion presence, salt concentration, etc.) can also substantially impact results~\cite{mobley2017predicting, mobley2017predictingb}, so unless modelers are confident they have these factors correct, sensitivity to these choices may also need to be examined.

\subsection{Are my simulations any good?}
\label{sec:are-they-good}
There are different easily measurable indicators that can test how well converged simulations are, and if all alchemical states have been sufficiently sampled for a rigorous analysis. Furthermore, once you have established that individual perturbations are well behaved, there are some tricks to ensure the overall perturbation network gives reliable results.

\begin{figure}
    \includegraphics[width=0.90\linewidth]{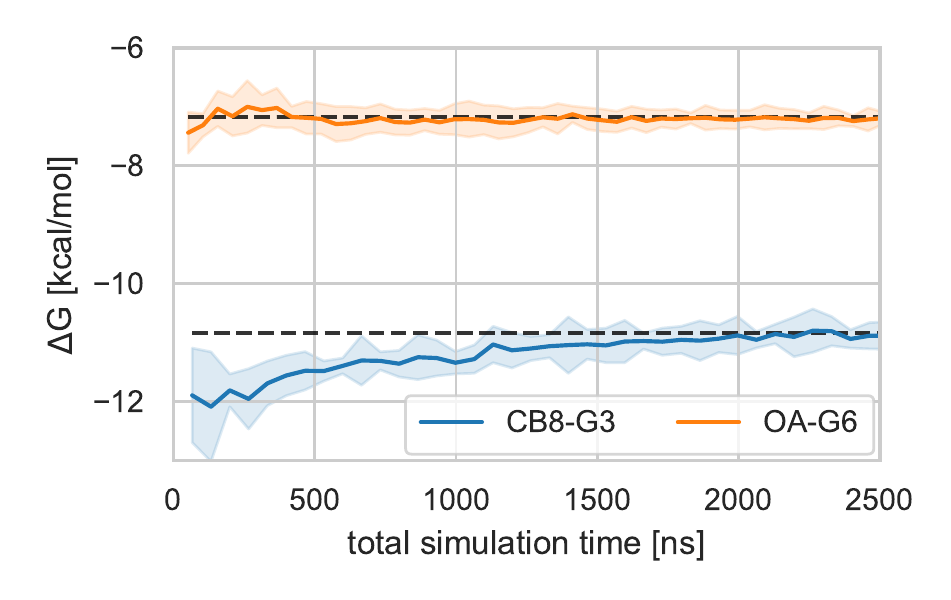}
    \caption{Average binding free energy of 5 replicate Hamiltonian replica exchange calculations as a function of total simulation time (i.e. the sum of the simulation time of all replicas) for the two host-guest systems CB8-G3 and OA-G3. Shaded areas represent 95\% confidence intervals around the mean computed from the 5 replicates data. The horizontal dashed lines show the final binding free energy prediction of the two calculations after a total of 5230 ns for OA-G3 and 6650 ns for CB8-G3. Longer correlation times in CB8-G3 cause the calculation to converge more slowly. The original data used to generate the plot can be found at \url{https://github.com/samplchallenges/SAMPL6/blob/master/host_guest/Analysis/SAMPLing/Data/reference_free_energies.csv}.
}
    \label{fig:freeenergytrajectories}
\end{figure}

\paragraph{Convergence of simulations}
Fig.~\ref{fig:freeenergytrajectories} illustrates how looking at the convergence of your data may be important. In this example, the guest G3 shows different convergence behaviour for two different hosts. The CB8 host with guest G3 has a longer correlation time than the octa acid (OA) host. In some cases, slow correlation time may not be expected and therefore not a feature known in advance. To this end, you should always look at all simulation data available check convergence behaviour for each free energy estimate and if need be extend existing simulations or try an approach that requires simulations in two separate binding modes where they interconvert at very slow timescales.  

\paragraph{Overlap matrix}
One way of assessing reliability of the calculations is checking the phase space overlap between neighboring $\vec{\lambda}$-windows~\cite{wu2005phasespaceb, wu2005phasespacec}. For this purpose, a so-called overlap matrix $\mathcal{O}$ can be used. $\mathcal{O}$ is a $K\times K$ matrix, with $K$ being the number of simulated states, i.e. values of $\vec{\lambda}$. Sufficient overlap is important for reweighting estimators such as BAR or MBAR, but cannot help assess reliability of estimates when using TI. 
These matrices are graphical representations of the phase space overlap, i.e. the average probability that a sample generated at state $\vec{\lambda}_{j}$ can be observed at state $\vec{\lambda_{i}}$. As this probability is computed considering the samples from all states (and not just the adjacent states), the values in each row and column add up to 1. In this analysis, the goal is to ensure every state has overlap with its neighbors in both directions -- so that off-diagonal elements are sufficiently larger than zero. For accurate calculations, the matrix should be at least tridiagonal.

\begin{figure}
\includegraphics[width=0.90\columnwidth]{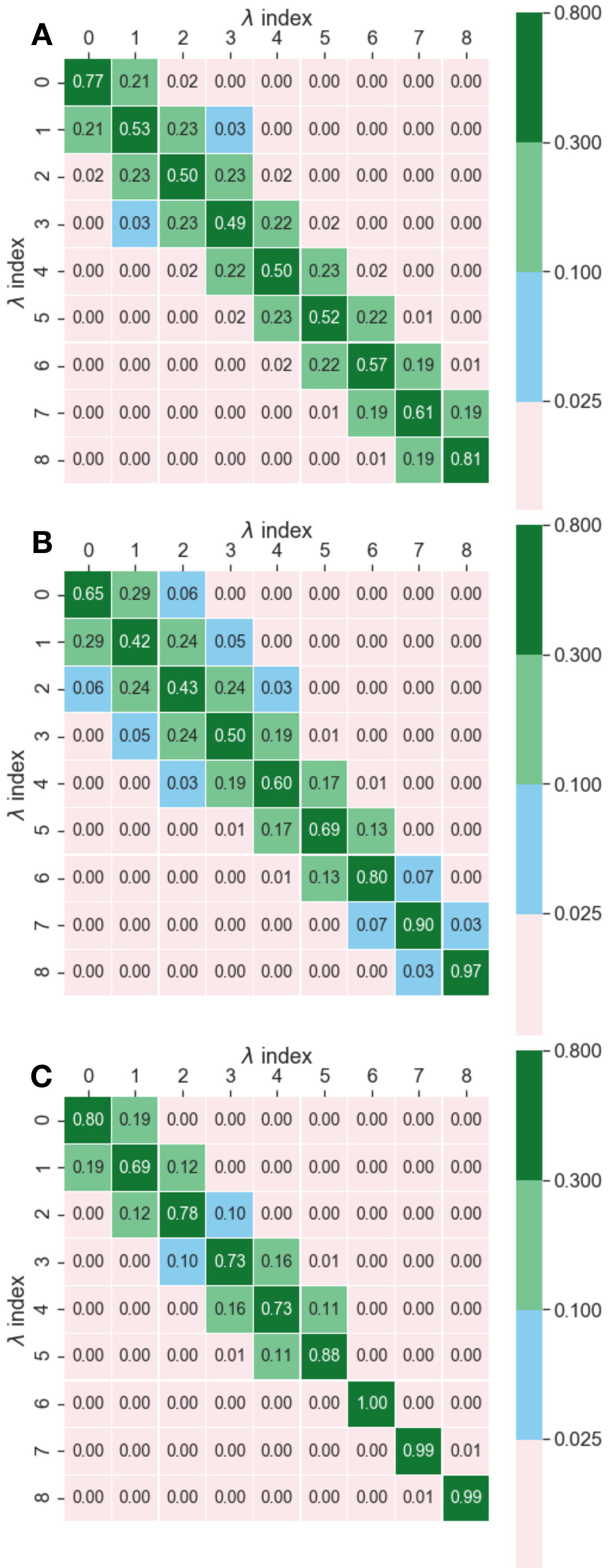}
\caption{\label{fig:overlap} \textbf{Overlap matrices:} Visualising overlap matrices can help with assessing the quality of simulation data. (\textbf{A}) shows good overlap with all first off-diagonal entries well above 0.03, the suggested threshold, (\textbf{B}) is an example of mediocre overlap with good overlap at lower $\vec{\lambda}$ values and poor overlap at high $\vec{\lambda}$ values. (\textbf{C}) shows poor overlap resulting in disconnected simulations with unreliable MBAR estimates.}
\end{figure}

Details on the calculation and properties of these matrices can be found elsewhere~\cite{klimovich2015guidelines}.
In an overlap matrix $\mathcal{O}$, the off-diagonal values (${O}_{i,j,i\ne j}$) are negatively correlated with the variance of the free energy difference. Accordingly, the uncertainty of the free energy difference between the states $i$ and $j$ will be smaller when ${O}_{i,j,i\ne j}$ is larger (and thus the values in the main diagonal (${O}_{i,j,i=j}$) are smaller). In order to obtain a reliable estimate of the free energy all neighbouring states must be connected, i.e. there must be sufficient overlap between the samples of these states (general description: ${O}_{i,j,i\ne j}\ge$ threshold).
However, due to the mathematical derivation it is difficult to explicitly describe the relation of the overlap matrix and the variance by formulae. Consequently, the threshold has to be derived empirically. It has been proposed that the values of the first off-diagonals (i.e. the diagonals above and below the main diagonal) should at least be 0.03 to obtain a reliable free energy estimate~\cite{klimovich2015guidelines}. Smaller values should be considered as a warning sign (see Fig.~\ref{fig:overlap} \textbf{C}), as the variance tends to be underestimated in case of poor overlap.

Fig.~\ref{fig:overlap} \textbf{A}, \textbf{B}, and \textbf{C} shows examples of good, mediocre, and poor overlap respectively. For Fig.~\ref{fig:overlap} \textbf{A}, the probability to find a sample from state $i$ in its neighbouring state $j$ is about 0.2 for all states adjacent to the main diagonal, and hence the overall connectivity is good. In the case of Fig.~\ref{fig:overlap} \textbf{B}, the overlap is strongly diminishing in the lower right corner, raising concerns regarding the reliability of the free energy estimate obtained. For Fig.~\ref{fig:overlap} \textbf{C}, the state at $\vec{\lambda}$ index = 6 is connected to neither of its neighbouring states. While this does not necessarily imply that the result for this perturbation is wrong, the energy estimate must at least be considered as highly unreliable.
In order to overcome the issue of poor overlap in this example, additional sampling should be performed by introducing additional states, i.e. $\vec{\lambda}$ values.

Interestingly, as the variance is inversely correlated with the number of states~\cite{klimovich2015guidelines}, it can in principle be reduced below any arbitrary threshold with enough simulation time and a large enough number of $\vec{\lambda}$ windows. However, decreasing the variance to a value close to 0 is not feasible, as this approach would significantly increase the calculation time. While variance can be decreased by increasing simulation length, if the overlap between states is known to be poor, increasing the number of $\vec{\lambda}$ values, or adjusting the spacing of those values to better cover regions of poor overlap will likely provide a larger immediate impact. Different approaches are described in Sec.~\ref{sec:simulation_protocol_choice} and more details can be found in the literature~\cite{dakka2018concurrent, hahn2019alchemical}.

\paragraph{Cycle closure error}
Relative free energy calculations, which compute the change in free energy on making a change to a molecule (e.g. adding a functional group to a ligand) may provide an additional opportunity for error/consistency checking. Particularly, such calculations are often done to span a graph or tree of free energy calculations~\cite{xu2019optimal,wang2013modeling,liu2013lead}. In some cases the free energy change to go between molecules A and B can be obtained via multiple transformation pathways. This allows a type of consistency checking where we assess how much the free energy change for that transformation in practice differs from equivalence. 

Significant deviations from this typically indicate insufficient configurational sampling along the lambda schedule of one or more of the transformations involved. This approach may be generalised to sets of connected transformations given the requirement that the sum of free energy changes along edges of a closed cycle should be zero. This analysis is called ``cycle closure''. In practice, such thermodynamic cycles do not actually sum to zero, and deviations become increasingly large as the size of the cycle increases owing to propagation of error. Though no firm guidelines have emerged, it may be judicious to perform additional configurational sampling along edges of a network that are involved in cycles closing poorly. This may be done by extending the duration of simulations, or by averaging free energy changes over multiple repeats. The latter approach may yield more reproducible free energy changes, but at the expense of a stronger bias on the estimated free energies due to repeated use of the same input coordinates.

A scheme to reduce cycle closure errors is used in FEP+ whereby calculated free energy changes along the nodes of the network are re-sampled assuming estimates of the calculated free energy change along a node may be obtained from a Gaussian distribution centered on the estimated free energy change and with a standard deviation equal to the estimated standard deviation of the free energy change. The procedure then uses a maximum likelihood method to find new sets of free energy changes that minimize cycle closure errors~\cite{wang2013modeling}. An alternative approach computes the free energy change between a target and reference compound as a weighted average over all unique paths in the network, with the weights derived from the propagated uncertainties of each node~\cite{mey2016blinded}. Approaches as illustrated by Yang et al. for perturbation map design can also be used to compute relative free energies between target and reference compounds~\cite{yang2020optimal}.

\paragraph{Reversible binding simulations}
An even more stringent test of the correctness of binding free energy calculations is to compare the results to the equilibrium binding constants derived from long timescale reversible binding simulations~\cite{pan2017quantitative}. For small ligands with millimolar affinities, repeated binding to and unbinding from the protein can occur for a large number of times in a sufficiently long unbiased MD simulation (10-100 $\mu$s), and the equilibrium binding constants can be computed from the ratio of bound to unbound fractions of the simulation time. The agreement between the binding free energy calculations and the reversible binding simulations---given the same system preparation and the same force field parameters---will strongly support the correctness of both calculations, as the same results are arrived at by two independent methods, and any discrepancy will suggest some systematic error in one, or both, of the two methods. As part of validation testing of alchemical free energy codes a benchmark set to compare alchemical and direct computation of equilibrium binding constant should become standard in future.

\subsection{Common issues to watch out for during analysis}

It is important to carefully examine output data for common problems. Some of the most important things to check for are:
\begin{itemize}
\item \textbf{Sampling of the binding site by the ligand:} Make sure the ligand samples the binding site reasonably tightly for its expected potency and fit, and that it does not depart out of binding site in the coupled end state if it is a moderate to strong binder. 
\item \textbf{Consistency of free energy estimates across different estimators} Significant discrepancies (further outside the error estimates than would be plausible) between free energies calculated with different free energy estimators such as TI, BAR, and MBAR. All of these estimators converge to the same results with sufficient sampling. Differences between them indicate poor overlap or errors in processing.
\item \textbf{Have replicas mixed well?} Poor replica mixing (for replica-exchange) or $\lambda$-space sampling for single-replica methods. If the system is not mixing between states, then the states are insufficiently close for mixing, or else there are bottlenecks in the configurational sampling that limit the accuracy.
\item \textbf{Behaviour of correlation times:} Correlation time that does not vary relatively smoothly as a function of $\vec{\lambda}$. Discontinuities indicate that the system is sampling significantly different configurations with only small changes to the Hamiltonian changes. This usually indicates sampling problems.
\item \textbf{Dependence of the free energy on initial conformation of the system.} Ensemble average properties should not depend on the starting point.
\item \textbf{Torsional sampling} Torsions with multiple low-energy minima where some of these minima are that are visited rarely or not at all. Which torsions have low energy minima can best be found by comparing to the simulation in the solvent. There should be clear physical reasons that simulation in the complex has different torsional distributions that the ligand in the solvent. 
\item \textbf{Free energy dependence on $\vec{\lambda}$} The free energy should vary relatively smoothly with $\vec{\lambda}$. If it varies drastically, then either there need to be finer sampling in $\vec{\lambda}$ in this region, or there are sampling problems there.
\item \textbf{Convergence of free energy} The free energy should clearly converge as a function or of simulation time (Fig. ~\ref{fig:convergence_forward_reverse}).
\item If using nonequilbirum methods, is the result independent of the speed at which the nonequilibrium change is performed? Nonequilibrium methods are in theory independent of the switching time in the limit of good sampling unless the switching time is simply too short. 
\item \textbf{Visualization of data} In general, inspect output data such as energies and visualize the simulation trajectories and assess if they match your expectations. Many issues can be spotted by a straight forward visualiztion. 
\end{itemize}

\subsection{Best practices for reporting data }
\label{sec:plot_data}
Following best practices for data generation and their analysis does not mean that data is reported in the optimal way. As a practitioner of alchemical free energy simulations you also should use best practices for reporting and plotting your results. We encourage the following standard set of analyses and ways to represent data. 
\paragraph{Statistics to include}
As with any modelling technique, misuse of statistical analysis can skew the perception of how well models perform in free energy predictions. First, error estimates should always be included on your predictions in whatever form you present your data (scatterplots, barplots, etc; see next paragraph). We recommend performing triplicates of your predictions at minimum, with starting points that are expected to be uncorrelated, to ensure some measure of reliability in your data. This replication may seem excessive, but uncertainty estimates often underestimate the true statistical uncertainty. Where performing multiple replicas of the simulation is not possible, an error estimate from e.g. MBAR can be used, though bearing in mind this is likely an underestimated error. 

As alchemical free energy methods are used in drug discovery to quantify and rationalise structure activity relationships (SAR), the models ability to (a) correlate well with experiment and (b) rank-order the molecules by affinity, should both be computed. Conventionally, this means including an R\textsuperscript{2} (or Pearson's R), where $R=+1$ means high correlation, $R=0$ means no correlation, and $R=-1$ means high anti-correlation) and a Kendall \texttau{} (with perfect ranking agreement when \texttau=1 and perfect disagreement when \texttau=-1) metric in your results. Additionally, practitioners may choose to include a Spearman \textrho{} as well. Brown et al.~\cite{brown2009healthy} have provided a useful analysis in terms of upper bounds of expected possible correlations between experiment and computation with a given potency range for the compounds. For example, for potency ranges of 2 log units it would be impossible to get a higher correlation in R than 0.8 because of experimental uncertainties~\cite{brown2009healthy}. What often is neglected to include is an error analysis on correlation statistics that arise from the errors of both experimental and computed data. One way to include such error analysis for correlation metrics is using bootstrapping on the datasets. The D3R community challenges follows best practices on their data evaluation with readily available python scripts online~\cite{2018drugdata}, based on work by Pat Walters~\cite{walters2013what}. Other analysis software also provide similar functionality for bootstrapping datasets~\cite{antonia2019michellab}. 

Mean unsigned error (MUE, also called mean absolute error/MAE) is another key statistic to include in your results. Even though some models' near-perfect correlation and ranking statistics might suggest excellent accuracy, MUE values can still have errors of show multiple kcal/mol of error, providing important additional insight into performance. Furthermore, MUE allows for unbiased comparisons between predictive models as it is less sensitive to dataset size. Other metrics such as Gaussian Random Affinity Model (GRAM)~\cite{cui2020gram}, Predictive Interval (PI) and Relative Absolute Error (RAE), attempt to correct for the inherent potency range of a dataset, which can aid in comparing success between different targets. We recommend further reading on evaluation of computational models~\cite{jain2008recommendations, walters2013what, brown2009healthy, walterthoughts}.

Reporting the results of relative free energy calculations requires care. As shown in Fig.~\ref{fig:fig_types_of_networks}, relative free energies can be performed arbitrarily as a forward or a reverse process, and thus relative free energies may be reported as either positively or negatively valued. The consequence of the two possible signs for relative free energies is that correlation statistics (such as Pearson's R and Kendall \texttau{}) can be skewed depending on which sign is analysed. The issue of this inconsistency can be circumvented by either plotting all datapoints within a consistent quadrant~\cite{perez-benito2019predicting}, or by avoiding the use of correlation statistics for assessment of relative free energy calculations and instead measuring accuracy using RMSE and MUE which are unaffected by choice of sign.

\begin{figure}
  \includegraphics[width=0.95\linewidth]{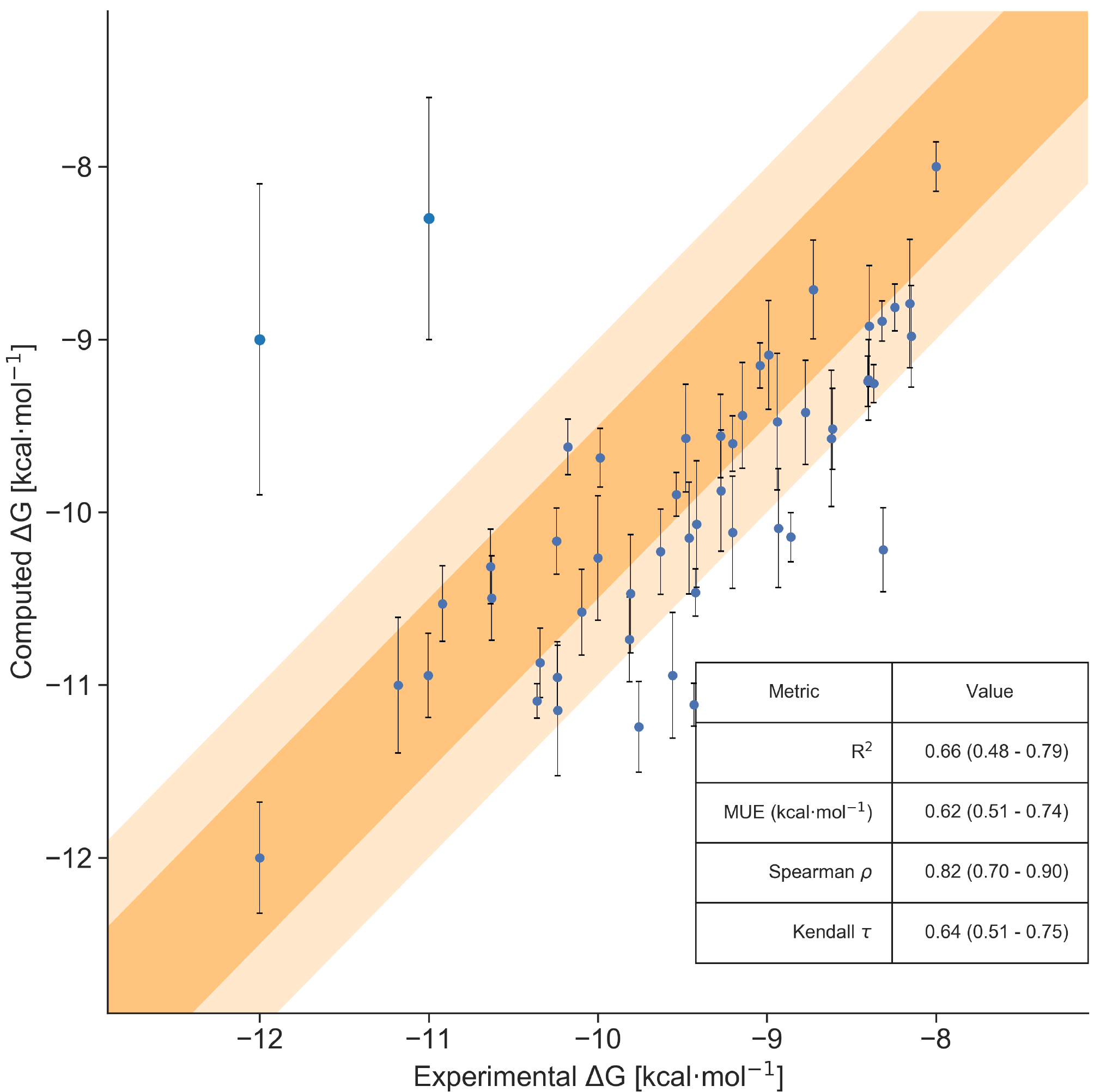}
  \caption{\textbf{An example of recommended practices for graphing alchemical free energy predictions.} This figure shows the relation between predicted and experimentally-determined Gibbs free energy in kcal/mol with standard errors as error bars. The dark and light-orange regions depict the 1- and 2-kcal/mol confidence bounds. Statistical metrics for the data are reported, with 95\% confidence intervals determined by bootstrapping analysis. Extra care should be taken when investigating potential outliers further.}
   \label{fig:scatterplot_analysis}
\end{figure}

\paragraph{Presenting your data}
As essentially all alchemical free energy prediction schemes are regression problems, the preferred type of plot is a scatter plot (see Fig.~\ref{fig:scatterplot_analysis}). Most alchemical free energy projects will look at 10-50 ligands; any study with \textless10 ligands is more suitable for bar plots (with inclusion of error bars), and will unlikely provide meaningful statistics. Any study with \textgreater50 ligands often contains multiple protein targets to which alchemical free energies may perform better on some targets than others. Because of this, it is bad practice to place multiple datasets on the same plot as this can suggest high model accuracy even though the individual models perform less well~\cite{walterthoughts}.

As we are interested mainly in the linear relationship between the alchemical free energy predictions and the experimentally -determined affinity values, plots should be depicted with the same range on both axes (i.e. $x=y$) with a 1:1 aspect ratio, with units for both experiment and simulation converted to be the same. If this skews the plot to a point where it is difficult to read of information, using the same dimensions, such that e.g. 1 cm is 1 kcal/mol is acceptable. Furthermore, bounds should be depicted for the 1- and 2-kcal/mol confidence regions. These regions can serve as tools to communicate your model performance: any predictions inside the 1 kcal/mol region can be seen as highly reliable, any predictions inside the 2 kcal/mol region should be seen as somewhat reliable, and any predictions outside the confidence regions should be expected to be unreliable (and handled as outliers). In a drug discovery context, this type of data depiction may suggest the reliability of alchemical FE predictions in the project, and can give an idea of how trustworthy predictions can be for synthesis ideas. It is also recommended to included experimental error bars in all plots.

An example of a best practice scatter comparison between computed and experimental values is shown in Fig.~\ref{fig:scatterplot_analysis}, highlighting outliers, error bars and confidence intervals. The data for this plot is artificially generated for illustration purposes.

\section{Conclusions}
\label{sec:conclusion}
Alchemical free energy calculations have seen a vast increase in popularity both in academic research as well as pharmaceutical industry applications in structure based drug discovery~\cite{schindler2020largescale, sherborne2016collaborating, wagner2017computational}. Commercial products such as FEP+ and Flare, which provide a convenient user interface make the setup and use of these methods a lot easier~\cite{wang2015accurate, kuhn2020assessment}, but this convenience comes with less flexibility in terms of choice of simulation protocols. It is also important to understand the current limitations of the methodology to recognise when automated workflow tools can be used effectively for a given protein target and when they are likely to fail still. Prospective prediction challenges such as the Drug Design Data resource grand challenges provide a community driven platform to evaluate different free energy protocols against each other on blinded targets~\cite{gaieb2018d3r, gaieb2019d3r}. Such efforts have highlighted that selection of seemingly identical or similar potential energy function or simulation package does not guarantee production of similar free energies owing to differences in simulation protocols. 
We hope that the best practice guide provides a set of tools that allow a better understanding of how to setup, run, and reliably interpret alchemical free energy calculations.

\section{Selection of available software packages}

\label{sec:software}
There are many different software solutions available for the setup, running, and analysis of alchemcial free energy calculations. These will vary in customizability and ways in which they are ran, e.g. graphical user interface versus command line tool or python script. The following provides a non-exhaustive list of commercial and noncommercial tools available for conducting alchemical free energy calculations. 
\begin{itemize}

\item [] \textbf{Simulation software: Commercial}
   \begin{itemize}
    \item \href{https://www.schrodinger.com/fep}{FEP+} is a tool offered by Schr\"{o}dinger Inc. under a commercial license. It has an intuitive GUI which makes it easier for non-experts to run alchemical free energy calculations and analyze the results. It runs the DESMOND MD package under the hood and hence parallelizes well on GPUs~\cite{wang2015accurate}. 
    \item \href{https://www.cresset-group.com/software/flare/}{Flare} is a commercial structure-based drug design software offered by Cresset. Similar to FEP+ it has an easily accessible graphical user interface and strives to facilitate free energy calculations for non-experts while offering advanced users full control via a Python API. It only runs on GPUs, using CUDA or OpenCL~\cite{kuhn2020assessment}. It is build on top of the open source software packages Sire and BioSimSpace (cf. below).
    \item The molecular operating environment (\href{https://www.chemcomp.com/Products.htm}{MOE}) offered by the Chemical Computing Group (CCG) has a tool for performing free energy calculations. It is built on AMBER-TI (cf. below).
    \end{itemize}
    \item[]All the above tools also provide a convenient setup and analysis suite and are really a one in all product. 
\item [] \textbf{Simulation software: Free/low cost academic and Commercial}
	\begin{itemize}
	\item \href{https://www.charmm.org/}{CHARMM} has a variety of tools developed over the years. The PERT module can be used to define initial and final states and define the intermediate lambda points. FREN and BAR modules can be used to analyze the data after the MD run. Lambda-dynamics-based free energy calculation can be carried out using the BLOCK module.  
	\item \href{https://ambermd.org/}{AMBER}, including its new pmemd.cuda version supports free energy calculations~\cite{salomon-ferrer2013overview}. 
	\end{itemize}
\item [] \textbf{Simulation software: Open Source}
	\begin{itemize}
	\item \href{https://www.plumed.org/}{PLUMED} is an open source tool which enables the usage of a variety of MD engines. It is designed as a plugin for MD packages such that it analyzes the trajectory on the fly. It also offers a VMD based plugin for the computation of collective variables~\cite{bonomi2019promoting}.   	
	\item \href{https://biosimspace.org/}{BioSimSpace} is a free, open source, multiscale molecular simulation framework, written to allow computational modellers to quickly prototype and develop new algorithms for molecular simulation and molecular design~\cite{hedges2019biosimspace}. 
	\item \href{https://siremol.org/}{Sire} is a multiscale, molecular simulation framework that provides several applications, including SOMD, an MD/MC code for performing FEP calculations via an interface to OpenMM. 
	\item \href{http://getyank.org/latest/index.html}{YANK} is a tool developed by John Chodera and group on the top of OpenMM MD package. It allows the users to write their inputs in easy-to-use YAML format.
	\item \href{http://www.gromacs.org/}{GROMACS} is a molecular simulation package with a significant number of free energy methods implementations. The LiveCOMS GROMACS tutorial includes an example free energy calculation~\cite{lemkul2018From}.
	\item \href{http://pmx.mpibpc.mpg.de/instructions.html}{PMX}, an add-on to GROMACS, offers a mutation free energy calculation module\cite{abraham2015gromacs}.
	\item \href{https://github.com/qusers/Q6}{Q} is MD code for performing FEP calculations using a variety of force fields~\cite{aqvistjohan2017q6}. 
	\end{itemize}
\item[] \textbf{Setup tools:}
	\begin{itemize}
	\item \href{http://pmx.mpibpc.mpg.de/instructions.html}{PMX}: GROMACS \url{https://github.com/deGrootLab/pmx}.
	\item \href{https://github.com/MobleyLab/Lomap}{Lomap/Lomap2} : Relative alchemical transformation graph planning for setting up perturbation networks~\cite{liu2013lead}.
	\item \href{http://www.charmm-gui.org/}{CHARMM-GUI} is a web-based tool for setting up a variety of MD simulations. It can be used to generate CHARMM scripts for solvation and ligand-binding free energy calculations~\cite{jo2008charmmgui}.
	\item \href{https://github.com/qusers/qligfep}{QligFEP} offers robust and fast setup of FEP calculations for the software package Q~\cite{jespers2019qligfep}.
	\item ProtoCaller, a setup tool for the automation of Gromacs free energy calculations. \href{https://github.com/protocaller/ProtoCaller}{ProtoCaller}~\cite{suruzhon2020protocaller}
	\item \href{https://fesetup.readthedocs.io/en/latest/introduction.html}{FESetup} has been developed primarily to setup calculations in AMBER, GROMACS and SIRE~\cite{loeffler2015fesetup}
	\end{itemize}
\item []\textbf{Analysis tools:}
	\begin{itemize}

	\item Alchemlyb: Multipackage free energy analysis
	\url{https://github.com/alchemistry/alchemlyb}~\cite{daviddotson2020alchemistry}.
	\item pymbar: MBAR implementation, but have to roll your own analysis wrapper      
	\url{https://github.com/choderalab/pymbar} \cite{shirts2008statisticallya}.
	\item Arsenic: Standardising alchemical free energy analysis \url{https://github.com/openforcefield/Arsenic}
	\item Free Energy Workflows: Sire-specific free energy map analysis using weighted path averages \url{https://github.com/michellab/freenrgworkflows}.
	\end{itemize}
\item[] Generally, commercial software will offer more complete pipelines in which standalone analysis applications are not necessarily needed; free and open source packages often require manual analysis.
\end{itemize}

\section{Alchemical free energy datasets: an overview}
\label{sec:benchmark}
The following contains a non-exhaustive summary of alchemical free energy datasets that can serve as a starting point to review approaches or test new implementations. The field is moving towards a more standardised way of generating protein-ligand benchmark datasets and the progress of these efforts can be tracked here: \url{https://github.com/openforcefield/FE-Benchmarks-Best-Practices}. Currently lacking an exhaustive set of benchmark datasets, the review by Williams-Noonan et al.~\cite{williams-noonan2018free} contains an overview of recently published alchemical free energy studies. For comparison of FEP+ and Gromacs (using the AMBER99SB-ILDN and GAFF2 force field), cf. the recently published study by Pérez-Benito et al.~\cite{perez-benito2019predicting}.
An overview of further suggested benchmark sets can be found in the review by Mobley and Gilson~\cite{mobley2017predicting} or on \url{alchemistry.org}~\cite{alchemistry}. These include cyclodextrins, the Cytochrome C peroxidase (CCP) protein model binding site, thrombin and bromodomains as well as solvation benchmark sets~\cite{paliwal2011benchmark}. Please refer to table~\ref{tab:benchmarks}, for a small overview of datasets, what forcefields they used, and what the original study was it came from. 

\begin{table}
\caption{Selection of example datasets}
\begin{tabular}{cccc}
\textbf{Publication} & \textbf{Targets} & \textbf{Ligands} & \textbf{Force Field} \\
\hline
\multicolumn{4}{|c|}{D3R Grand Challenges~\cite{D3R}} \\
\hline
GC3~\cite{gaieb2019d3r} & 6 & 266 & various \\
GC2~\cite{gaieb2018d3r} & 1 & 102 & various \\
GC2015~\cite{gathiaka2016d3r} & 2 & 215 & various \\
\hline
\multicolumn{4}{|c|}{SAMPL Challenges~\cite{SAMPL}} \\
\hline
SAMPL6~\cite{rizzi2018overview} & 3 & 21 & various \\
SAMPL5~\cite{yin2017overview} & 3 & 22 & various \\
SAMPL4~\cite{muddana2014sampl4} & 2 & 23 & various \\
\hline
\multicolumn{4}{|c|}{Schrödinger Datasets} \\
\hline
FEP+ Dataset~\cite{wang2015accurate} & 8 & 199 & OPLS2.1 \\
FEP+ Dataset~\cite{harder2016opls3} & 8 & 199 & OPLS3 \\
FEP+ Dataset~\cite{roos2019opls3e} & 8 & 199 & OPLS3e \\
FEP+ Dataset~\cite{song2019using} & 8 & 199 & GAFF 1.8 \\
FEP+ Dataset~\cite{gapsys2020large} & 8 & 199 & various \\
FEP+ Dataset~\cite{kuhn2020assessment} & 8 & 199 & GAFF2.1 \\
Fragments ~\cite{steinbrecher2015accurate} & 8 & 96 & OPLS2.1 \\
Scaffold Hopping~\cite{wang2017accurate} & 6 & 21 & OPLS3 \\
Scaffold Hopping~\cite{kuhn2020assessment} & 6 & 21 & GAFF2.1 \\
Macrocycles~\cite{yu2017accurate} & 7 & 33 & OPLS3 \\
\hline
\multicolumn{4}{|c|}{Further Suggested Datasets} \\
\hline
Cucurbit[7]uril (CB7)~\cite{mobley2017predicting} & 1 & 15 & NA \\
Deep cavity cavitand~\cite{mobley2017predicting} & 2 & 19 & NA \\
T4 Lysozyme~\cite{mobley2017predicting} & 2 & 20 & NA \\
Merck set~\cite{MCompChem2019Sep} & 5 & 169 & OPSL3 \\

\hline
\label{tab:benchmarks}
\end{tabular}
\end{table}

\section{Checklist}
\label{sec:checklist}
\begin{Checklists*}
\begin{checklist}{ Know what you want to simulate}
    \textbf{Initial questions you should ask before you set up an alchemical free energy calculation using molecular dynamics simulations}
\begin{itemize}
    \item Do I understand the biology, chemistry and physics of my system?
    \item Have I properly prepared my protein and ligand systems?
    \item Does my system contain any structures that require custom parameters?
    \item What simulation protocol will provide the most evidence to answer my hypothesis?
    \item Are the projected computational expense and runtime realistic for my goals?
    \item Will my protocol be reproducible? 
    \item Will my statistics be reliable? If not, would more replicates solve the problem? 
    \item Can I open-source my data?
\end{itemize}
\end{checklist}

\begin{checklist}{Preparing your simulations}
\textbf{Steps to getting started setting up your alchemical free energy calculation}
\begin{itemize}
    \item Make sure you know why you have picked your (combination of) force field(s)
    \item Minimize your system
    \item Equilibrate your system with your choice of thermodynamic ensemble
    \item Check the stability of your system and whether it behaves the way you believe it should
\end{itemize}
\end{checklist}

\begin{checklist}{Running absolute simulations}
        \textbf{Steps to running your absolute alchemical free energy calculations}
\begin{itemize}
 \item Check your ligands have the same, biologically correct binding pose
        \item Make sure your \textlambda-scheduling is set appropriately
        \item Check if your ligands are discharging and decoupling correctly
        \item Set up your restraints correctly
        \item Make sure you subsample the data in your free energy estimation protocol
        \item Apply the appropriate correction terms
\end{itemize}
\end{checklist}

\begin{checklist}{Running relative simulations}
        \textbf{Steps to running your relative alchemical free energy calculations}
\begin{itemize}
   \item Check your ligands  have the same, biologically correct binding pose
        \item Make sure your λ-scheduling is set correctly
        \item Make sure your molecular transformations are realistic (1-5 heavy atoms for reliable computations)
        \item Generate a perturbation network by your method of choice; check whether you have enough cycle closures to check consistency in the results
        \item Check whether dummy atoms were assigned correctly
        \item Consider subsampling the data in your free energy estimation protocol
        \item Apply the 
        appropriate correction terms
\end{itemize}
\end{checklist}
\end{Checklists*}

\begin{Checklists*}
\begin{checklist}{How do I know which simulations are unreliable?}
    \textbf{Situations suggesting your relative alchemical free energy calculations have not run properly (assuming absence of experimental affinities)}
        \begin{itemize}
                \item Standard error (\textsigma) should not be \textgreater1 kcal·mol$^{-1}$ 
    \item Simulated systems have not converged - trajectories should be manually checked for consistency; other methods such as generating RMSD plots are also recommended
    \newline\newline\textit{Relative:}
    \item If you observe hysteresis in perturbations and incorrect cycle closures
    \item Energy differences \textgreater$\sim$15 kcal$\cdot$mol$^{-1}$  are likely unreliable
    \newline\newline\textit{Absolute:}
    \item Energies \textless$\sim$-15 kcal$\cdot$mol$^{-1}$  are likely unreliable
    \item The ligand has not sampled most of the intended region after the decoupling step
    \item The ligand is drifting out of the intended region after the decoupling step
        \end{itemize}
\end{checklist}

\begin{checklist}{Why are they not reliable?}
    \textbf{Suggestions for finding out why your alchemical free energy calculations may not be reliable}
\begin{itemize}
    \item Check again whether dummy atoms were assigned correctly
    \item Inspect the trajectories across the λ-schedule (particularly the endpoints) for problems described in the text
    \item Inspect the overlap matrices for lack of overlap
\end{itemize}
\end{checklist}

\begin{checklist}{Data Analysis}
    \textbf{Steps to analyzing your output data correctly}
\begin{itemize}
    \item Make sure you have run enough replicates to ensure statistical reliability (\textgreater3)
    \item Compute both correlation and ranking coefficients and ranking statistics (e.g. r, \textrho, MUE and \texttau)
    \item Include error bars in all your visual analyses
\end{itemize}
\end{checklist}
\end{Checklists*}
\clearpage

\section*{Author Contributions}
%
\textbf{ASJSM}: Coordinated the document, contributed to most sections, and co-designed Figs.~\ref{fig:fig_binding_thermodynamic_cycle},~\ref{fig:fig_topology},~\ref{fig:fig_mcss},~\ref{fig:fig_types_of_networks},~\ref{fig:fig_absolute_thermodynamic_cycle},~\ref{fig:scatterplot_analysis},and created Figs.~\ref{fig:fig_what_is_lambda},~\ref{fig:fig_what_is_alchemy},~\ref{fig:overlap},~\ref{fig:pmf} and replotted~\ref{fig:automatic-equilibration-detection} and~\ref{fig:fig_types_of_networks}.\\
\textbf{JS}: Created Figs.~ \ref{fig:fig_what_is_alchemy},~\ref{fig:fig_binding_thermodynamic_cycle},~\ref{fig:fig_topology},~\ref{fig:fig_mcss},~\ref{fig:fig_absolute_thermodynamic_cycle},~\ref{fig:scatterplot_analysis}, and an initial draft of~\ref{fig:fig_types_of_networks}. Wrote Sec.~\ref{sec:plot_data}, the checklist Sec.~\ref{sec:checklist}, and contributed to general formatting discussions and editing.\\
\textbf{MK}: Contributed to Sec.~\ref{sec:data_analysis}, provided the data for figure~\ref{fig:overlap}, compiled the dataset for Sec.~\ref{sec:benchmark} and helped edit the paper.\\
\textbf{DLM}: Contributed to the outline, drafted some of the sections, gave ideas on figures, and helped edit the paper.\\
\textbf{GT}: Contributed to Sec.~\ref{sec:intro} and~\ref{sec:drugdiscovery}, and helped edit the paper.\\
\textbf{AR}: Created figure~\ref{fig:freeenergytrajectories}, contributed to sections~\ref{sec:theory} and~\ref{sec:simulation_protocol_choice}, and helped edit the paper.\\
\textbf{MRS}: Helped create figure~\ref{fig:fig_what_is_lambda}, wrote Sec.~\ref{sec:important_path} describing choices for alchemical pathways and parts of~\ref{sec:data_analysis} on the analysis for free energy calculations. Reviewed and edited text throughout.\\
\textbf{LNN}: Helped write the simulation length, stopping conditions, and information saving section. Edited and reviewed alchemical path section.\\
\textbf{BA}: Helped write the uncertainty estimation, stopping conditions, and output analysis sections and created figure ~\ref{fig:convergence_forward_reverse}.\\
\textbf{JM}: Contributed to Sec.~\ref{subsec:reproducible},~\ref{sec:prerequisites},~\ref{sec:important_path},~\ref{subsec:estimators},~\ref{subsec:uncertainty}, and~\ref{sec:conclusion}\\
\textbf{JDC}: Wrote Sec.~\ref{sec:decorrelating-samples} and~\ref{sec:automatic-equilibration-detection} discussed structure and design of the whole document, suggested Figs.~\ref{fig:fig_what_is_alchemy} and~\ref{fig:fig_sampling_scheme} \\
\textbf{HX}: Contributed Sec.~\ref{subsec:accuracy}, to Sec.~\ref{sec:relative-fe-protocol}, and to Sec.~\ref{sec:are-they-good}.\\
\textbf{HBM} Contributed to Sec.~\ref{sec:plot_data} and Fig.~ \ref{fig:scatterplot_analysis} and helped edit the paper.\\
For a more detailed description of author contributions,
see the GitHub issue tracking and changelog at \githubrepository.

\section*{Other Contributions}
%
Julia E. Rice participated in the original discussion of the document at the Best Practices in Molecular Simulation Workshop Hosted by at NIST, Gaithersburg, MD, August 24th-25th, 2017

For a more detailed description of contributions from the community and others, 
see the GitHub issue tracking and changelog at \githubrepository.

\section*{Potentially Conflicting Interests}
JM is a current member of the Scientific Advisory Board of Cresset. 
MK is employed by Cresset who commercially distribute a software for performing alchemical free energy calculations. MRS is a Open Science Fellow and consultant for Silicon Therapeutics.

\section*{Funding Information}
ASJSM and JM acknowledge funding through an EPSRC flagship software grant: EP/P022138/1
MK and JM acknowledge funding through Innovate UK by KTP partnership 011120.
AR acknowledges partial support from the Tri-Institutional Program in Computational Biology and Medicine and the Sloan Kettering Institute.
HEBM acknowledges support from a Molecular Sciences Software Institute Investment Fellowship and Relay Therapeutics.
JDC is a current member of the Scientific Advisory Board of OpenEye Scientific Software and a consultant to Foresite Laboratories. A complete funding history for the Chodera lab can be found at \url{http://choderalab.org/funding}. DLM appreciates financial support from the National Institutes of Health (1R01GM108889-01, 1R01GM124270-01A1 and R01GM132386), and the National Science Foundation (CHE 1352608).

\section*{Author Information}
\makeorcid
\bibliography{alchemical,manual}



\end{document}